\documentclass[symmetry,article,accept,pdftex,oneauthor]{Definitions/mdpi} 

\firstpage{1} 
\pubvolume{1}
\issuenum{1}
\articlenumber{0}
\pubyear{2024}
\copyrightyear{2024}
\datereceived{ } 
\daterevised{ } 
\dateaccepted{ } 
\datepublished{ } 
\hreflink{https://doi.org/} 

\usepackage{bm}
\usepackage[cal=cm,scr=dutchcal]{mathalfa}
\DeclareFontEncoding{LGR}{}{}
\DeclareSymbolFont{sfgreek}{LGR}{cmss}{m}{n}
\DeclareMathSymbol{\sfgamma}{\mathord}{sfgreek}{`g}
\DeclareMathSymbol{\sfxi}{\mathord}{sfgreek}{`x}
\DeclareMathSymbol{\sfchi}{\mathord}{sfgreek}{`q}
\DeclareMathSymbol{\sfPi}{\mathord}{sfgreek}{`P}
\newcommand{\mymathscr}[1]{\mbox{$\mathscr{#1}$}}

\providecommand{\ADD}[1]{{\protect\color{black} #1}}
\providecommand{\ADDMORE}[1]{{\protect\color{black} #1}}

\Title{A Unified Perspective on Poincar\'e and Galilei Relativity: \\ II. General Relativity: A. Kinematics}

\TitleCitation{A Unified Perspective on Poincar\'e and Galilei Relativity: \\ II. General Relativity: A. Kinematics}

\Author{Christian Y. Cardall $^{1,\dagger}$\orcidA{}}

\AuthorNames{Christian Y. Cardall}

\AuthorCitation{Cardall, C. Y.}

\address{%
$^{1}$ \quad Physics Division, Oak Ridge National Laboratory, Oak Ridge, TN 37831-6354, USA; cardallcy@ornl.gov}

\corres{cardallcy@ornl.gov}

\abstract{Building on the first paper in this series (Paper~I), a unified perspective on Poincar\'e and Galilei physics in a 5-dimensional spacetime setting is further pursued by consideration of the kinematics of general relativity, with the gravitational dynamics to be addressed separately.
The metric of the 5-dimensional affine spacetimes governed by the Bargmann groups considered in Paper~I (central extensions of the Poincar\'e and Galilei groups) is generalized to curved spacetime by extending the usual 1+3 (traditionally `3+1') formalism of general relativity on 4-dimensional spacetime to a 1+3+1 formalism, whose spacetime kinematics is shown to be consistent with that of the usual 1+3 formalism. 
Spacetime tensor laws governing the motion of an elementary classical material particle and the dynamics of a simple fluid are presented, along with their 1+3+1 decompositions; these reference the foliation of spacetime in a manner that partially reverts the Einstein perspective (accelerated fiducial observers, and geodesic material particles and fluid elements) to a Newton-like perspective (geodesic fiducial observers, and accelerated material particles and fluid elements subject to a gravitational force). 
These spacetime laws of motion for particles and fluids also suggest that a strong-field Galilei general relativity would involve a limit in which not only $c \rightarrow \infty$, but also $G \rightarrow \infty$, such that $G / c^2$ remains constant.  
}


\begin{document}

\section{Introduction}
\label{sec:Introduction}

This paper continues the development of a unified perspective on Poincar\'e and Galilei relativity begun in the first paper in this series (hereafter Paper~I) \cite{Cardall2024A-Unified-Persp}.
As used here, `relativity' refers to the invariance of physical laws under the action of a symmetry group that mixes time and space.
The Poincar\'e group mixes time into space and space into time, whereas the Galilei group only mixes time into space while leaving time invariant.
The profound consequences of the Poincar\'e group's mixing of space into time---such as the dependence of the time interval between two events on the motion of the observer---are the phenomena traditionally labeled `relativistic'.
However, the dependence of the space interval between two events on the motion of the observer also renders physics governed by the Galilei group `relativistic' as far as space is concerned.
Therefore Paper~I begins by arguing that the traditional terms `non-relativistic physics' and `relativistic physics' should be replaced by the more precise terms `Galilei physics' (or `Galilei relativity') and `Poincar\'e physics' (or `Poincar\'e relativity') respectively.

Physics governed by the Poincar\'e group or the Galilei group both being `relativistic' in this sense, 
Paper~I argues further that the terms `special relativity' and `general relativity' ideally would be released from their traditional association with `physics according to Einstein' to denote more generally `physics on flat spacetime' and `physics on curved spacetime' respectively, regardless of whether physics is governed by the Galilei group or the Poincar\'e group---globally in the case of flat spacetime, but only locally in the case of curved spacetime.
Thus on flat spacetime one would speak of both `Galilei special relativity' and `Poincar\'e special relativity', the latter being `special relativity' as traditionally understood following Einstein.
On curved spacetime, one similarly might consider a `Galilei general relativity' alongside `Poincar\'e general relativity', the latter being `general relativity' as traditionally understood following Einstein.

The relativistic invariance of Poincar\'e physics is manifest, indeed automatic, when expressed in terms of equations governing tensor fields on 4-dimensional spacetime.
Einstein's proposal of Poincar\'e special relativity as the solution to theoretical and empirical puzzles posed by Maxwell's electrodynamics was formulated in terms of time-dependent fields on 3-dimensional position space.
Minkowski subsequently introduced the concept of spacetime as a flat 4-dimensional Lorentz manifold (pseudo-Riemann manifold with metric tensor of signature (-,+,+,+)) unifying time and space; this allowed him to express electrodynamics in terms of the electromagnetic 4-potential (a covector, or linear form) and electromagnetic field tensor (an antisymmetric bilinear form) unifying the electric and magnetic fields, and material particle dynamics in terms of the 4-momentum, whose vector version unifies mass with vector 3-momentum, and whose covector version unifies energy with covector 3-momentum.
In short order von Laue unified the energy density, energy flux, momentum density, and momentum flux of a material continuum in the energy-momentum 4-flux tensor, and this---along with Minkowski's introduction of spacetime---was key to Einstein's development of Poincar\'e general relativity in order to accommodate gravity as spacetime curvature.
(References to the original literature of the early 20th century can be found in historical notes in the relevant sections of \cite{Gourgoulhon2013Special-Relativ}.)

One can also try to shoehorn Galilei physics into a 4-dimensional spacetime perspective (e.g. \cite{de-Saxce2016Galilean-Mechan,Cardall2019Minkowski-and-G,Cardall2020Combining-3-Mom}, Paper~I, and historical references therein), but the fit is uneasy and incomplete.
The fit is uneasy because the spacetime that arises from the speed of light $c \rightarrow \infty$ limit of the Einstein metric and its inverse is not a Lorentz manifold: there is no spacetime metric, and therefore no metric duality of spacetime tensors (raising and lowering of indices); no Levi-Civita connection on spacetime (natural covariant derivative determined by the metric); and no Levi-Civita tensor on spacetime (natural volume form associated with the metric).  
The fit is incomplete because of the strict separation of mass and energy implied by Galilei physics: a tensor formalism on 4-dimensional spacetime makes manifest the relativistic invariance of conservation of matter and balance of 3-momentum, but not balance of energy.
A 4-velocity vector and a matter-momentum 4-flux tensor are natural enough, but there is no satisfactory 4-momentum covector or energy-momentum 4-flux tensor.
The root of the problem is that transformations of time and space yield corresponding transformations of inertia and  3-vector momentum; only by metric duality do these directly correspond also to transformations of energy and  3-covector momentum.
Both Galilei physics and Poincar\'e physics do include a position space 3-metric that relates 3-vector momentum to 3-covector momentum, so that balance of 3-momentum can be expressed indifferently in terms of either.
And the spacetime metric of Poincar\'e physics implies the equivalence of mass and energy by which the relativistic invariance of balance of inertia is also the relativistic invariance of balance of energy.
But Galilei transformations leave the mass of a material particle invariant, and do not directly exhibit the transformation of kinetic energy implied by the transformation of its 3-momentum.
And for a material continuum, Galilei transformations do not allow the first law of thermodynamics to be integrated into a mass-momentum 4-flux tensor, as happens in Poincar\'e physics. 
Without a spacetime metric and in failing to manifestly include energy, the practical and aesthetic appeal of a spacetime tensor formalism is significantly compromised in the case of Galilei physics on a 4-dimensional spacetime manifold.

However, a tensor formalism for Galilei physics that manifestly exhibits the transformation of kinetic energy can be accommodated in a 5-dimensional spacetime setting governed by the Bargmann group, a central extension of the Galilei group (e.g. \cite{de-Saxce2012Bargmann-group-,de-Saxce2016Galilean-Mechan,de-Saxce20175-Dimensional-T}, Paper~I, and references therein).  
The Bargmann group emerged in connection with representation theory in quantum mechanics, with the Galilei group as a notable example (\cite{Bargmann1954On-Unitary-Ray-}, see also e.g. \cite{Levy-Leblond1971Galilei-Group-a,Levy-Leblond1974The-pedagogical,Levy-Leblond1976Quantum-fact-an}).
Still with a focus on quantum physics, the Bargmann group inspired a tensor formalism for Galilei physics on a 5-dimensional extended spacetime \cite{Pinski1968Galilean-Tensor,Omote1989Galilean-Covari}, and its relevance for classical (i.e. non-quantum) Galilei physics was subsequently recognized as well \cite{Duval1985Bargmann-struct}.
\ADD{It is worth emphasizing at the outset that in this well-established reformulation of Galilei relativity, the extended spacetime introduces no new physical degrees of freedom, and serves only as a mathematical device to better express Galilei physics in a tensor formalism.}

In part---but only in part---the move from a 4-dimensional spacetime setting to a 5-dimensional extended spacetime setting is conceptually similar to the move from a 3-dimensional position space setting to a 4-dimensional spacetime setting.
Consider a classical material particle with no internal degrees of freedom, described completely by its position $(x(t), y(t), z(t))$ in 3-dimensional position space as a function of time $t$ according to a fiducial (that is, Eulerian or `lab frame') observer.
These points trace out the particle's trajectory in position space, a (not necessarily injective, or 1-to-1) curve parametrized by $t$ with tangent vector field $\bm{v}$, the coordinate 3-velocity.
Described in terms of 4-dimensional spacetime, the particle's location $(t(\tau), x(\tau), y(\tau), z(\tau))$ according to a fiducial observer can be given as a function of the proper time $\tau$ measured by a comoving (that is, Lagrangian or `material frame') observer moving along with the particle.
These points in spacetime trace the particle's worldline, a (now definitely injective) curve parametrized by $\tau$ with tangent vector field $\bm{U}$, the 4-velocity.
But while the particle is now regarded as a `history' in 4-dimensional spacetime, from a kinematical and dynamical perspective it is still characterized by only three degrees of freedom.
Thus the 4-velocity $\bm{U}$, and the worldline it determines, are subject to a constraint: $\bm{g}\left( \bm{U}, \bm{U} \right) = -c^2$ in the case of Poincar\'e physics, where $\bm{g}$ is the Einstein metric; or $\bm{\tau}\left( \bm{U} \right) = 1$ in the case of Galilei physics, where $\bm{\tau}$ (not to be confused with proper time $\tau$, a scalar) is the time 1-form normal to the spacelike position-space leaves associated with absolute time.
In a 5-dimensional extended spacetime motivated by the Bargmann group, the particle location $(t(\tau), x(\tau), y(\tau), z(\tau), \eta(\tau))$ now includes an additional `action coordinate' $\eta$ related to kinetic energy per unit mass, in such a way that, not only for Poincar\'e physics but now also for Galilei physics, the extended spacetime is a pseudo-Riemann (indeed, Lorentz) manifold with metric $\bm{G}$.
In this new setting the forms $\bm{g}$ (for Poincar\'e physics) and $\bm{\tau}$ (for Galilei physics) are still invariant structures governing causality, and the same constraint $\bm{g}\left( \bm{\mathscr{U}}, \bm{\mathscr{U}} \right) = -c^2$ or $\bm{\tau}\left( \bm{\mathscr{U}} \right) = 1$ applies, where $\bm{\mathscr{U}}$ is the 5-velocity tangent to the particle worldline in Bargmann-extended spacetime. 
The Bargmann metric $\bm{G}$ governs the extended spacetime geometry, and also provides an additional constraint $\bm{G}\left( \bm{\mathscr{U}}, \bm{\mathscr{U}} \right) = 0$ ensuring that the particle continues to be characterized by only three degrees of freedom. 
But unlike the time coordinate $t$ associated with the move to 4-dimensional spacetime, the additional coordinate $\eta$ associated with the move to extended 5-dimensional spacetime is afforded no independent physical significance, and no explicit field dependence on it is allowed (all partial derivatives with respect to $\eta$ vanish).
The utility of the additional dimension in the present context is purely to allow Galilei physics to be expressed in a manifestly invariant formalism in terms of spacetime tensor fields on a pseudo-Riemann manifold. 

The purpose of this series is to develop a more unified perspective on Poincar\'e and Galilei relativity, including by exploring the possibility of a strong-field Galilei general relativity that could serve as a useful approximation in astrophysical scenarios such as core-collapse supernovae.
The basic strategy is to reexpress standard Poincar\'e physics on 4-dimensional spacetime in a 5-dimensional setting more congenial to Galilei physics, and then deduce the corresponding Galilei-invariant theory by taking the $c \rightarrow \infty$ limit.
Focusing on flat spacetime, Paper~I took a first step by elucidating and emphasizing the fact that the Poincar\'e group can also be centrally extended to what might be called the Bargmann-Poincar\'e group, in a manner analogous to the more familiar central extension of the Galilei group to what might now be called the Bargmann-Galilei group (traditionally simply the `Bargmann group').

This installment in the series picks up where Paper~I left off: Sec.~\ref{sec:Metric} (re)introduces a curved spacetime version of the Bargmann metric $\bm{G}$, derived using a procedure similar to that employed in Paper I in flat spacetime, but using the 1+3 (traditionally `3+1') formalism of Poincar\'e general relativity as the starting point.
\ADD{(Useful as it may be for the purpose of numerically solving the Einstein equations as an initial value problem, a 1+3 spacetime foliation of course is not fundamental to a spacetime perspective on Poincar\'e physics. 
But the absolute time of Galilei physics \textit{does} require a 1+3 spacetime foliation, making it necessary as a common setting enabling a more unified perspective on Poincar\'e and Galilei physics.)}
The projection operator $\overleftarrow{\bm{\gamma}}^*$ needed for 1+3 and 1+3+1 tensor decompositions is the subject of Sec.~\ref{sec:Decomposition}, after which Sec.~\ref{sec:LeviCivita} relates the spacetime Levi-Civita connections $\bm{\nabla}$ (associated with the 4-metric $\bm{g}$) and $\bm{\mathscr{D}}$ (associated with the 5-metric $\bm{G}$) to the position space Levi-Civita connection $\bm{D}$ associated with the 3-metric $\bm{\gamma}$. 

The gravitational `kinematics' referred to in the title of this paper comes into focus in Sec.~\ref{sec:ExtrinsicGeometry}, where it is demonstrated that the extrinsic curvature tensor $\bm{K}$ of the usual 1+3 formalism of Poincar\'e relativity (e.g. \cite{Gourgoulhon201231-Formalism-in}) is all that is needed to also completely characterize the extrinsic geometry of the position space spacelike leaves in the 1+3+1 foliation of the Bargmann spacetimes considered here.
The label `kinematics' refers to the fact that the usual 1+3 formalism of Poincar\'e relativity locates the gravitational degrees of freedom in the 3-metric $\bm{\gamma}$, with the extrinsic curvature $\bm{K}$ describing the evolution of $\bm{\gamma}$ between neighboring position space leaves, thus constituting a kind of `velocity' of the gravitational degrees of freedom. 
The gravitational `dynamics'---the Einstein equations relating the spacetime metric to the energy-momentum content on spacetime, and the evolution of $\bm{K}$ emerging therefrom (in effect, the `acceleration' of the gravitational degrees of freedom)---will be addressed in the next paper in this series.

The remaining sections in the present installment serve as preparation for the gravitational dynamics to be considered in the sequel, and also provide an initial application of the geometry of the curved Bargmann spacetimes elucidated here to the motion of an elementary material particle and a simple fluid.
As mentioned above, the initial move in the overarching strategy employed in this series is to reexpress standard Poincar\'e physics on 4-dimensional spacetime in a 5-dimensional setting more congenial to Galilei physics.
Key to this `encoding' of familiar 4-dimensional physics in a 5-dimensional setting is a `decoding' operator $\overleftarrow{\bm{g}}^*$ introduced in Sec.~\ref{sec:PhysicalLawsBargmann}.
Its utility for obtaining tensor laws on 5-dimensional Bargmann spacetime from tensor laws on 4-dimensional spacetime is illustrated in Sec.~\ref{sec:DynamicsParticles} on the dynamics of an elementary particle and Sec.~\ref{sec:DynamicsFluids} on the dynamics of a simple fluid.
The latter section introduces the kinetic-energy--momentum--mass-density 5-flux tensor $\bm{\mathscr{T}}$, 
the 5-dimensional encoding of the total-energy--momentum 4-flux tensor $\bm{T}$; this can be expected to appear in the Einstein equations on the 5-dimensional Bargmann spacetimes to be considered in a subsequent installment.
Along with a concluding summary, Sec.~\ref{sec:Conclusion} includes remarks about the relationships between this work and previous instantiations of Newton gravity in a 5-dimensional setting \cite{Duval1985Bargmann-struct,de-Saxce2016Galilean-Mechan,de-Saxce20175-Dimensional-T}, and generalizations of Newton-Cartan gravity in 4-dimensional spacetime in which the connection includes torsion (e.g. \cite{Geracie2015Curved-non-rela,Van-den-Bleeken2017Torsional-Newto,van-den-Bleeken2022Torsional-Newto,Hansen2019Action-Principl,Cariglia2018General-theory-,Hansen2020Non-relativisti}).

\section{The Metric of Foliated Curved Spacetimes}
\label{sec:Metric}

Consider the metric components associated with a foliation of spacetime into spacelike `position space' leaves.
These are shown first for the usual Einstein spacetime, and then for Bargmann spacetimes.
In addition to the 3-metric on each leaf of a foliation, the lapse function and shift vector relating neighboring leaves are interpreted in terms of vector fields tangent to the worldlines of `fiducial observers', which are everywhere normal to the leaves of the foliation.
The classical (i.e. non-quantum) kinematics of a free material particle is also summarized, as this enables a simple derivation of the Bargmann metric in curved spacetime.

\subsection{The Metric of Foliated Einstein Spacetime $\mathcal{E}$}
\label{sec:MetricE}

An `Einstein spacetime' $\mathcal{E}$ is a 4-dimensional Lorentz manifold, a pseudo-Riemann manifold endowed with a metric $\bm{g}$ of signature $(-,+,+,+)$.
This is the setting of Poincar\'e general relativity as traditionally understood.
Assume that a spacetime $\mathcal{E}$ is such that it admits a Cauchy surface, a spacelike hypersurface (i.e. a submanifold of codimension 1, and therefore of dimension 3) such that each timelike or null curve intersects the surface only once. 
Then $\mathcal{E}$ admits a foliation into a family $\left( \mathcal{S}_t \right)_{t \in \mathbb{R}}$ of spacelike hypersurfaces, where each leaf or slice $\mathcal{S}_t$ of the foliation is a level surface of a scalar field $t$ (e.g. \cite{Gourgoulhon201231-Formalism-in}).
That is, there exists an atlas of $\mathcal{E}$ in which $t$ serves as a global time coordinate for every chart (coordinate patch) in the atlas.
For convenience, let $\mathcal{S}$ denote $\mathcal{S}_t$ for some value of $t$.
Let $(U, X)$ be a chart on  
$\mathcal{E}$,
with $U$ an open subset of $\mathcal{E}$,
and $X  = ( t, x )$ coordinates adapted to the foliation.
Then $(U_\mathcal{S}, x)$ is a chart on $\mathcal{S}$, where $U_\mathcal{S} = U \cap \mathcal{S}$ and $x = (x^1, x^2, x^3)$ are local position space coordinates.
The time coordinate index is $0$, that is, $X^0 = t$.

Write the coordinate basis vector fields associated with the coordinates $X$ as $\left( \bm{\partial} X_\nu \right) = \left( \bm{\partial} t, \bm{\partial} x_j \right)$, and its dual basis of 1-forms as $\left( \bm{\mathrm{d}} X^\mu \right)  = \left( \bm{\mathrm{d}} t, \bm{\mathrm{d}} x^i \right)$.
Here the notation $\bm{\partial} X_\nu = \partial / \partial{X^\nu}$ for a coordinate basis vector field is introduced as a simplified alternative intended to be more visually parallel to the standard notation $\bm{\mathrm{d}} X^\mu$ for a basis 1-form (the latter being also the exterior derivative of the coordinate function $X^\mu$).
Greek indices take values in $\{0,1,2,3\}$ (the conventional spacetime coordinate indices), with letters $\mu, \nu, \dots $ near the middle of the alphabet preferred for free indices, and letters $\alpha, \beta, \dots $ near the beginning of the alphabet preferred for dummy indices.
Lower-case latin indices take values in $\{1,2,3\}$ (the position space coordinate indices), with letters $i, j, \dots $ near the middle of the alphabet preferred for free indices, and letters $a, b, \dots $ near the beginning of the alphabet preferred for dummy indices.

In the 1+3 formalism of Poincar\'e general relativity associated with the foliation of spacetime $\mathcal{E}$ into spacelike hypersurfaces $\mathcal{S}_t$ (traditionally known as the `3+1' formalism, e.g. \cite{Gourgoulhon201231-Formalism-in}), the components of the metric $\bm{g}$ with respect to the coordinates $X$ are given by
\begin{equation}
\mathsf{g} = \left[ g_{\mu\nu} \right] =
	\begin{bmatrix}
		-c^2 \, \alpha^2 + \beta_a \beta^a &  \beta_j  \\[5pt]
		\beta_i & \gamma_{ij}
	\end{bmatrix}
	\quad \quad (\mathrm{on \ } \mathcal{E}), 
\label{eq:Metric4}
\end{equation}
where $c$ is the speed of light, while the inverse matrix
\begin{equation}
\overleftrightarrow{\mathsf{g}} = \left[ g^{\mu\nu} \right] =
	\begin{bmatrix}
		-\frac{1}{c^2  \alpha^2} & \frac{1}{c^2  \alpha^2} \, \beta^j  \\[5pt]
		 \frac{1}{c^2  \alpha^2} \, \beta^i 
				& \gamma^{ij} -  \frac{1}{c^2  \alpha^2} \, \beta^i \beta^j
	\end{bmatrix}
	\quad \quad (\mathrm{on \ } \mathcal{E})
\label{eq:InverseMetric4} 
\end{equation}
collects the components of the inverse metric $\overleftrightarrow{\bm{g}}$.
The determinant $g$ of the matrix $\mathsf{g}$ of metric components is
\begin{equation}
g = \det \mathsf{g} = - c^2 \, \alpha^2 \, \gamma,
\label{eq:MetricDeterminant4}
\end{equation}
where $\gamma$ is the determinant of $\left[ \gamma_{i j} \right]$.
Notably, a $c \rightarrow \infty$ limit of $\bm{g}$ and its matrix determinant $g$ do not exist due to the presence of $c^2$ in $g_{tt} = g_{00}$, precluding a Galilei general relativity featuring a 4-dimensional Lorentz manifold as its spacetime.
However, the inverse metric $\overleftrightarrow{\bm{g}}$ of Poincar\'e physics limits sensibly to the degenerate inverse `metric' $\overleftrightarrow{\bm{\gamma}}$ as $c \rightarrow \infty$, the latter being a viable (and indeed fundamental) tensor object in Galilei physics, as discussed at length for instance in Paper~I. 

The meaning of the functions $\gamma_{ij}$ in Eq.~(\ref{eq:Metric4}) follows from the observation that the metric $\bm{g}$ on $\mathcal{E}$ induces a 3-metric $\bm{\gamma}$ on each leaf $\mathcal{S}$ of the spacetime foliation.
That is, $\bm{\gamma}(\bm{u}, \bm{v}) = \bm{g}(\bm{u}, \bm{v})$ for vector fields $\bm{u}, \bm{v}$ tangent to $\mathcal{S}$.
The leaves $\mathcal{S}$ are spacelike, that is, the signature of the metric $\bm{\gamma}$ on $\mathcal{S}$ is $(+,+,+)$.
Thus each leaf $\mathcal{S}$ of the foliation is a Riemann manifold in its own right.
The components of the inverse metric $\overleftrightarrow{\bm{\gamma}}$ on $\mathcal{S}$ are $\gamma^{i j}$, appearing in Eq.~(\ref{eq:InverseMetric4}).

The functions $\alpha$ and $\beta_j$ appearing in the components of the metric $\bm{g}$ in Eq.~(\ref{eq:Metric4}) have to do with the relationship between neighboring spacelike leaves of the foliation of $\mathcal{E}$.
A key idea illuminating this relationship is the notion of fiducial observers, whose worldlines orthogonally thread the leaves $\mathcal{S}$ of the foliation.
To discuss this, adopt notation for metric duality used also in Paper~I: in particular, $\underline{\bm{v}} = \bm{g} \cdot \bm{v}$ is the 1-form associated by metric duality with a vector field $\bm{v}$, and $\overleftarrow{\bm{\omega}} = \overleftrightarrow{\bm{g}} \cdot \bm{\omega} = \bm{\omega} \cdot \overleftrightarrow{\bm{g}} = \overrightarrow{\bm{\omega}}$ is the vector field associated by metric duality with the 1-form $\bm{\omega}$.
The dot operator only denotes contraction via an obvious `pairing of covariant and contravariant indices', and not the scalar product of two vectors; the latter will be written only in terms of a metric.

Begin with the gradient 1-form $\bm{\nabla} t$, which will lead to the fiducial observer vector field $\bm{n}$ and the dual fiducial observer vector field $\bm{\chi}$.
(For scalar fields, the Levi-Civita connection $\bm{\nabla}$ associated with $\bm{g}$ is simply the exterior derivative $\bm{\mathrm{d}}$.) 
This gradient 1-form is normal to $\mathcal{S}$ in the sense that $\bm{\nabla} t \cdot \bm{v} = 0$ for any vector field $\bm{v}$ tangent to $\mathcal{S}$.
Define the fiducial observer 1-form
\begin{equation}
\underline{\bm{n}}  = - c^2 \, \alpha \, \bm{\nabla} t = - c^2 \, \alpha \, \bm{\mathrm{d}} t, \quad \quad
	\underline{\mathsf{n}} = \left[ n_\nu \right] = \begin{bmatrix} - c^2 \, \alpha & 0_j \end{bmatrix}
	\quad \quad (\text{on } \mathcal{E}).
\label{eq:FiducialObserverCovectorE} 
\end{equation}
Define also the dual fiducial observer 1-form
\begin{equation}
\underline{\bm{\chi}} = - \frac{1}{c^2} \, \underline{\bm{n}} 
\quad \quad (\text{on } \mathcal{E}),
\label{eq:DualCollinearE}
\end{equation}
with
\begin{equation}
\underline{\bm{\chi}}  = \alpha \, \bm{\nabla}t = \alpha \, \bm{\mathrm{d}}t, \quad \quad
	\underline{\sfchi} = \left[ \chi_\nu \right] = \begin{bmatrix} \alpha & 0_j  \end{bmatrix}
	\quad \quad (\text{on } \mathcal{E}).
\label{eq:DualFiducialObserverCovectorE}
\end{equation}
Raising the index by contraction with $\overleftrightarrow{\bm{g}}$ yields the fiducial observer vector field
\begin{equation}
\bm{n} = - c^2 \, \alpha \, \overleftarrow{\bm{\nabla} t} 
	= \frac{1}{\alpha} \left( \bm{\partial}t - \beta^a \, \bm{\partial} x_a \right), \quad \quad 
	\mathsf{n} = \left[ n^\mu \right] = \begin{bmatrix} \frac{1}{\alpha} \\[5pt] -\frac{1}{\alpha}\beta^i  \end{bmatrix}
	\quad \quad (\text{on } \mathcal{E}),
\label{eq:FiducialObserverVectorE}
\end{equation}
manifestly normal to $\mathcal{S}$ with respect to $\bm{g}$ since it derives from $\bm{\nabla} t$, and the dual fiducial observer vector field
\[
\bm{\chi} = \alpha \, \overleftarrow{\bm{\nabla}t} =
		- \frac{1}{c^2 \, \alpha} \left( \bm{\partial}t - \beta^a \, \bm{\partial} x_a \right), 
\quad \quad \quad
\sfchi = \left[ \chi^\mu \right] = 
		\begin{bmatrix} -\frac{1}{c^2 \, \alpha} \\[5pt] \frac{1}{c^2 \, \alpha}\beta^i  \end{bmatrix} 
			\quad \quad (\text{on } \mathcal{E}).
\]
The name `dual fiducial observer 1-form' for $\underline{\bm{\chi}}$ comes from the fact that 
\[
\underline{\bm{\chi}} \cdot \bm{n} = 1.
\]
The observer vector field $\bm{n}$ for $\mathcal{E}$ in Eq.~(\ref{eq:FiducialObserverVectorE}) is interpreted as the 4-velocity field of the congruence of curves comprising the worldlines of the fiducial observers, and indeed it satisfies the normalization 
\[
\bm{g} \left( \bm{n}, \bm{n} \right) = \underline{\bm{n}} \cdot \bm{n} = - c^2 
\]
expected of a 4-velocity.

Introduce also a `normal evolution vector field' collinear with the fiducial observer vector field:
\begin{equation}
\bm{m} = \alpha \, \bm{n},
\label{eq:NormalEvolutionVectorE}
\end{equation}
which satisfies
\[
\bm{\nabla}t \cdot \bm{m} = \alpha \, \bm{\nabla}t \cdot \bm{n} = 1
\]
as verified from the component expressions above.
Under a displacement $\delta t \, \bm{m}$ for infinitesimal $\delta t$, a leaf $\mathcal{S}_t$ is carried (or `Lie dragged') to $\mathcal{S}_{t + \delta t}$: for any point $\mathscr{p} \in \mathcal{S}_t$,
\[
t ( \mathscr{p} + \delta t \, \bm{m} ) = t(\mathscr{p}) + ( \bm{\nabla}t \cdot \bm{m} ) \, \delta t  = t(\mathscr{p}) + \delta t.
\]
This is the reason $\bm{m}$ is called the normal evolution vector field. 

Together with the tensor $\bm{g}$ that governs proper time on $\mathcal{E}$, the normal evolution vector field $\bm{m}$ yields the meaning of the lapse function $\alpha$.
Consider the proper time increment $\delta \tau$ corresponding to a displacement $\delta t \, \bm{m}$ on $\mathcal{E}$:
\[
c \, \delta \tau = \sqrt{ - \bm{g} ( \, \delta t \, \bm{m}, \, \delta t \, \bm{m} \, ) } = c \, \alpha \, \delta t.
\]
Thus the lapse function $\alpha$ relates a time coordinate interval between two neighboring points encountered by a fiducial observer to the proper time measured by that observer.

The meaning of the shift vector components $\beta^i$ follows from rewriting Eq.~(\ref{eq:FiducialObserverVectorE}) as
\begin{equation}
\bm{\partial}t = \bm{m} + \bm{\beta},
\label{eq:ShiftVector}
\end{equation}
where $\bm{\beta} = \beta^a \, \bm{\partial} x_a$.
Because $\bm{\nabla} t = \bm{\mathrm{d}}t$ is an element of the 1-form basis dual to the coordinate basis (so that $\bm{\nabla}t \cdot \bm{\partial} t = 1$), the vector field $\bm{\partial} t$ is, like $\bm{m}$, a time evolution vector field by which neighboring spacelike leaves $\mathcal{S}$ are Lie-dragged from one to another.
But unlike $\bm{m}$, the vector field $\bm{\partial} t$, aligning at each point with the local $t$ coordinate axis, is not in general normal to the leaves $\mathcal{S}$.
Consider a point $\mathscr{p}$ in slice $\mathcal{S}_t$, along with points $\mathscr{p} + \delta t \, \bm{m}$ and $\mathscr{p} + \delta t \, \bm{\partial} t$ in neighboring slice $\mathcal{S}_{t + \delta t}$.
Then according to Eq.~(\ref{eq:ShiftVector}), the vector $\delta t \, \bm{\beta}$ points from $\mathscr{p} + \delta t \, \bm{m}$ to $\mathscr{p} + \delta t \, \bm{\partial} t$ in $\mathcal{S}_{t + \delta t}$.
Thus the shift vector field $\bm{\beta}$, everywhere tangent to the leaves $\mathcal{S}$, is the coordinate 3-velocity with which points of constant coordinate position $\left( x^i \right)$ move relative to the fiducial observers.
The 1-form components $\beta_j$ are given by $\beta_j = \gamma_{j a} \, \beta^a$.

Consider next a material particle on $\mathcal{E}$, whose coordinates along its worldline are $X(t) = (t, x(t) )$.
Its 4-velocity is
\[
\begin{aligned}
\bm{U} = \frac{\mathrm{d}}{\mathrm{d}\tau} 
	&= \frac{\mathrm{d}t}{\mathrm{d}\tau} \frac{\partial}{\partial t} 
		+ \frac{\mathrm{d}t}{\mathrm{d}\tau} \frac{\mathrm{d} x^a}{\mathrm{d} t} 
			\frac{\partial}{\partial x^a} \\[5pt]
	&= \frac{\mathrm{d}t}{\mathrm{d}\tau} \left( \bm{\partial} t + \bm{v} \right),
\end{aligned}
\]
where $\mathrm{d}\tau$ is an increment of proper time along the worldline, and $v^i = \mathrm{d} x^i / \mathrm{d} t$ are the components of the coordinate 3-velocity $\bm{v} = v^a \, \bm{\partial}x_a$.
The increment $\mathrm{d}\tau$ is related to the coordinate increments $\mathrm{d}X^\mu$ by the line element along the worldline:
\[
\begin{aligned}
-c^2 \, \mathrm{d}\tau^2 
	&= g_{\alpha\beta} \, \mathrm{d}X^\alpha \, \mathrm{d}X^\beta \\[5pt]
	&= -c^2 \, \alpha^2 \, \mathrm{d}t^2 
		+ \gamma_{ab} ( \mathrm{d}x^a + \beta^a \mathrm{d}t )
			( \mathrm{d}x^b + \beta^b \mathrm{d}t ) \\[5pt]
	&= -c^2 \, \alpha^2 \, \mathrm{d}t^2
		\left( 1 - \frac{1}{c^2} \, \bm{\gamma}(\bm{V},\bm{V}) \right)
\end{aligned}
\]
where
\begin{equation}
\bm{V} = V^a \, \bm{\partial} x_a = \frac{1}{\alpha} \left( \bm{v} + \bm{\beta} \right)
\label{eq:ThreeVelocity}
\end{equation}
is tangent to $\mathcal{S}$.
Defining $\Lambda_{\bm{V}}$ by
\[
\Lambda_{\bm{V}}^{-1} = \sqrt{1 - \frac{1}{c^2} \, \bm{\gamma}(\bm{V},\bm{V})}
\]
and substituting into the line element along the worldline gives
\begin{equation}
\Lambda_{\bm{V}} = \frac{\alpha \, \mathrm{d}t}{\mathrm{d}\tau}.
\label{eq:dtdtau}
\end{equation}
Using this together with Eq.~(\ref{eq:FiducialObserverVectorE}) in the above expression for the 4-velocity $\bm{U}$ then yields
\[
\bm{U} = \Lambda_{\bm{V}} \left( \bm{n} + \bm{V} \right), \quad \quad
\mathsf{U} = \left[ U^\mu \right] = 
		 \begin{bmatrix} \frac{1}{\alpha} \, \Lambda_{\mathsf{V}} \\[5pt]
			\Lambda_{\mathsf{V}}  \left( V^i - \frac{1}{\alpha} \, \beta^i \right) \end{bmatrix}.
\]
It is evident that $\bm{V}$ is the physical 3-velocity of the particle measured by a fiducial observer with 4-velocity $\bm{n}$ whose worldline crosses that of the particle at a given point.
Note that $- \bm{g}(\bm{n},\bm{V}) / c^2 = \underline{\bm{\chi}} \cdot \bm{V} = 0$, and that
$\Lambda_{\bm{V}} = - \bm{g}(\bm{n},\bm{U}) / c^2 = \underline{\bm{\chi}} \cdot \bm{U}$ is the Lorentz factor of their relative motion, and that $\bm{U} \rightarrow \bm{n}$ as $\bm{V} \rightarrow 0$.

As to momentum, the vector version for a free particle of mass $m$ is the inertia-momentum
\[
\overleftarrow{\bm{P}} = m \, \bm{U} = m \, \Lambda_{\bm{V}} \left( \bm{n} + \bm{V} \right).
\]
Arguably more fundamental is the covector version, the total-energy--momentum
\[
\begin{aligned}
\bm{P} = m \, \underline{\bm{U}} 
	& = m \, \Lambda_{\bm{V}} \left( \underline{\bm{n}} + \underline{\bm{V}} \right) \\[5pt]
	& = -m c^2 \Lambda_{\bm{V}} \, \underline{\bm{\chi}} 
		+ m \, \Lambda_{\bm{V}} \, \underline{\bm{V}}.
\end{aligned}
\]
Defining the physical 3-momentum $\bm{p}$ and total energy $\mathscr{E}_{\bm{p}}$, both measured by the fiducial observer, via
\[
\overleftarrow{\bm{p}} = m \, \Lambda_{\bm{V}} \, \bm{V},
\quad \quad \mathscr{E}_{\bm{p}} = m c^2 \Lambda_{\bm{V}}, 
\]
one can write instead
\begin{equation}
\bm{P} = -\mathscr{E}_{\bm{p}} \,\underline{\bm{\chi}} + \bm{p}, \quad \quad
\mathsf{P} = \left[ P_\nu \right] 
	= \begin{bmatrix} -\alpha \, \mathscr{E}_{\bm{p}} + p_a \beta^a  & p_j \end{bmatrix},
\label{eq:EnergyMomentumE}
\end{equation}
along with
\[
\mathscr{E}_{\bm{p}} = \sqrt{ m^2 c^4 + c^2 \, \overleftrightarrow{\bm{\gamma}}(\bm{p},\bm{p})}.
\]
Note that the covector $\bm{p}$ is tangent to $\mathcal{S}$ in the sense that $\bm{p} \cdot \bm{n} = 0$.
Moreover
\begin{equation}
\bm{U}_{\bm{p}} = \frac{\mathscr{E}_{\bm{p}}}{m c^2} \left( \bm{n} + \bm{V}_{\bm{p}}\right),
\quad \quad \bm{V}_{\bm{p}} = \frac{c^2 \overleftarrow{\bm{p}}}{\mathscr{E}_{\bm{p}}},
\quad \quad \Lambda_{\bm{p}} = \sqrt{ 1 + \frac{1}{m^2 c^2} \overleftrightarrow{\bm{\gamma}}(\bm{p},\bm{p})}
\label{eq:FourVelocityThreeMomentum}
\end{equation}
express the 4-velocity, physical 3-velocity, and Lorentz factor in terms of $\bm{p}$.
Thus the inertia-momentum
\begin{equation}
\overleftarrow{\bm{P}} = m \, \bm{U}_{\bm{p}}
	=  \frac{\mathscr{E}_{\bm{p}}}{c^2} \left( \bm{n} + \bm{V}_{\bm{p}}\right)
	= \frac{\mathscr{E}_{\bm{p}}}{c^2} \, \bm{n} + \overleftarrow{\bm{p}}
\label{eq:InertiaMomentumE}
\end{equation}
is represented by the column
\[
\overleftarrow{\mathsf{P}} =
\begin{bmatrix} \frac{1}{\alpha} ( \mathscr{E}_{\bm{p}} / c^2 ) \\[5pt]
			  p^i -  \frac{1}{\alpha} ( \mathscr{E}_{\bm{p}} / c^2 ) \beta^i		 
\end{bmatrix}.
\]
While the inertia-momentum vector $\overleftarrow{\bm{P}}$ has a meaningful $c \rightarrow \infty$ limit as $\mathscr{E}_{\bm{p}} / c^2 \rightarrow m$, the total-energy--momentum covector $\bm{P}$ does not, preventing a fully satisfactory 4-dimensional spacetime treatment of Galilei physics as discussed in Paper~I.

\subsection{The Metric of Foliated Bargmann Spacetimes $B\mathcal{E}$ and $B\mathcal{G}$}
\label{sec:MetricB}

A `Bargmann spacetime' $B\mathcal{E}$ or $B\mathcal{G}$ is a 5-dimensional Lorentz manifold, a pseudo-Riemann manifold endowed with a metric $\bm{G}$ of signature $(-,+,+,+,+)$.
In the case of $B\mathcal{E}$ this will turn out to be a reformulation of Poincar\'e general relativity in a 5-dimensional setting.
To be explored is whether a limit of this 5-dimensional reformulation in which $c \rightarrow \infty$ might yield a strong-field theory of gravitation consistent with Galilei physics that is not available, or is less readily available, in a 4-dimensional setting---a `Galilei general relativity' with pseudo-Riemann spacetime $B\mathcal{G}$.
Assume that a spacetime $B\mathcal{E}$ or $B\mathcal{G}$ is such that it admits a foliation into a family $\left( \mathcal{S}_{(t,\eta)} \right)_{(t,\eta) \in \mathbb{R}^2}$ of position space leaves of dimension 3 (as with the leaves $\mathcal{S}_t$ of $\mathcal{E}$) and therefore of codimension 2 (unlike the the leaves $\mathcal{S}_t$ of $\mathcal{E}$, which are of codimension 1).
Each leaf $\mathcal{S}_{(t,\eta)}$ of the foliation is the locus defined by constant values of scalar fields $t$ and $\eta$.
That is, there exists an atlas of $B\mathcal{E}$ or $B\mathcal{G}$ in which $t$ and $\eta$ serve as global time and `action' coordinates respectively for every chart in the atlas. 
The origin, meaning, and significance of the action coordinate $\eta$ are described in Paper~I in the context of the affine (and therefore flat) Bargmann-Minkowski and Bargmann-Galilei spacetimes $B\mathbb{M}$ and $B\mathbb{G}$, and will be summarized in the context of the curved spacetimes $B\mathcal{E}$ and $B\mathcal{G}$ below. 
For convenience, let $\mathcal{S}$ denote $\mathcal{S}_{(t,\eta)}$ for some $(t,\eta)$.
Let $(U, \mathscr{X})$ be a chart on $B\mathcal{E}$ or $B\mathcal{G}$,
with $U$ an open subset, and $\mathscr{X} = ( X, \eta ) = ( t, x, \eta )$ coordinates adapted to the foliation.
Then $(U_\mathcal{S}, x)$ is a chart on $\mathcal{S}$, where $U_\mathcal{S} = U \cap \mathcal{S}$ and $x = (x^1, x^2, x^3)$ are local position space coordinates.
The time coordinate index is $0$ and the action coordinate index is $4$, that is, $\mathscr{X}^0 = X^0 = t$ and $\mathscr{X}^4 = \eta$.

Write the coordinate basis vector fields associated with the coordinates $\mathscr{X}$ as $\left( \bm{\partial} \mathscr{X}_J \right) = \left( \bm{\partial} X_\nu, \bm{\partial} \eta \right) = \left( \bm{\partial} t, \bm{\partial} x_j, \bm{\partial} \eta \right)$, and its dual basis of 1-forms as $\left( \bm{\mathrm{d}} \mathscr{X}^I \right) = \left( \bm{\mathrm{d}} X^\mu, \bm{\mathrm{d}} \eta \right) = \left( \bm{\mathrm{d}} t, \bm{\mathrm{d}} x^i, \bm{\mathrm{d}} \eta \right)$.
Upper-case latin indices take values in $\{0,1,2,3,4\}$ (the extended spacetime coordinate indices), with letters $I, J, \dots $ near the middle of the alphabet preferred for free indices, and letters $A, B, \dots $ near the beginning of the alphabet preferred for dummy indices.

As discussed in Paper~I \ADDMORE{and \cite{de-Saxce2016Galilean-Mechan}}, the action coordinate $\mathscr{X}^4 = \eta$ is defined with reference to the kinetic energy per unit mass of a material particle. 
\ADDMORE{This heuristic derivation of the differential relationship between the new coordinate $\eta$ and the other coordinates leads to a `Bargmann metric' $\bm{G}$; this structure constrains the nature of the extended spacetime in a way that allows the transformation of kinetic energy to be manifest in a tensor formalism.}
With coordinates $\mathscr{X}(t) = (t, x(t), \eta(t) )$ along its worldline, the 5-velocity of the particle is  
\[
\bm{\mathscr{U}} = \frac{\mathrm{d}}{\mathrm{d}\tau} 
	= \frac{\mathrm{d}t}{\mathrm{d}\tau} \left( \bm{\partial} t + \bm{v} \right)
		 + \mathscr{U}^4 \, \bm{\partial} \eta 
	\quad\quad (\text{on } B\mathcal{E} \text{ or } B\mathcal{G}),
\]
the terms except the last being the same as for the 4-velocity $\bm{U}$ on $\mathcal{E}$. 
The metric on a Bargmann spacetime is deduced by imposing the condition
\begin{equation}
\mymathscr{U}^4 
	= \frac{ \mathrm{d} \eta }{ \mathrm{d} \tau } 
	=  \frac{ \mathrm{d} t }{ \mathrm{d} \tau } \frac{ \mathrm{d} \eta }{ \mathrm{d} t }
	=
	\left\{
	\begin{aligned}
		 c^2 \left( \Lambda_{\bm{V}} - 1 \right) & & (\text{on } B\mathcal{E}) \\
	\frac{1}{2} \, \bm{\gamma}( \bm{V}, \bm{V} ) & & (\text{on } B\mathcal{G}).
	\end{aligned}
	\right. 
\label{eq:FifthVelocityComponent}
\end{equation}
For $B\mathcal{E}$, substitute into this equation the same expressions $-c^2 \, \mathrm{d}\tau^2 = g_{\alpha\beta} \, \mathrm{d}X^\alpha \, \mathrm{d}X^\beta$ and $\Lambda_{\bm{V}} = \alpha \, \mathrm{d}t / \mathrm{d}\tau$ as on $\mathcal{E}$ and find
\[
0 = \beta_a \beta^a \, \mathrm{d}t^2 
		+ 2 \, \mathrm{d}t \, \beta_a \mathrm{d}x^a  
		+ \mathrm{d} x^a \, \gamma_{a b} \, \mathrm{d} x^b
		 - 2 \, \alpha \, 	\mathrm{d} \eta \, \mathrm{d} t + \frac{1}{c^2} \, \mathrm{d} \eta^2
	\quad (\mathrm{on \ } B\mathcal{E}).
\]
(The corresponding equation in the Conclusion of Paper~I has a sign error in the $2 \, \mathrm{d}t \, \beta_a \mathrm{d}x^a$ term even though the matrix $\mathsf{G}$ of metric components presented there is correct.)
For $B\mathcal{G}$, substitute $\mathrm{d}\tau = \alpha \, \mathrm{d}t$ and $\bm{\gamma}( \bm{V}, \bm{V} ) = \gamma_{ab} ( \mathrm{d}x^a + \beta^a \mathrm{d}t ) ( \mathrm{d}x^b + \beta^b \mathrm{d}t ) / \alpha^2 \, \mathrm{d}t^2$ and find
\[
0 = \beta_a \beta^a \, \mathrm{d}t^2 
		+ 2 \, \mathrm{d}t \, \beta_a \mathrm{d}x^a  
		+ \mathrm{d} x^a \, \gamma_{a b} \, \mathrm{d} x^b
		 - 2 \, \alpha \, 	\mathrm{d} \eta \, \mathrm{d} t 
	\quad (\mathrm{on \ } B\mathcal{G}).
\]
These are in the form of line elements, suggesting a metric $\bm{G}$ whose components with respect to the coordinates $\mathscr{X}$ are given by
\begin{equation}
\mathsf{G} = \left[ G_{IJ} \right] =
	\begin{cases}
		 \begin{bmatrix} 
		 	\beta_a \beta^a && \beta_j && -\alpha \\[5pt]
			 \beta_i && \gamma_{ij} && 0_i \\[5pt]
			  -\alpha && 0_j && \frac{1}{c^2}
		\end{bmatrix} 
			& (\mathrm{on \ } B\mathcal{E}) \\[30pt]
		 \begin{bmatrix} \beta_a \beta^a && \beta_j && -\alpha \\[5pt]
			 \beta_i && \gamma_{ij} && 0_i \\[5pt]
			  -\alpha && 0_j && 0
		\end{bmatrix} 
			& (\mathrm{on \ } B\mathcal{G}),
	\end{cases}
\label{eq:Metric5}
\end{equation}
while the inverse matrix
\begin{equation}
\overleftrightarrow{\mathsf{G}} = \left[ G^{IJ} \right] =
	\begin{cases}
		\begin{bmatrix} 
			-\frac{1}{c^2  \alpha^2} && \frac{1}{c^2  \alpha^2} \, \beta^j  
				&& -\frac{1}{\alpha} \\[5pt]
			\frac{1}{c^2  \alpha^2} \, \beta^i 
				&& \gamma^{ij} -  \frac{1}{c^2  \alpha^2} \, \beta^i \beta^j
				&& \frac{1}{\alpha} \, \beta^i \\[5pt]
			-\frac{1}{\alpha} && \frac{1}{\alpha} \, \beta^j  && 0 
		\end{bmatrix} 
			& (\mathrm{on \ } B\mathcal{E}) \\[30pt]
		\begin{bmatrix} 
			0 && 0^j && -\frac{1}{\alpha} \\[5pt]
			0^i && \gamma^{ij} && \frac{1}{\alpha} \, \beta^i \\[5pt]
			-\frac{1}{\alpha} && \frac{1}{\alpha} \, \beta^j  && 0 
		\end{bmatrix} 
			& (\mathrm{on \ } B\mathcal{G})
	\end{cases}
\label{eq:InverseMetric5}
\end{equation}
collects the components of the inverse metric $\overleftrightarrow{\bm{G}}$.
The determinant $G$ of the matrix $\mathsf{G}$ of metric components is
\begin{equation}
G = \det \mathsf{G} = - \alpha^2 \, \gamma
\label{eq:MetricDeterminant5}
\end{equation}
for both $B\mathcal{E}$ and $B\mathcal{G}$, to be compared with Eq.~(\ref{eq:MetricDeterminant4}).
In comparing Eq.~(\ref{eq:Metric5}) with Eq.~(\ref{eq:Metric4}), note that while a $c \rightarrow \infty$ limit of $\bm{g}$ and its matrix determinant $g$ do not exist, the speed of light $c$ and the lapse function $\alpha$ appearing in $g_{tt} = g_{00}$ have been shifted to $G_{\eta\eta} = G_{44}$ and $G_{t \eta} = G_{04}$ respectively in the case of $B\mathcal{E}$, in such a way that a $c \rightarrow \infty$ limit makes sense, allowing a metric on $B\mathcal{G}$.
This is what allows for a formulation of Galilei general relativity featuring a 5-dimensional pseudo-Riemann spacetime.
But in comparing Eq.~(\ref{eq:InverseMetric4}) with Eq.~(\ref{eq:InverseMetric5}), it is interesting to note that 
\[
G^{\mu\nu} = g^{\mu\nu} \quad \quad (\text{on }B\mathcal{E})
\] 
for the usual spacetime components $\mu, \nu \in \{0,1,2,3\}$; this seems to be connected with the fact that the inverse metric $\overleftrightarrow{\bm{g}}$ of Poincar\'e physics limits sensibly to the degenerate inverse `metric' $\overleftrightarrow{\bm{\gamma}}$ as $c \rightarrow \infty$ as noted previously.

The meaning of the functions $\gamma_{ij}$ in Eq.~(\ref{eq:Metric5}) for the components of $\bm{G}$ on $B\mathcal{E}$ or $B\mathcal{G}$ is the same as in Eq.~(\ref{eq:Metric4}) for the components of $\bm{g}$ on $\mathcal{E}$.
That is, the metric $\bm{G}$ on $B\mathcal{E}$ or $B\mathcal{G}$ induces a 3-metric $\bm{\gamma}$ on each leaf $\mathcal{S}$ of the foliation, as expressed by the fact that $\bm{\gamma}(\bm{u}, \bm{v}) = \bm{G}(\bm{u}, \bm{v})$ for vector fields $\bm{u}, \bm{v}$ tangent to $\mathcal{S}$.
Again the leaves $\mathcal{S}$ are spacelike, that is, the signature of the metric $\bm{\gamma}$ on $\mathcal{S}$ is $+++$, and each leaf $\mathcal{S}$ is a Riemann manifold in its own right.
And again the components of the inverse metric $\overleftrightarrow{\bm{\gamma}}$ on $\mathcal{S}$ are $\gamma^{i j}$, appearing in Eq.~(\ref{eq:InverseMetric5}).

The functions $\alpha$ and $\beta_j$ appearing in the components of the metric $\bm{G}$ in Eq.~(\ref{eq:Metric5}) also have the same meaning in relating neighboring spacelike leaves of the foliation of $B\mathcal{E}$ or $B\mathcal{G}$ as they have on $\mathcal{E}$.
Once again there are fiducial observers whose worldlines orthogonally thread the leaves $\mathcal{S}$ of the foliation.
In discussing vector fields and 1-forms on $B\mathcal{E}$ or $B\mathcal{G}$, the underbar and overarrow notation introduced previously refer to metric duality with respect to $\bm{G}$; that is, $\underline{\bm{v}} = \bm{G} \cdot \bm{v}$ is the 1-form associated by metric duality with a vector field $\bm{v}$, and $\overleftarrow{\bm{\omega}} = \overleftrightarrow{\bm{G}} \cdot \bm{\omega} = \bm{\omega} \cdot \overleftrightarrow{\bm{G}} = \overrightarrow{\bm{\omega}}$ is the vector field associated by metric duality with the 1-form $\bm{\omega}$.
It is worth emphasizing again that in this paper the dot operator only denotes contraction via an obvious `pairing of covariant and contravariant indices', and not the scalar product of two vectors; the latter will be written only in terms of a metric.

On $B\mathcal{E}$ or $B\mathcal{G}$, begin not only with the gradient 1-form $\bm{\mathscr{D}} t$ as on $\mathcal{E}$, but also with the gradient 1-form $\bm{\mathscr{D}} \eta$, in order to obtain the fiducial observer vector field $\bm{\mathscr{n}}$ and the dual fiducial observer vector field $\bm{\chi}$.
(Again, for scalar fields, the Levi-Civita connection $\bm{\mathscr{D}}$ associated with $\bm{G}$ is simply the exterior derivative $\bm{\mathrm{d}}$.)
For a reason that will be clear momentarily---and for an additional reason explained in Sec.~\ref{sec:PhysicalLawsBargmann}---the dual fiducial observer vector field $\bm{\chi}$ on $B\mathcal{E}$ or $B\mathcal{G}$ uses the same symbol as on $\mathcal{E}$, but the fiducial observer vector field is represented by the script character $\bm{\mathscr{n}}$ on $B\mathcal{E}$ or $B\mathcal{G}$ instead of the italic character $\bm{n}$ as on $\mathcal{E}$.
The gradient 1-forms $\bm{\mathscr{D}} t$ and $\bm{\mathscr{D}} \eta$ are normal to $\mathcal{S}$ in the sense that $\bm{\mathscr{D}} t \cdot \bm{v} = 0$ and $\bm{\mathscr{D}} \eta \cdot \bm{v} = 0$ for any vector field $\bm{v}$ tangent to $\mathcal{S}$.
On both $B\mathcal{E}$ and $B\mathcal{G}$ define the fiducial observer 1-form
\begin{equation}
\underline{\bm{\mathscr{n}}}  = - \bm{\mathscr{D}}\eta = -\bm{\mathrm{d}}\eta, \quad \quad
	\underline{\mathscr{n}} = \left[ \mathscr{n}_J \right] = \begin{bmatrix} 0 & 0_j & -1 \end{bmatrix}
	\quad \quad (\text{on } B\mathcal{E} \text{ or } B\mathcal{G})
\label{eq:FiducialObserverCovectorB}
\end{equation}
and the dual fiducial observer 1-form
\begin{equation}
\underline{\bm{\chi}}  = \alpha \, \bm{\mathscr{D}}t = \alpha \, \bm{\mathrm{d}}t, \quad \quad
	\underline{\sfchi} = \left[ \chi_J \right] = \begin{bmatrix} \alpha & 0_j & 0 \end{bmatrix}
	\quad \quad (\text{on } B\mathcal{E} \text{ or } B\mathcal{G}).
\label{eq:DualFiducialObserverCovectorB}
\end{equation}
That the fiducial observer 1-form is related to $\bm{\mathscr{D}}\eta$ on $B\mathcal{E}$ or $B\mathcal{G}$ rather than $\bm{\mathscr{D}}t$ as on $\mathcal{E}$ is the present justification for using a different symbol $\underline{\bm{\mathscr{n}}}$ rather than $\underline{\bm{n}}$.
Raising the index by contraction with $\overleftrightarrow{\bm{G}}$ yields the fiducial observer vector field
\begin{equation}
\bm{\mathscr{n}} = -\overleftarrow{\bm{\mathscr{D}}\eta} 
	= \frac{1}{\alpha} \left( \bm{\partial}t - \beta^a \, \bm{\partial} x_a \right), \quad \quad 
	\mathscr{n} = \left[ \mathscr{n}^I \right] = \begin{bmatrix} \frac{1}{\alpha} \\[5pt] -\frac{1}{\alpha}\beta^i \\[5pt] 0 \end{bmatrix}
	\quad \quad (\text{on } B\mathcal{E} \text{ or } B\mathcal{G})
\label{eq:FiducialObserverVectorB}
\end{equation}
and the dual fiducial observer vector field
\begin{equation}
\bm{\chi} = \alpha \, \overleftarrow{\bm{\mathscr{D}}t} =
\left\{
	\begin{aligned}
		- \frac{1}{c^2 \, \alpha} \left( \bm{\partial}t - \beta^a \, \bm{\partial} x_a \right) 
			- \bm{\partial}\eta \\[5pt]
			- \bm{\partial}\eta,
	\end{aligned}
\right.
\quad \quad \quad
\sfchi = \left[ \chi^I \right] = 
	\begin{cases}
		\begin{bmatrix} -\frac{1}{c^2 \, \alpha} \\[5pt] \frac{1}{c^2 \, \alpha}\beta^i \\ -1 \end{bmatrix} 
			& (\text{on } B\mathcal{E}) \\[20pt]
		\begin{bmatrix} 0 \\ 0^i \\ -1 \end{bmatrix} 
			& (\text{on } B\mathcal{G}).
	\end{cases}
\label{eq:DualObserverVectorB}
\end{equation}
The name `dual fiducial observer 1-form' for $\underline{\bm{\chi}}$ is appropriate since 
\[
\underline{\bm{\chi}} \cdot \bm{\mathscr{n}} = 1
\]
similar to the analogous relation on $\mathcal{E}$.
The observer vector field $\bm{\mathscr{n}}$ for $B\mathcal{E}$ or $B\mathcal{G}$ in Eq.~(\ref{eq:FiducialObserverVectorB}) is interpreted as the 5-velocity field of the congruence of curves comprising the worldlines of the fiducial observers; its first four components coincide with those of the fiducial observer 4-velocity field $\bm{n}$ in Eq.~(\ref{eq:FiducialObserverVectorE}) on $\mathcal{E}$.
The fifth component in Eq.~(\ref{eq:FiducialObserverVectorB}) vanishes because, as worked out in Paper~I and reiterated above, the entire machinery of the Bargmann setting derives from the fact that the fifth component of the 5-velocity represents kinetic energy per unit mass relative to the fiducial observer, and of course the fiducial observer has no kinetic energy relative to itself.
As to $\underline{\bm{\chi}}$, the first  four components of Eq.~(\ref{eq:DualFiducialObserverCovectorB}) agree with Eq.~(\ref{eq:DualFiducialObserverCovectorE}) on $\mathcal{E}$.
The major difference concerning these vector fields and 1-forms on the extended spacetimes $B\mathcal{E}$ or $B\mathcal{G}$ compared with the spacetime $\mathcal{E}$ is that the 4-vectors $\bm{n}$ and $\bm{\chi}$ are collinear on $\mathcal{E}$ according to Eq.~(\ref{eq:DualCollinearE}), while on $B\mathcal{E}$ or $B\mathcal{G}$ the 5-vectors $\bm{\mathscr{n}}$ and $\bm{\chi}$ (and their corresponding 5-covectors) are not. 
This is related to the formal separation of inertia from kinetic energy enabled by the 5-dimensional Bargmann setting.

The analysis in Paper~I supports the interpretation of $\bm{\mathscr{n}}$ as the 5-velocity of fiducial observers.
As shown there, one requirement on a vector $\bm{\mathcal{U}}$ purporting to be the 5-velocity tangent to some observer's worldline in the 5-dimensional Bargmann setting of $B\mathcal{E}$ or $B\mathcal{G}$ is that it be null with respect to $\bm{G}$, that is, $\bm{G}\left( \bm{\mathcal{U}}, \bm{\mathcal{U}} \right) = 0$.
At first this is disconcerting relative to experience with the traditional 4-dimensional spacetime $\mathcal{E}$; but it is not $\bm{G}$ that governs causality on $B\mathcal{E}$ or $B\mathcal{G}$, and this leads to the second requirement on a vector purporting to be a tangent vector to an observer's (necessarily timelike) worldline.
This second consideration, causality, differs between $B\mathcal{E}$ and $B\mathcal{G}$, as also described in Paper~I.
On $B\mathcal{E}$, while $\bm{g}$ is no longer the spacetime metric, when regarded as a tensor  on $B\mathcal{E}$ normal to the action axis, it still governs proper time and the timelike vs. spacelike nature of vectors, such that a vector $\bm{\mathcal{U}}$ qualifies as a 5-velocity only if it is future-directed and $\bm{g}\left( \bm{\mathcal{U}}, \bm{\mathcal{U}} \right) = -c^2$.
On $B\mathcal{G}$ the tensor governing proper time and the timelike vs. spacelike nature of vectors is the 1-form $\bm{\tau} = \underline{\bm{\chi}}$, via the requirement that a vector $\bm{\mathcal{U}}$ qualifies as a 5-velocity only if $\bm{\tau}\left( \bm{\mathcal{U}} \right) = 1$.
On $B\mathcal{E}$ and $B\mathcal{G}$ these two requirements are met by $\bm{\mathscr{n}}$, but not by $\bm{\chi}$ (see Eqs.~(\ref{eq:NormsB}) and (\ref{eq:MutualContractionsB}) below). 
The status of $\bm{\mathscr{n}}$ as not just any observer vector field, but as the `fiducial' or `reference' observer vector field, is cemented by the fact that it is also everywhere normal to the leaves $\mathcal{S}$, that is, $\bm{G}\left( \bm{\mathscr{n}}, \bm{v} \right) = 0$ for any vector field $\bm{v}$ tangent to $\mathcal{S}$.
This is analogous to the corresponding condition for the fiducial observer vector field $\bm{n}$ on $\mathcal{E}$, namely $\bm{g}\left( \bm{n}, \bm{v} \right) = 0$.

As on $\mathcal{E}$, introduce also a `normal evolution vector field' collinear with the fiducial observer vector field:
\[
\bm{\mathscr{m}} = \alpha \, \bm{\mathscr{n}},
\]
which on $B\mathcal{E}$ or $B\mathcal{G}$ satisfies the two relations
\[
\begin{aligned}
\bm{\mathscr{D}}t \cdot \bm{\mathscr{m}} &= \alpha \, \bm{\mathscr{D}}t \cdot \bm{\mathscr{n}} = 1, \\[5pt]
\bm{\mathscr{D}}\eta \cdot \bm{\mathscr{m}}  &= \alpha \, \bm{\mathscr{D}}\eta \cdot \bm{\mathscr{n}} = 0
\end{aligned}
	\quad \quad (\text{on } B\mathcal{E} \text{ or } B\mathcal{G})
\]
as verified from the component expressions above.
As with $\bm{\mathscr{n}}$ on $B\mathcal{E}$ or $B\mathcal{G}$ vs. $\bm{n}$ on $\mathcal{E}$, note also here the use of the script character $\bm{\mathscr{m}}$ on $B\mathcal{E}$ or $B\mathcal{G}$ instead of the italic character $\bm{m}$ on $\mathcal{E}$.
Under a displacement $\delta t \, \bm{\mathscr{m}}$ for infinitesimal $\delta t$, a leaf $\mathcal{S}_{(t,\eta)}$ is carried (or `Lie dragged') to ${S}_{(t + \delta t,\eta)}$: for any point $\mathscr{p} \in \mathcal{S}_{(t,\eta)}$,
\[
\begin{aligned}
t ( \mathscr{p} + \delta t \, \bm{\mathscr{m}} ) 
	&= t(\mathscr{p}) + ( \bm{\mathscr{D}}t \cdot \bm{\mathscr{m}} ) \, \delta t  
		= t(\mathscr{p}) + \delta t, \\[5pt]
\eta ( \mathscr{p} + \delta t \, \bm{\mathscr{m}} ) 
	&= \eta(\mathscr{p}) + ( \bm{\mathscr{D}}\eta \cdot \bm{\mathscr{m}} ) \, \delta t  
		= \eta(\mathscr{p}).
\end{aligned}
	\quad \quad (\text{on } B\mathcal{E} \text{ or } B\mathcal{G})
\]
Thus the interpretation of $\bm{\mathscr{m}}$ as the normal evolution vector field on $B\mathcal{E}$ or $B\mathcal{G}$ is justified in a manner analogous to the interpretation of $\bm{m}$ on $\mathcal{E}$. 

As on $\mathcal{E}$, the normal evolution vector field $\bm{\mathscr{m}}$, together with the tensors $\bm{g}$ and $\bm{\tau}$ that continue to govern proper time on $B\mathcal{E}$ and $B\mathcal{G}$ respectively as described in Paper~I, yield the meaning of the lapse function $\alpha$.
Consider the proper time increment $\delta \tau$ corresponding to a displacement $\delta t \, \bm{\mathscr{m}}$.
On $B\mathcal{E}$, as on $\mathcal{E}$,
\[
c \, \delta \tau = \sqrt{ - \bm{g} ( \, \delta t \, \bm{\mathscr{m}}, \, \delta t \, \bm{\mathscr{m}} \, ) } = c \, \alpha \, \delta t
	\quad \quad (\text{on } B\mathcal{E}).
\]
On $B\mathcal{G}$,
\[
\delta \tau = \bm{\tau} ( \, \delta t \, \bm{\mathscr{m}} \, ) = \alpha \, \delta t
	\quad \quad (\text{on } B\mathcal{G}).
\]
Thus, on both $B\mathcal{E}$ and $B\mathcal{G}$, the lapse function $\alpha$ once again relates a time coordinate interval between two neighboring points encountered by a fiducial observer to the proper time measured by that observer.

The shift vector components $\beta^i$ also have the same meaning as they do on $\mathcal{E}$, for Eq.~(\ref{eq:FiducialObserverVectorB}) on $B\mathcal{E}$ or $B\mathcal{G}$ yields the analogous relation 
\[
\bm{\partial}t = \bm{\mathscr{m}} + \beta^a \, \bm{\partial} x_a.
\]
Moreover, because $\bm{\mathscr{D}} t = \bm{\mathrm{d}}t$ is an element of the 1-form basis dual to the coordinate basis, not only $\bm{\mathscr{D}}t \cdot \bm{\partial} t = 1$, but also $\bm{\mathscr{D}}\eta \cdot \bm{\partial} t = 0$.
Thus once again the vector field $\bm{\partial} t$ is, like $\bm{\mathscr{m}}$, a time evolution vector field by which neighboring spacelike leaves $\mathcal{S}$ are Lie-dragged from one to another---just not the normal evolution vector field $\bm{\mathscr{m}}$.
And again $\bm{\beta} = \beta^a \, \bm{\partial} x_a$ is the coordinate 3-velocity, tangent to $\mathcal{S}$, with which points of constant coordinate position $\left( x^i \right)$ move relative to the fiducial observers.

Having already discussed the 5-velocity $\bm{\mathscr{U}}$ of a material particle in order to deduce a Bargmann spacetime and its metric, close this section by turning to momentum.
The vector version is the inertia--momentum--kinetic-energy
\begin{equation}
\begin{aligned}
\overleftarrow{\bm{\Pi}} &= m \, \bm{\mathscr{U}}_{\bm{p}} \\[5pt]
	&= \frac{\mathscr{E}_{\bm{p}}}{c^2} \, \bm{\mathscr{n}} + \overleftarrow{\bm{p}} 
		+ \epsilon_{\bm{p}} \, \bm{\partial}\eta, 
\end{aligned}
\quad \quad
\overleftarrow{\sfPi} = \left[ \Pi^I \right] 
= 
\left\{
\begin{aligned}
\begin{bmatrix} \frac{1}{\alpha} ( \mathscr{E}_{\bm{p}} / c^2 ) \\[5pt]
			p^i -   \frac{1}{\alpha} \, ( \mathscr{E}_{\bm{p}} / c^2 ) \beta^i \\[5pt]
		         \epsilon_{\bm{p}} 			 
\end{bmatrix} & & (\text{on } B\mathcal{E}) \\[5pt]
\begin{bmatrix} \frac{1}{\alpha} \, m \\[5pt]
			p^i -  \frac{1}{\alpha} \, m \, \beta^i \\[5pt]
		         \epsilon_{\bm{p}} 			 
\end{bmatrix} & & (\text{on } B\mathcal{G}) 
\end{aligned}
\right.
\label{eq:InertiaMomentumB}
\end{equation}
where
\[
\epsilon_{\bm{p}}
	=
	\left\{
	\begin{aligned}
		 m c^2 \left( \Lambda_{\bm{V}} - 1 \right) & & (\text{on } B\mathcal{E}) \\
	\frac{m}{2} \, \bm{\gamma}( \bm{V}, \bm{V} ) & & (\text{on } B\mathcal{G})
	\end{aligned}
	\right. 
\]
is the particle kinetic energy.
The terms except the last of the vector version $\overleftarrow{\bm{\Pi}}$ of momentum on $B\mathcal{E}$ or $B\mathcal{G}$ are the same as the vector version $\overleftarrow{\bm{P}}$ of momentum on $\mathcal{E}$.

Here the overarrow represents index raising with respect to $\bm{G}$, not $\bm{g}$, and this has a profound consequence for momentum in Bargmann spacetimes: 
taking $\bm{\Pi} = \bm{G} \cdot \overleftarrow{\bm{\Pi}}$, one finds that instead of a total-energy--momentum covector, one has a relative-energy--momentum--mass or kinetic-energy--momentum--mass covector
\begin{equation}
\bm{\Pi} = -\epsilon_{\bm{p}} \,\underline{\bm{\chi}} + \bm{p} + m \, \underline{\bm{\mathscr{n}}},
	 \quad \quad
\sfPi = \left[ \Pi_J \right] 
	= \begin{bmatrix} -\alpha \, \epsilon_{\bm{p}} + p_a \beta^a  & p_j & -m \end{bmatrix}
\quad \quad (\text{on } B\mathcal{E} \text{ or } B\mathcal{G}).
\label{eq:EnergyMomentumB}
\end{equation}
Notably, $\bm{\Pi}$ does not simply add a component to $\bm{P}$.
The remarkable nature of a Bargmann spacetime $B\mathcal{E}$ or $B\mathcal{G}$ is that mass is disentangled from kinetic energy by removing it from the first component and moving it to the fifth component without a factor of $c^2$.
This is what allows for kinetic energy to be handled in a tensor formalism while remaining strictly separated from mass as required by Galilei physics, as discussed in Paper~I.

\section{Tensor Decomposition on Foliated Curved Spacetimes}
\label{sec:Decomposition}

Comparison with experiment requires that tensor fields on spacetime be decomposed into pieces consistent with the way humans experience time evolution in position space.
This is done by using vector fields and 1-forms normal to the spacelike leaves of the foliation to construct a projection operator $\overleftarrow{\bm{\gamma}}^*$ closely related to the induced metric $\bm{\gamma}$ on those leaves.

\subsection{Tensor Decomposition on Foliated Einstein Spacetime $\mathcal{E}$}
\label{sec:DecompositionE}

The 1+3 splitting of a spacetime $\mathcal{E}$ embodied in its foliation into 3-dimensional hypersurfaces---leaves $\mathcal{S}$ of codimension 1---is accompanied by a 1+3 decomposition of tensor fields at each point into pieces parallel to the fiducial observer's worldline and tangent to $\mathcal{S}$.
This is accomplished with a projection tensor constructed using the fiducial observer vector field $\bm{n}$ and the dual fiducial observer 1-form $\underline{\bm{\chi}} = - \underline{\bm{n}} / c^2$ introduced in Sec.~\ref{sec:MetricE}.
These are normal to $\mathcal{S}$ in the sense that they respectively have vanishing contractions with 1-forms and vector fields tangent to $\mathcal{S}$.
Their norms with respect to the spacetime metric $\bm{g}$ of $\mathcal{E}$ are
\[
\begin{aligned}
\bm{g}( \bm{n}, \bm{n} ) = \underline{\bm{n}} \cdot \bm{n} 
&=  -c^2, \\
\bm{g}( \bm{\chi}, \bm{\chi} ) = \underline{\bm{\chi}} \cdot \bm{\chi} 
&= 
- \frac{1}{c^2}
\end{aligned}
\quad \quad (\text{on } \mathcal{E}),
\]
and their mutual contraction is
\[
\bm{g}( \bm{\chi}, \bm{n} ) = \underline{\bm{\chi}} \cdot \bm{n} 
= 1
\quad \quad (\text{on } \mathcal{E}).
\]
These properties ensure that
\begin{equation}
\overleftarrow{\bm{\gamma}}	
= \bm{\delta} - \bm{n} \otimes \underline{\bm{\chi}}, 
	\quad \quad \overleftarrow{\mathsf{\sfgamma}} = \left[ {\gamma^\mu}_\nu \right] =
	 \begin{bmatrix} 
		 	0 && 0_j \\[5pt]
			 \beta^i && {\delta^i}_j \\[5pt]
	\end{bmatrix}  \quad \quad (\text{on } \mathcal{E}),
\label{eq:ProjectionFiducialE} 
\end{equation}
where $\bm{\delta} = \overleftarrow{\bm{g}}$ is the identity tensor on $\mathcal{E}$, is the projection tensor satisfying
\begin{align*}
0 &= \overleftarrow{\bm{\gamma}} \cdot \bm{n} 
= \overleftarrow{\bm{\gamma}} \cdot \bm{\chi}, \\[5pt]
0 &= \underline{\bm{n}} \cdot \overleftarrow{\bm{\gamma}}
= \underline{\bm{\chi}} \cdot \overleftarrow{\bm{\gamma}}  
\end{align*}
on $\mathcal{E}$ as desired.
For the decomposition of a vector field on $\mathcal{E}$, the components normal and tangent to $\mathcal{S}$ are given by contraction with $\underline{\bm{\chi}}$ and $\overleftarrow{\bm{\gamma}}$ respectively.
For the decomposition of a 1-form on $\mathcal{E}$, the components normal and tangent to $\mathcal{S}$ are given by contraction with $\bm{n}$ and $\overleftarrow{\bm{\gamma}}$ respectively.

Lowering the first index and raising the second index of the projection operator yield extensions of the 3-metric $\bm{\gamma}$ and its inverse $\overleftrightarrow{\bm{\gamma}}$ on $\mathcal{S}$ to $\mathcal{E}$:
\begin{equation}
\bm{\gamma} = \bm{g} - \underline{\bm{n}} \otimes \underline{\bm{\chi}}, 
	\quad \quad \sfgamma = \left[ \gamma_{\mu\nu} \right] =
	 \begin{bmatrix} 
		 	\beta_a \beta^a && \beta_j \\[5pt]
			 \beta_i && \gamma_{ij}
	\end{bmatrix}  \quad \quad (\text{on } \mathcal{E})
\label{eq:ExtendedThreeMetricE}
\end{equation}
and
\[
\overleftrightarrow{\bm{\gamma}} = \overleftrightarrow{\bm{g}} - \bm{n} \otimes \bm{\chi},
	\quad \quad \overleftrightarrow{\sfgamma} = \left[ \gamma^{\mu\nu} \right] =
	 \begin{bmatrix} 
		 	0 && 0^j \\[5pt]
			 0^i && \gamma^{ij}
	\end{bmatrix}  \quad \quad (\text{on } \mathcal{E}). 
\]
The same symbols $\bm{\gamma}$ and$ \overleftrightarrow{\bm{\gamma}}$ are used for both the original tensors on $\mathcal{S}$ and their extensions to $\mathcal{E}$, with the understanding that for instance $\bm{\gamma}\left( \bm{u}, \bm{v} \right) = \bm{\gamma} \left( \overleftarrow{\bm{\gamma}}(\bm{u}), \overleftarrow{\bm{\gamma}}(\bm{v}) \right)$ defines the extension for vector fields $\bm{u}$ and $\bm{v}$ on $\mathcal{E}$. 

For a vector field $\bm{v}$ on $\mathcal{E}$, the contraction $\overleftarrow{\bm{\gamma}} \cdot \bm{v} = \overleftarrow{\bm{\gamma}} ( \bm{v} )$ with the projection operator $\overleftarrow{\bm{\gamma}}$ yields a vector field tangent to $\mathcal{S}$;
and dual to this, for a 1-form $\bm{\omega}$ defined on $\mathcal{S}$, the evaluation $\bm{\omega} \left( \overleftarrow{\bm{\gamma}} ( \bm{v} ) \right)$ defines an extension $\bm{\omega} \cdot \overleftarrow{\bm{\gamma}}$ of $\bm{\omega}$ to $\mathcal{E}$.
These observations lead to a general projection operator $\overleftarrow{\bm{\gamma}}^*$ for all tensors on $\mathcal{E}$. 
For a $(p,q)$ tensor field $\bm{T}$, meaning that it is $p$ times contravariant and $q$ times covariant, its components are simply given by contracting with $\overleftarrow{\bm{\gamma}}$ on all indices; that is,
\begin{equation}
{\left( \overleftarrow{\bm{\gamma}}^* \bm{T} \right)^{\mu_1 \dots \mu_p}}_{\nu_1 \dots \nu_q}
	= {\gamma^{\mu_1}}_{\alpha_1} \dots {\gamma^{\mu_p}}_{\alpha_p} \; {T^{\alpha_1 \dots \alpha_p}}_{\beta_1 \dots \beta_q}
		\; {\gamma^{\beta_1}}_{\nu_1} \dots {\gamma^{\beta_q}}_{\nu_q}
		\quad \quad (\text{on } \mathcal{E})
\label{eq:GeneralProjectionOperatorE}
\end{equation}
projects $\bm{T}$ from $\mathcal{E}$ to $\mathcal{S}$.
A vector field $\bm{v}$ and a 1-form $\bm{\omega}$, as $(1,0)$ and $(0,1)$ tensor fields respectively, are of course special cases.

\subsection{Tensor Decomposition on Foliated Bargmann Spacetimes $B\mathcal{E}$ and $B\mathcal{G}$}
\label{sec:DecompositionB}

In the case of $B\mathcal{E}$ or $B\mathcal{G}$, instead of the 1+3 splitting introduced on $\mathcal{E}$, a 1+3+1 decomposition of tensors is desired at each point, yielding pieces parallel to the fiducial observer's worldline, tangent to $\mathcal{S}$, and parallel to the action axis.
A projection tensor to $\mathcal{S}$ does make use of the fiducial observer vector field $\bm{\mathscr{n}}$ and the dual fiducial observer 1-form $\underline{\bm{\chi}}$ introduced in Sec.~\ref{sec:MetricB}.
However, unlike the situation on 4-dimensional $\mathcal{E}$, the vector fields $\bm{\mathscr{n}}$ and $\bm{\chi}$ (and the metric dual 1-forms $\underline{\bm{\mathscr{n}}}$ and $\underline{\bm{\chi}}$) are not collinear on $B\mathcal{E}$ or $B\mathcal{G}$.
As in Paper~I, it is useful to define an `action vector' field $\bm{\xi}$ that characterizes the noncollinearity of $\bm{\mathscr{n}}$ and $\bm{\chi}$, and also happens to be collinear with (but opposite in direction to) the $\eta$ axis.
It is defined by
\begin{equation}
\bm{\xi} = - \bm{\partial}\eta =
\left\{
\begin{aligned}
\frac{1}{c^2} \, \bm{\mathscr{n}} + \bm{\chi}  	& & (\text{on } B\mathcal{E}) \\ 
\bm{\chi}  	& & (\text{on } B\mathcal{G}),
\end{aligned}
\right.
\quad \quad \quad
\sfxi = \left[ \xi^I \right] = \begin{bmatrix} 0 \\ 0^i \\ -1 \end{bmatrix}
	\quad \quad (\text{on } B\mathcal{E} \text{ or } B\mathcal{G}).
\label{eq:ActionVector}
\end{equation}
The corresponding 1-form is
\begin{equation}
\underline{\bm{\xi}} = 
\left\{
\begin{aligned}
\frac{1}{c^2} \, \underline{\bm{\mathscr{n}}} + \underline{\bm{\chi}} \\
\underline{\bm{\chi}}, 
\end{aligned}
\right.
\quad \quad \quad
\underline{\sfxi} = \left[ \xi_J \right] = 
\begin{cases}
\begin{bmatrix} \alpha & 0_j & -\frac{1}{c^2} \end{bmatrix} & (\text{on } B\mathcal{E}) \\[10pt]
\begin{bmatrix} \alpha & 0_j & 0 \end{bmatrix}			& (\text{on } B\mathcal{G}).
\end{cases}
\label{eq:ActionCovector}
\end{equation}
Notice that $\bm{\xi}$ and $\bm{\chi}$ coincide on $B\mathcal{G}$.
Since there is no action coordinate on $\mathcal{E}$ there is no action vector either; or rather, it degenerates to the zero vector in accord with Eq.~(\ref{eq:DualCollinearE}).

As with $\mathcal{E}$, but now involving also the action vector $\bm{\xi}$ in addition to $\bm{\mathscr{n}}$ and $\bm{\chi}$, the mutual contractions of these vector fields and 1-forms point towards their appearance in a projection operator $\overleftarrow{\bm{\gamma}}$ needed for 1+3+1 tensor decompositions.
The norms of these vector fields are
\begin{equation}
\begin{aligned}
\bm{G}( \bm{\mathscr{n}}, \bm{\mathscr{n}} ) = \underline{\bm{\mathscr{n}}} \cdot \bm{\mathscr{n}} 
&=  0 \quad \quad \quad (\text{on } B\mathcal{E} \text{ or } B\mathcal{G}), \\[5pt]
\bm{G}( \bm{\chi}, \bm{\chi} ) = \underline{\bm{\chi}} \cdot \bm{\chi} 
&= \left\{ \begin{aligned} 
- \frac{1}{c^2} & & (\text{on } B\mathcal{E})  \\[5pt]
\phantom{-} 0 &  & (\text{on } B\mathcal{G}),
\end{aligned} \right. \\[5pt]
\bm{G}( \bm{\xi}, \bm{\xi} ) = \underline{\bm{\xi}} \cdot \bm{\xi} 
&= \left\{ \begin{aligned}
\phantom{-} \frac{1}{c^2} & & (\text{on } B\mathcal{E}) \\[5pt]
\phantom{-} 0 	&  & (\text{on } B\mathcal{G}). 
\end{aligned} \right.
\end{aligned}
\label{eq:NormsB}
\end{equation}
The mutual contractions are
\begin{equation}
\begin{aligned}
\bm{G}( \bm{\chi}, \bm{\mathscr{n}} ) = \underline{\bm{\chi}} \cdot \bm{\mathscr{n}} 
&= 1,  \\[5pt]
\bm{G}( \bm{\mathscr{n}}, \bm{\xi} ) = \underline{\bm{\mathscr{n}}} \cdot \bm{\xi} 
&= 1,  \\[5pt]
\bm{G}( \bm{\chi}, \bm{\xi} ) = \underline{\bm{\chi}} \cdot \bm{\xi} 
&= 0
\end{aligned}
	\quad \quad (\text{on } B\mathcal{E} \text{ or } B\mathcal{G}).
\label{eq:MutualContractionsB}
\end{equation}
These properties ensure that
\begin{equation}
\overleftarrow{\bm{\gamma}}	
= \bm{\delta} - \bm{\mathscr{n}} \otimes \underline{\bm{\chi}} - \bm{\xi} \otimes \underline{\bm{\mathscr{n}}}, 
	\quad \quad \overleftarrow{\mathsf{\sfgamma}} = \left[ {\gamma^I}_J \right] =
	 \begin{bmatrix} 
		 	0 && 0_j && 0 \\[5pt]
			 \beta_i && {\delta^i}_j && 0_i \\[5pt]
			  0 && 0_j && 0
	\end{bmatrix}
	\quad \quad (\text{on } B\mathcal{E} \text{ or } B\mathcal{G}),
\label{eq:ProjectionFiducialB}
\end{equation}
where in this context $\bm{\delta} = \overleftarrow{\bm{G}}$ is the identity tensor on $B\mathcal{E}$ or $B\mathcal{G}$, is the projection tensor satisfying
\[
\begin{aligned}
0 &= \overleftarrow{\bm{\gamma}} \cdot \bm{\mathscr{n}} 
= \overleftarrow{\bm{\gamma}} \cdot \bm{\chi}
= \overleftarrow{\bm{\gamma}} \cdot \bm{\xi}, \\[5pt]
0 &= \underline{\bm{\mathscr{n}}} \cdot \overleftarrow{\bm{\gamma}}  
= \underline{\bm{\chi}} \cdot \overleftarrow{\bm{\gamma}} 
= \underline{\bm{\xi}} \cdot \overleftarrow{\bm{\gamma}} 
\end{aligned}
	\quad \quad (\text{on } B\mathcal{E} \text{ or } B\mathcal{G})
\]
on $B\mathcal{E}$ or $B\mathcal{G}$ as desired.
For the decomposition of a vector field on $B\mathcal{E}$ or $B\mathcal{G}$, the components parallel to the fiducial observer worldline, tangent to $\mathcal{S}$, and parallel to the action axis are given by contraction with $\underline{\bm{\chi}}$, $\overleftarrow{\bm{\gamma}}$, and $-\underline{\bm{\mathscr{n}}}$, respectively.
For the decomposition of a 1-form on $B\mathcal{E}$ or $B\mathcal{G}$, the components parallel to the fiducial observer worldline, tangent to $\mathcal{S}$, and parallel to the action axis are given by contraction with $\bm{\mathscr{n}}$, $\overleftarrow{\bm{\gamma}}$, and $-\bm{\xi}$, respectively.

Lowering the first index or raising the second index of the projection operator yields extensions of the 3-metric $\bm{\gamma}$ and its inverse $\overleftrightarrow{\bm{\gamma}}$on $\mathcal{S}$ to $B\mathcal{E}$ or $B\mathcal{G}$:
\begin{equation}
\bm{\gamma} = \bm{G} - \underline{\bm{\mathscr{n}}} \otimes \underline{\bm{\chi}} 
		- \underline{\bm{\xi}} \otimes \underline{\bm{\mathscr{n}}}, 
	\quad \quad \sfgamma = \left[ \gamma_{IJ} \right] =
	 \begin{bmatrix} 
		 	\beta_a \beta^a && \beta_j && 0 \\[5pt]
			 \beta_i && \gamma_{ij} && 0_i \\[5pt]
			  0 && 0_j && 0
	\end{bmatrix} 
	\quad \quad (\text{on } B\mathcal{E} \text{ or } B\mathcal{G})
\label{eq:ExtendedThreeMetricB}
\end{equation}
and
\[
\overleftrightarrow{\bm{\gamma}} = \overleftrightarrow{\bm{G}} - \bm{\mathscr{n}} \otimes \bm{\chi} 
		- \bm{\xi} \otimes \bm{\mathscr{n}},
	\quad \quad \overleftrightarrow{\sfgamma} = \left[ \gamma^{IJ} \right] =
	 \begin{bmatrix} 
		 	0 && 0^j && 0 \\[5pt]
			 0^i && \gamma^{ij} && 0^i \\[5pt]
			  0 && 0^j && 0
	\end{bmatrix}
	\quad \quad (\text{on } B\mathcal{E} \text{ or } B\mathcal{G}).
\]
Again, the same symbols $\bm{\gamma}$ and$ \overleftrightarrow{\bm{\gamma}}$ are used for both the original tensors on $\mathcal{S}$ and their extensions to $B\mathcal{E}$ or $B\mathcal{G}$.
The components of $\overleftarrow{\bm{\gamma}}$, $\bm{\gamma}$, and $\overleftrightarrow{\bm{\gamma}}$ on $\mathcal{E}$ in Sec.~\ref{sec:DecompositionE} all agree with the time and position space components here on $B\mathcal{E}$ or $B\mathcal{G}$.

As on $\mathcal{E}$, the projection tensor $\overleftarrow{\bm{\gamma}}$ on $B\mathcal{E}$ or $B\mathcal{G}$ that gives vector fields tangent to $\mathcal{S}$ and extensions of 1-forms on $\mathcal{S}$ also provides for a general projection operator $\overleftarrow{\bm{\gamma}}^*$ for all tensors on $B\mathcal{E}$ or $B\mathcal{G}$.
In particular
\begin{equation}
{\left( \overleftarrow{\bm{\gamma}}^* \bm{T} \right)^{I_1 \dots I_p}}_{J_1 \dots J_q}
	= {\gamma^{I_1}}_{A_1} \dots {\gamma^{I_p}}_{A_p} \; {T^{A_1 \dots A_p}}_{B_1 \dots B_q}
		\; {\gamma^{B_1}}_{J_1} \dots {\gamma^{B_q}}_{J_q}
	\quad \quad (\text{on } B\mathcal{E} \text{ or } B\mathcal{G})
\label{eq:GeneralProjectionOperatorB}
\end{equation}
is the counterpart of Eq.~(\ref{eq:GeneralProjectionOperatorE}).

For $\alpha \rightarrow 1$ and $\beta^i \rightarrow 0$, all of this agrees with the expressions obtained for the flat spacetimes $B\mathbb{M}$ and $B\mathbb{G}$ in Sec.~5.3 of Paper~I.

\section{Levi-Civita Connections on Foliated Curved Spacetimes}
\label{sec:LeviCivita}

Noted here are the Levi-Civita connections $\bm{\nabla}$ on spacetime $\mathcal{E}$ and $\bm{\mathscr{D}}$ on spacetimes $B\mathcal{E}$ or $B\mathcal{G}$---pseudo-Riemann manifolds all---and the Levi-Civita connection $\bm{D}$ on the Riemann submanifolds $\mathcal{S}$ that constitute the position space leaves of their foliations.
The acceleration of fiducial observers and other important directional derivatives are also considered. 
Finally, projective relationships are given between certain tensor gradients on $\mathcal{E}$, or $B\mathcal{E}$ or $B\mathcal{G}$, and tensor gradients on $\mathcal{S}$.

\subsection{Levi-Civita Connection on Foliated Einstein Spacetime $\mathcal{E}$}
\label{sec:LeviCivitaE}

As a pseudo-Riemann manifold, the geometry of a 4-dimensional Einstein spacetime $\mathcal{E}$ is determined by its spacetime metric $\bm{g}$.
Let $\bm{\nabla}$ denote the Levi-Civita connection, the covariant derivative operator on $\mathcal{E}$ satisfying $\bm{\nabla}\bm{g} = 0$. 
The variations $\nabla_\nu \, \bm{\partial} X_\mu = {^{\bm{g} \!}{\Gamma^\alpha}_{\mu \nu}} \, \bm{\partial} X_\alpha$ of the coordinate basis vectors define the connection coefficients
\[
 ^{\bm{g}} \! {\Gamma^\rho}_{\mu \nu} = \frac{1}{2} \, g^{\rho \alpha} 
 	\left( \frac{\partial g_{\alpha \nu} }{\partial X^\mu} 
		+ \frac{\partial g_{\mu \alpha}}{\partial X^\nu}
		- \frac{\partial g_{\mu \nu}}{\partial X^\alpha}\right)
			\quad \quad (\text{on } \mathcal{E}).
\]
By relating the values of basis vector fields (and, by implication, basis 1-forms) at neighboring points, the connection $\bm{\nabla}$ enables the definition of the gradient of a tensor field of arbitrary type, the gradient $\bm{\nabla} \bm{T}$ of a $(p,q)$ tensor field $\bm{T}$ being a tensor field of type $(p,q+1)$.
This is accomplished via a Leibniz rule: in addition to the partial derivatives of tensor component functions, the Levi-Civita connection $\bm{\nabla}$ adds an additional term for each component resulting from the variation of the corresponding basis vector field or basis 1-form.
The connection coefficients are also called Christoffel symbols of the second kind; the Christoffel symbols of the first kind,
\[
 ^{\bm{g}} \! \Gamma_{ \rho \mu \nu} 
 	= g_{\rho\alpha} \, ^{\bm{g}} \! {\Gamma^\alpha}_{\mu \nu}
 	= \frac{1}{2}  \left( \frac{\partial g_{\rho \nu} }{\partial X^\mu} 
			+ \frac{\partial g_{\mu \rho}}{\partial X^\nu}
			- \frac{\partial g_{\mu \nu}}{\partial X^\rho}\right)
\quad \quad (\text{on } \mathcal{E}),
\]
will also be useful on occasion.

%
Contraction of the gradient of a tensor field with a vector field, with the contraction taking place on the new `tensor slot' opened by the gradient operation, results in a directional derivative.
An example of interest here is the 4-acceleration $c^2 \bm{a}$ of fiducial observers on the 1+3 foliated spacetime $\mathcal{E}$, given by the directional derivative of $\bm{n}$ along itself, and related directional derivatives called here `generalized accelerations', for example involving $\bm{\chi}$.
Such generalized accelerations will play a role in the extrinsic geometry of the foliation, discussed in Sec.~\ref{sec:ExtrinsicGeometry}, where the 1-form versions of such directional derivatives are more relevant.
For the directional derivative $\bm{\nabla}_{\! \bm{n}}  \underline{\bm{n}}$, one can use the definition of the Christoffel symbols to find
\[
\begin{aligned}
n^\alpha \nabla_\alpha n_\nu 
	&= n^\alpha \left( \partial_\alpha n_\nu 
		- \, ^{\bm{g}\!}{\Gamma^\beta}_{\nu\alpha} \, n_\beta \right) \\[5pt]
	&= n^\alpha \, \partial_\alpha n_\nu
		+  \frac{1}{2} \, \partial_\nu g^{\alpha\beta} \, n_\alpha \, n_\beta \\[5pt]
	&= c^2 \, a_\nu,
\end{aligned}
\]
where
\begin{equation}
a_\nu = \partial_\nu \ln \alpha 
	- \chi_\nu \, n^\alpha \, \partial_\alpha \ln \alpha.
\label{eq:AccelerationPartialDerivativesE}
\end{equation}
Metric duality and the antisymmetry of two of the terms in $^{\bm{g}\!}\Gamma_{\rho\mu\nu}$ have been exploited,
and Eq.~(\ref{eq:FiducialObserverCovectorE}) for $\underline{\mathsf{n}}$ and Eq.~(\ref{eq:DualFiducialObserverCovectorE}) for $\underline{\sfchi}$ have been employed; the latter implies the handy relation
\begin{equation}
\partial_\mu \chi_\nu = \chi_\nu \, \partial_\mu \ln \alpha,
\label{eq:HandyRelationChi}
\end{equation}
which will be of use later on.
In this case all that is needed from the metric is the single component $g^{00} = - 1 / c^2 \alpha^2$.
Alternatively, explicit use of connection coefficients can be avoided altogether by making use of the definition of $\underline{\bm{n}}$ in Eq.~(\ref{eq:FiducialObserverCovectorE}) in terms of the gradient of the coordinate function $t$, and the fact that the Levi-Civita connection $\bm{\nabla}$ is torsion-free ($\nabla_{\! \mu} \nabla_{\! \nu} f = \nabla_{\! \nu} \nabla_{\! \mu} f$ for a scalar field $f$).
In any case, the (1-form version of the) 4-acceleration of the fiducial observers on $\mathcal{E}$ can also be written in more geometric form as
\begin{equation}
\bm{\nabla}_{\! \bm{n}}  \underline{\bm{n}} = c^2 \, \underline{\bm{a}} 
	\quad \quad (\text{on } \mathcal{E}),
\label{eq:FiducialAccelerationE}
\end{equation}
where 
\begin{equation}
\underline{\bm{a}} = \bm{D} \ln \alpha, \quad \quad
\underline{\mathsf{a}} = \left[ a_\nu \right] = \begin{bmatrix} \beta^a \, \partial_a \ln \alpha  
	& \partial_j \ln \alpha \end{bmatrix}
	\quad \quad (\text{on } \mathcal{E}),
\label{eq:AccelerationComponentsE}
\end{equation}
with $\bm{D}$ being the (extension to $\mathcal{E}$ of the) Levi-Civita connection on spacelike leaves $\mathcal{S}$ discussed below in Sec.~\ref{sec:LeviCivitaS} (here, for a scalar field, simply an exterior derivative).
Note that $\underline{\bm{a}}$ is tangent to $\mathcal{S}$ (since $\underline{\bm{a}} \cdot \bm{n} = 0$) and exhibits no time derivatives.
That the variation of $\bm{n}$, which is normal to the spacelike leaves $\mathcal{S}$, along the fiducial observers' worldlines is determined by the position space variation of the lapse function $\alpha$ is a manifestation of the fact that the scalar field $\alpha$ determines the spacetime foliation. 
The equivalent generalized accelerations
\begin{equation}
\bm{\nabla}_{\! \bm{\chi}} \underline{\bm{n}} = \bm{\nabla}_{\! \bm{n}} \underline{\bm{\chi}} =
	-  \underline{\bm{a}} 
	\quad \quad (\text{on } \mathcal{E})
\label{eq:AlternativeAccelerationE}
\end{equation}
follow from Eq.~(\ref{eq:DualCollinearE}) expressing the collinearity of $\bm{n}$ and $\bm{\chi}$ on $\mathcal{E}$.

\subsection{Levi-Civita Connection on Foliated Bargmann Spacetimes $B\mathcal{E}$ and $B\mathcal{G}$}
\label{sec:LeviCivitaB}

A 5-dimensional Bargmann spacetime $B\mathcal{E}$ or $B\mathcal{G}$ is also a pseudo-Riemann manifold, with geometry determined by its spacetime metric $\bm{G}$.
Let $\bm{\mathscr{D}}$ denote the Levi-Civita connection satisfying $\bm{\mathscr{D}}\bm{G} = 0$ on $B\mathcal{E}$ or $B\mathcal{G}$.
The connection coefficients
\[
 ^{\bm{G}} \! {\Gamma^K}_{I J} = \frac{1}{2} \, G^{K A} 
 	\left( \frac{\partial G_{A J} }{\partial \mathscr{X}^I} 
		+ \frac{\partial G_{I A}}{\partial \mathscr{X}^J}
		- \frac{\partial G_{I J}}{\partial \mathscr{X}^A}\right)
			\quad \quad (\text{on } B\mathcal{E} \text{ or } B\mathcal{G})
\]
give the variations $\mathscr{D}_J \, \bm{\partial} \mathscr{X}_I = {^{\bm{G} \!}{\Gamma^A}_{I J}} \, \bm{\partial} \mathscr{X}_A$ of the coordinate basis vectors.
Part of what it means to minimally embed Poincar\'e or Galilei physics in a Bargmann-extended spacetime is that no field---representing the metric, a material continuum, or anything else---on $B\mathcal{E}$ or $B\mathcal{G}$ is allowed to depend explicitly on the action coordinate $\mathscr{X}^4 = \eta$; thus in the above and all other expressions in this paper involving partial derivatives, any partial derivative with respect to $\mathscr{X}^4 = \eta$ vanishes.

The Christoffel symbols of the first kind are
\[
 ^{\bm{G}} \! \Gamma_{K I J} 
 	= G_{K A} \, ^{\bm{G}} \! {\Gamma^A}_{I J}
 	= \frac{1}{2}  \left( \frac{\partial G_{K J} }{\partial \mathscr{X}^I} 
			+ \frac{\partial G_{I K}}{\partial \mathscr{X}^J}
			- \frac{\partial G_{I J}}{\partial \mathscr{X}^K}\right)
\quad \quad (\text{on } B\mathcal{E} \text{ or } B\mathcal{G}).
\]
For the conventional spacetime indices $\mu, \nu, \rho \in \left\{ 0,1,2,3 \right\}$, it will be useful later to relate $^{\bm{g}} \! \Gamma_{ \rho \mu \nu}$ to $^{\bm{G}} \! \Gamma_{ \rho \mu \nu}$.
Noting from Eqs.~(\ref{eq:DualFiducialObserverCovectorE}) and (\ref{eq:DualFiducialObserverCovectorB}) that the components $\chi_\nu$ are the same on $B\mathcal{E}$ or $B\mathcal{G}$ as on $\mathcal{E}$, and comparing Eq.~(\ref{eq:Metric4}) for $\mathsf{g}$ with the upper left $4 \times 4$ block of Eq.~(\ref{eq:Metric5}) for $\mathsf{G}$, for the conventional spacetime indices one has
\[
g_{\mu\nu} = G_{\mu\nu} - c^2 \, \chi_\mu \, \chi_\nu.
\]
Making use of Eq.~(\ref{eq:HandyRelationChi}), the desired relation is
\begin{equation}
^{\bm{g}} \! \Gamma_{ \rho \mu \nu} 
	= {^{\bm{G}\!} \Gamma}_{ \rho \mu \nu}
	 	- c^2 ( \chi_\rho \, \chi_\nu \, \partial_\mu \ln \alpha 
				+ \chi_\mu \, \chi_\rho \, \partial_\nu \ln \alpha 
				-  \chi_\mu \, \chi_\nu \, \partial_\rho \ln \alpha ).
\label{eq:Christoffel_g_G}
\end{equation}
The dependence on the lapse function $\alpha$ and its derivatives is explicitly separated, as the components $G_{\mu\nu}$ and therefore ${^{\bm{G}\!} \Gamma}_{ \rho \mu \nu}$ depend only on $\beta_i$ and $\gamma_{ij}$. 

%
Directional derivatives of interest include the 5-acceleration of fiducial observers on $B\mathcal{E}$ or $B\mathcal{G}$ and related `generalized accelerations'.
In contrast to Eq.~(\ref{eq:FiducialAccelerationE}) for the 4-acceleration of fiducial observers on on $\mathcal{E}$, the 5-acceleration of fiducial observers on B$\mathcal{E}$ or B$\mathcal{G}$ vanishes:
\begin{equation}
\bm{\mathscr{D}}_{\! \bm{\mathscr{n}}} \underline{\bm{\mathscr{n}}} = 0
			\quad \quad (\text{on } B\mathcal{E} \text{ or } B\mathcal{G}). 
\label{eq:FiducialAccelerationB}
\end{equation}
This can be shown using the matrix expressions for $\underline{\mathsf{\mathscr{n}}}$ in Eq.~(\ref{eq:FiducialObserverCovectorB}) and $\overleftrightarrow{\mathsf{G}}$ in Eq.~(\ref{eq:InverseMetric5}), noting that $G^{44} = 0$ is the only metric component needed, to obtain
\[
\mathscr{n}^A \,  \mathscr{D}_A  \mathscr{n}_J 
	= -  \, \mathscr{n}^A \, {^{\bm{G}} \! \Gamma^B}_{JA} \, \mathscr{n}_B
	= -{^{\bm{G}} \! \Gamma}_{BJA} \, \mathscr{n}^B \, \mathscr{n}^A
	= \frac{1}{2} \, \partial_J G^{AB} \, \mathscr{n}_A \, \mathscr{n}_B
	= 0.
\] 
That the 5-velocity $\bm{\mathscr{n}}$ of fiducial observers is geodesic (vanishing 5-acceleration $\bm{\mathscr{D}}_{\! \bm{\mathscr{n}}}  \bm{\mathscr{n}}$) with respect to $\bm{G}$ on $B\mathcal{E}$ or $B\mathcal{G}$, while the 4-acceleration of fiducial observers is not geodesic with respect to $\bm{g}$ on $\mathcal{E}$, may be a tantalizing hint that there is something deeply natural or at least interesting about the Bargmann construction.
However, a result for $B\mathcal{E}$ that does coincide with what would be found on $\mathcal{E}$, obtained with a similar but slightly more involved calculation using Eq.~(\ref{eq:DualFiducialObserverCovectorB}) for $\underline{\sfchi}$ and $G^{00} = - 1/ c^2 \alpha^2$ as the only inverse metric component needed, is  
\[
\bm{\mathscr{D}}_{\! \bm{\chi}} \underline{\bm{\chi}} = 
\left\{ \begin{aligned}
\frac{1}{c^2} \,\underline{\bm{a}} & & (\mathrm{on \ } B\mathcal{E}), \\[5pt]
0 & & (\mathrm{on \ } B\mathcal{G}),
\end{aligned} \right. 
\]
where the same acceleration 1-form as on $\mathcal{E}$ (now with additional vanishing action component) appears:
\begin{equation}
\underline{\bm{a}} = \bm{D} \ln \alpha, \quad \quad
\underline{\mathsf{a}} = \left[ a_J \right] = \begin{bmatrix} \beta^a \, \partial_a \ln \alpha  
	& \partial_j \ln \alpha & 0 \end{bmatrix}
			\quad \quad (\text{on } B\mathcal{E} \text{ or } B\mathcal{G}). 
\label{eq:AccelerationComponentsB}
\end{equation}
Other generalized accelerations of interest include
\begin{equation}
\bm{\mathscr{D}}_{\!\bm{\xi}} \underline{\bm{\mathscr{n}}} = - \frac{1}{2} \,\underline{\bm{a}}
	\quad \quad (\text{on } B\mathcal{E} \text{ or } B\mathcal{G}),
	\quad \quad \quad  		
	\bm{\mathscr{D}}_{\! \bm{\xi}} \underline{\bm{\chi}} = 
\left\{ \begin{aligned}
\frac{1}{2c^2} \,\underline{\bm{a}} & & (\mathrm{on \ } B\mathcal{E}), \\[5pt]
0 & & (\mathrm{on \ } B\mathcal{G}).
\end{aligned} \right.
\label{eq:AdditionalAccelerationB}
\end{equation}
Using Eq.~(\ref{eq:ActionVector}) for the column $\sfxi$, these can be obtained as
\[
\begin{aligned}
\xi^A \,  \mathscr{D}_A  \mathscr{n}_J 
	&= -{^{\bm{G}} \! \Gamma^4}_{J4} = - \frac{1}{2} \, a_J, \\[5pt]
\xi^A \,  \mathscr{D}_A  \chi_J 
	&= \alpha \, {^{\bm{G}} \! \Gamma^0}_{J4} = \frac{1}{2c^2} \, a_J, \\[5pt]
\end{aligned}
\]
where the calculations of ${^{\bm{G}} \! \Gamma^4}_{J4}$ and ${^{\bm{G}} \! \Gamma^0}_{J4}$ are streamlined by noting that 
\[
G^{4I} = - \mathscr{n}^I, 
	\quad \quad G_{4J} = G_{J4} = - \xi_J = - \chi_J - \frac{1}{c^2} \mathscr{n}_J,
	\quad \quad G^{0I} = \frac{1}{\alpha} \, \chi^I 
		= \frac{1}{\alpha} \left( \xi^I - \frac{1}{c^2} \mathscr{n}^I \right).   
\]
Finally, the above directional derivatives can be combined using $\bm{\xi} = \bm{\chi} + \bm{\mathscr{n}} / c^2$ in Eq.~(\ref{eq:ActionVector}) and the linearity of the directional derivative $\bm{\mathscr{D}}_{\!\bm{u}}$ in $\bm{u}$ to obtain
\begin{equation}
\bm{\mathscr{D}}_{\! \bm{\chi}} \underline{\bm{\mathscr{n}}} = \bm{\mathscr{D}}_{\! \bm{\mathscr{n}}} \underline{\bm{\chi}} =
	-\frac{1}{2} \, \underline{\bm{a}}
			\quad \quad (\text{on } B\mathcal{E} \text{ or } B\mathcal{G}), 
\label{eq:AlternativeAccelerationB}
\end{equation}
which differ from their counterparts on $\mathcal{E}$ in Eq.~(\ref{eq:AlternativeAccelerationE}) by a factor of $1/2$.

\subsection{Levi-Civita Connection on Spacelike Leaves $\mathcal{S}$}
\label{sec:LeviCivitaS}

Each 3-dimensional spacelike leaf $\mathcal{S}$ of a foliated spacetime $\mathcal{E}$, or $B\mathcal{E}$ or $B\mathcal{G}$, is a Riemann manifold whose intrinsic geometry is governed by the induced metric $\bm{\gamma}$.
Let $\bm{D}$ denote the Levi-Civita connection on $\mathcal{S}$ satisfying $\bm{D}\bm{\gamma} = 0$. 
The connection coefficients
\[
 {\Gamma^k}_{i j} = \frac{1}{2} \, \gamma^{k a} 
 	\left( \frac{\partial \gamma_{a j} }{\partial x^i} 
		+ \frac{\partial \gamma_{i a}}{\partial x^j}
		- \frac{\partial \gamma_{i j}}{\partial x^a}\right)
\]
give the variations $D_j \, \bm{\partial} x_i = {\Gamma^a}_{i j} \, \bm{\partial} x_a$ of the coordinate basis vectors.
In the particular case of the intrinsic geometry of the leaves $\mathcal{S}$, no additional prefixing superscript identifying the metric is added to the connection coefficients.

\subsection{Projections of Tensor Gradients}
\label{sec:ProjectionsTensorGradients}

It will prove useful to consider projective relationships between certain tensor gradients on $\mathcal{E}$, or $B\mathcal{E}$ or $B\mathcal{G}$, and on $\mathcal{S}$.
Consider a tensor $\bm{T}$ on $\mathcal{E}$, or on $B\mathcal{E}$ or $B\mathcal{G}$, that is tangent to a position space leaf $\mathcal{S}$, that is, such that $\overleftarrow{\bm{\gamma}}^* \bm{T} = \bm{T}$, so that for $\bm{T}$ on $\mathcal{E}$, all possible contractions of $\bm{T}$ with $\bm{n}$ and with $\underline{\bm{\chi}}$ vanish; or for $\bm{T}$ on $B\mathcal{E}$ or $B\mathcal{G}$, all possible contractions of $\bm{T}$ with $\bm{\mathscr{n}}$ and $\bm{\xi}$, and with $\underline{\bm{\chi}}$ and $\underline{\bm{\mathscr{n}}}$, vanish.
What is the relationship of the projection of the spacetime gradient $\bm{\nabla}\bm{T}$ on $\mathcal{E}$, or of $\bm{\mathscr{D}}\bm{T}$ on $B\mathcal{E}$ or $B\mathcal{G}$, to the position space gradient $\bm{D}\bm{T}$?
It turns out to be the intuitively pleasing relation
\begin{equation}
\bm{D}\bm{T} =
\left\{ 
\begin{aligned}
\overleftarrow{\bm{\gamma}}^* \bm{\nabla}\bm{T} & & (\text{on } \mathcal{E}) \\[5pt]\overleftarrow{\bm{\gamma}}^* \bm{\mathscr{D}}\bm{T} & & (\text{on } B\mathcal{E} \text{ or } B\mathcal{G}),
\end{aligned}
\right.
\label{eq:DerivativeProjection}
\end{equation}
given the respective generalized projection operators $\overleftarrow{\bm{\gamma}}^*$ for $\mathcal{E}$ in Eq.~(\ref{eq:GeneralProjectionOperatorE}), or for $B\mathcal{E}$ or $B\mathcal{G}$ in Eq.~(\ref{eq:GeneralProjectionOperatorB}), built from the fundamental projection operators $\overleftarrow{\bm{\gamma}}$ in Eqs.~(\ref{eq:ProjectionFiducialE}) or (\ref{eq:ProjectionFiducialB}) respectively.
That is, for a tensor field tangent to $\mathcal{S}$, the projection of the spacetime covariant derivative to $\mathcal{S}$ is just the position space covariant derivative.
The essence of the proof is to show that the key property of a Levi-Civita connection---that the gradient of the metric vanishes---is satisfied by $\overleftarrow{\bm{\gamma}}^* \bm{\nabla}\bm{\gamma} = 0$ on $\mathcal{E}$ and by $\overleftarrow{\bm{\gamma}}^* \bm{\mathscr{D}}\bm{\gamma}$ on $B\mathcal{E}$ or $B\mathcal{G}$; then the uniqueness of the Levi-Civita connection implies that $\overleftarrow{\bm{\gamma}}^* \bm{\nabla} = \bm{D}$ and $\overleftarrow{\bm{\gamma}}^* \bm{\mathscr{D}} = \bm{D}$ respectively.
Using Eq.~(\ref{eq:ExtendedThreeMetricE}) for $\bm{\gamma}$ on $\mathcal{E}$,
\[
\left( \overleftarrow{\bm{\gamma}}^* \bm{\nabla}\bm{\gamma} \right)_{\rho\mu\nu} 
	= {\gamma^\gamma}_\rho \, {\gamma^\alpha}_\mu \, {\gamma^\beta}_\nu \,
		\nabla_\gamma \left( g_{\alpha\beta} - n_\alpha \chi_\beta \right)
	= 0 
\quad \quad (\text{on } \mathcal{E})
\]
because $\bm{\nabla} \bm{g} = 0$ and $\underline{\bm{n}} \cdot \overleftarrow{\bm{\gamma}} = 0$ and $\underline{\bm{\chi}} \cdot \overleftarrow{\bm{\gamma}} = 0$.
Similarly using Eq.~(\ref{eq:ExtendedThreeMetricB}) for $\bm{\gamma}$ on $B\mathcal{E}$ or $B\mathcal{G}$,
\[
\left( \overleftarrow{\bm{\gamma}}^* \bm{\mathscr{D}}\bm{\gamma} \right)_{K I J} 
	= {\gamma^C}_K \, {\gamma^A}_I \, {\gamma^B}_J \,
		\mathscr{D}_C \left( G_{A B} - \mathscr{n}_A \chi_B - \xi_A \mathscr{n}_B  \right)
	= 0 
\quad \quad (\text{on } B\mathcal{E})
\]
because $\bm{\mathscr{D}} \bm{G} = 0$ and $\underline{\bm{\mathscr{n}}} \cdot \overleftarrow{\bm{\gamma}} = 0$ and $\underline{\bm{\chi}} \cdot \overleftarrow{\bm{\gamma}} = 0$ and $\underline{\bm{\xi}} \cdot \overleftarrow{\bm{\gamma}} = 0$.

\section{Extrinsic Geometry of Spacelike Leaves}
\label{sec:ExtrinsicGeometry}

An extrinsic curvature tensor carries information about the way the spacelike leaves of a foliation are embedded in the ambient spacetime. 
This can be understood in at least three ways.
First, extrinsic curvature can be understood as relating a tensor gradient on a spacelike leaf $\mathcal{S}$ to a tensor gradient on the ambient spacetime.
Second, and closely related, an extrinsic curvature tensor can be understood as giving the variation along $\mathcal{S}$ of a 1-form or vector field normal to $\mathcal{S}$.
And third, less obviously related to the other two perspectives, is that an extrinsic curvature tensor describes the change of the induced metric along a normal vector connecting neighboring leaves.

\subsection{Extrinsic Geometry of Spacelike Leaves of Einstein Spacetime $\mathcal{E}$}
\label{sec:ExtrinsicGeometryE}

The relationship between the covariant derivatives $\bm{D}_{\bm{u}} \bm{v}$ and $\bm{\nabla}_{\!\bm{u}} \bm{v}$ of a vector field $\bm{v}$ tangent to $\mathcal{S}$, in a direction $\bm{u}$ also tangent to $\mathcal{S}$, provides one perspective on the extrinsic geometry of a leaf $\mathcal{S}$ of a foliation of spacetime.
Recall that $\bm{\nabla}$ is the Levi-Civita connection associated with the 4-metric $\bm{g}$ on $\mathcal{E}$, and that $\bm{D}$ is the Levi-Civita connection associated with the 3-metric $\bm{\gamma}$ on $\mathcal{S}$.

Apply Eq.~(\ref{eq:DerivativeProjection}) to the directional derivative $\bm{D}_{\bm{u}} \bm{v}$ where $\bm{u}$ and $\bm{v}$ are both vector fields tangent to $\mathcal{S}$.
Use Eq.~(\ref{eq:ProjectionFiducialE}), as well as the fact that $\underline{\bm{\chi}} \cdot \bm{v} = 0$ for $\bm{v}$ tangent to $\mathcal{S}$.
The result can be expressed
\begin{equation}
\bm{D}_{\bm{u}}  \bm{v} = \bm{\nabla}_{\!\bm{u}} \bm{v} 
	-  \bm{L}\left( \bm{u}, \bm{v} \right) \, \bm{n}
	\quad \quad (\text{on } \mathcal{E}),
\label{eq:CompareDerivativesE}
\end{equation}
where
\[
\bm{L}\left( \bm{u}, \bm{v} \right) = - \bm{g}\left( \bm{u}, \bm{\nabla}_{\!\bm{v}}  \bm{\chi} \right) 
	=  - \bm{u} \cdot  \bm{\nabla}_{\!\bm{v}} \underline{\bm{\chi}}
	\quad \quad (\text{on } \mathcal{E})
\]
is the extrinsic curvature tensor associated with $\underline{\bm{\chi}}$, describing its variation on $\mathcal{S}$.
The fact that $\bm{L}$ is symmetric with respect to arguments tangent to $\mathcal{S}$,
\[
\bm{L}\left( \bm{v}, \bm{u} \right) = \bm{L}\left( \bm{u}, \bm{v} \right),
\]
has also been used; this can be proved using the fact that $\underline{\bm{\chi}}$ is defined in terms of the gradient of the scalar function $t$ and the fact that the Levi-Civita connection $\bm{\nabla}$ is torsion-free, along with the fact that the contractions of $\bm{\nabla} t$ with $\bm{u}$ and $\bm{v}$ vanish when the latter are tangent to $\mathcal{S}$.

So far $\bm{L}$ has been defined as a tensor on $\mathcal{S}$,
but it will prove
useful to extend it to all vector fields $\bm{u}$, $\bm{v}$ on $\mathcal{E}$ in order to relate it to all the spacetime components of $\bm{\nabla}\underline{\bm{\chi}}$.
This is done by evaluating 
\[
\bm{L}\left( \bm{u}, \bm{v} \right) 
	= \bm{L} \left( \overleftarrow{\bm{\gamma}}(\bm{u}), \overleftarrow{\bm{\gamma}}(\bm{v}) \right) 
	= - \overleftarrow{\bm{\gamma}}(\bm{u}) \cdot  \bm{\nabla}_{\!\overleftarrow{\bm{\gamma}}(\bm{v})} \underline{\bm{\chi}}.
\]
Because of the structure of the projection tensor $\overleftarrow{\bm{\gamma}}$ in Eq.~(\ref{eq:ProjectionFiducialE}), in the course of this computation the `generalized acceleration' 
$\bm{\nabla}_{\! \bm{n}} \underline{\bm{\chi}}$ of Eq.~(\ref{eq:AlternativeAccelerationE}) appears, with the result
\[
\bm{L} = 
	- \bm{\nabla}\underline{\bm{\chi}} - \underline{\bm{a}} \otimes \underline{\bm{\chi}}
	\quad \quad (\text{on } \mathcal{E}),
\]
or in components
\[
L_{\mu \nu} = 
	- \nabla_{\! \nu}  \chi_\mu - a_\mu \chi_\nu
	\quad \quad (\text{on } \mathcal{E}).
\]
The trace is 
\[
L = - \bm{\nabla} \cdot \bm{\chi} = - \nabla_{\! \alpha}  \chi^\alpha
	\quad \quad (\text{on } \mathcal{E}).
\]
Again, this expression for $\bm{L}$ extended from $\mathcal{S}$ to $\mathcal{E}$ is valid for evaluation on any vector fields $\bm{u}$ and $\bm{v}$ on $\mathcal{E}$.

Because $\underline{\bm{\chi}} = - \underline{\bm{n}} / c^2$ on $\mathcal{E}$, one can define an alternative extrinsic curvature tensor $\bm{K}$ based on $\underline{\bm{n}}$ instead of $\underline{\bm{\chi}}$ via
\begin{equation}
\bm{L} = -\frac{1}{c^2} \, \bm{K}
	\quad \quad (\text{on } \mathcal{E}),
\label{eq:ExtrinsicCurvatureRelationE}
\end{equation}
where
\[
\bm{K}\left( \bm{u}, \bm{v} \right) = - \bm{g}\left( \bm{u}, \bm{\nabla}_{\!\bm{v}}  \bm{n} \right) 
	=  - \bm{u} \cdot  \bm{\nabla}_{\!\bm{v}} \underline{\bm{n}}
	\quad \quad (\text{on } \mathcal{E})
\]
for $\bm{u}$ and $\bm{v}$ tangent to $\mathcal{S}$, and rewrite the above equations accordingly, including
\begin{equation}
\bm{D}_{\bm{u}} \bm{v} = \bm{\nabla}_{\!\bm{u}} \bm{v} 
	-  \bm{K}\left( \bm{u}, \bm{v} \right) \, \bm{\chi}
\quad \quad (\text{on } \mathcal{E})
\label{eq:CompareDerivativesE_2}
\end{equation}
for the comparison of space and spacetime gradients,
\[
\bm{K} = 
	- \bm{\nabla}\underline{\bm{n}} - \underline{\bm{a}} \otimes \underline{\bm{n}}
	\quad \quad (\text{on } \mathcal{E}),
\]
for the relation between extrinsic curvature and $\bm{\nabla}\underline{\bm{n}}$, and
\[
K = - \bm{\nabla} \cdot \bm{n} = - \nabla_{\! \alpha}  n^\alpha
	\quad \quad (\text{on } \mathcal{E})
\]
for the trace.
Regarding $\bm{K}$ as primary, the expressions
\begin{equation}
\begin{aligned}
\bm{\nabla}\underline{\bm{\chi}} &= \frac{1}{c^2} \, \bm{K} 
	 - \underline{\bm{a}} \otimes \underline{\bm{\chi}}, \\[5pt]
\bm{\nabla}\underline{\bm{n}} &= - \bm{K} 
	 - \underline{\bm{a}} \otimes \underline{\bm{n}}
\end{aligned}
	\quad \quad (\text{on } \mathcal{E})
\label{eq:GradientsFiducialCovectorsE}
\end{equation}
will prove to be useful expressions for the gradients of $\underline{\bm{\chi}}$ and $\underline{\bm{n}}$.

This latter version $\bm{K}$ of the extrinsic curvature shows up in the time evolution of the 3-metric $\bm{\gamma}$---its variation from leaf to leaf of the foliation, given by its Lie derivative along the normal time evolution vector $\bm{m}$ introduced in Sec.~\ref{sec:MetricE}:
\[
\mathscr{L}_{\!\bm{m}}   \gamma_{\mu \nu} = 
	m^\alpha \, \nabla_\alpha \gamma_{\mu \nu} +
	\gamma_{\alpha \nu} \, \nabla_\mu m^\alpha +
	\gamma_{\mu \alpha} \, \nabla_\nu m^\alpha
	\quad \quad (\text{on } \mathcal{E}).
\]
Using Eqs.~(\ref{eq:DualCollinearE}), (\ref{eq:NormalEvolutionVectorE}), (\ref{eq:ExtendedThreeMetricE}), (\ref{eq:FiducialAccelerationE}), (\ref{eq:AccelerationComponentsE}), (\ref{eq:AlternativeAccelerationE}), and (\ref{eq:GradientsFiducialCovectorsE}) one finds
\begin{equation}
\bm{\mathscr{L}}_{\!\bm{m}}  \bm{\gamma} = -2 \, \alpha \, \bm{K}
	\quad \quad (\text{on } \mathcal{E}).
\label{eq:MetricEvolutionE}
\end{equation}
This provides another perspective on the meaning of the extrinsic curvature.
In contrast, a related calculation for the projection tensor $\overleftarrow{\bm{\gamma}}$ shows that
\begin{equation}
\bm{\mathscr{L}}_{\!\bm{m}}  \overleftarrow{\bm{\gamma}} = 0
	\quad \quad (\text{on } \mathcal{E}),
\label{eq:ProjectionEvolutionE}
\end{equation}
with the important consequence that for any tensor field $\bm{T}$ tangent to $\mathcal{S}$, the evolved tensor field $\bm{\mathscr{L}}_{\!\bm{m}} \bm{T}$ is also tangent to $\mathcal{S}$.

Looking ahead to a comparison with $B\mathcal{E}$ and $B\mathcal{G}$, an explicit component expression for $\bm{K}$ will be useful. From Eq.~(\ref{eq:GradientsFiducialCovectorsE}) and Eqs.~(\ref{eq:AccelerationPartialDerivativesE}) and (\ref{eq:HandyRelationChi}),
\[
K_{\mu\nu} = 
	- c^2 \alpha \, ^{\bm{g}} \! {\Gamma^0}_{\mu \nu}
	+ c^2 ( \chi_\mu \, \partial_\nu \ln \alpha +  \chi_\nu \, \partial_\mu \ln \alpha 
			- \chi_\mu \, \chi_\nu \, n^\alpha \, \partial_\alpha \ln \alpha ).
\]
Noticing that
\[
^{\bm{g}} \! {\Gamma^0}_{\mu \nu} 
	=  g^{0\alpha} \, ^{\bm{g}} \! {\Gamma}_{\alpha \mu \nu}
	= -\frac{1}{c^2\alpha} \, n^\alpha \, ^{\bm{g}} \! {\Gamma}_{\alpha \mu \nu}
\]
and using Eq.~(\ref{eq:Christoffel_g_G}) along with $n^\alpha  \chi_\alpha = 1$ yields the compact result
\begin{equation}
K_{\mu\nu} =  n^\alpha \, ^{\bm{G}} \! {\Gamma}_{\alpha \mu \nu}
\quad \quad (\text{on }\mathcal{E}).
\label{eq:ExtrinsicCurvatureComponentsE}
\end{equation}
This is independent of both the speed of light and derivatives of the lapse function.
Incidentally, the symmetry of the extension of $\bm{K}$ from $\mathcal{S}$ to $\mathcal{E}$ is also manifest.

\subsection{Extrinsic Geometry of Spacelike Leaves of Bargmann Spacetimes $B\mathcal{E}$ and $B\mathcal{G}$}
\label{sec:ExtrinsicGeometryB}

On $B\mathcal{E}$ or $B\mathcal{G}$ begin again with the relationship between the covariant derivatives $\bm{D}_{\bm{u}} \bm{v}$ and $\bm{\mathscr{D}}_{\!\bm{u}} \bm{v}$ of a vector field $\bm{v}$ tangent to $\mathcal{S}$, in a direction $\bm{u}$ also tangent to $\mathcal{S}$, where 
$\bm{\mathscr{D}}$ is the Levi-Civita connection associated with the 5-metric $\bm{G}$ and once again $\bm{D}$ is the Levi-Civita connection associated with the 3-metric $\bm{\gamma}$ on $\mathcal{S}$.
Applying Eq.~(\ref{eq:DerivativeProjection}) to the directional derivative $\bm{D}_{\bm{u}} \bm{v}$ where $\bm{u}$ and $\bm{v}$ are both vector fields tangent to $\mathcal{S}$, but now using Eq.~(\ref{eq:ProjectionFiducialB}), as well as the fact that $\underline{\bm{\chi}} \cdot \bm{v} = 0$ and $\underline{\bm{\mathscr{n}}} \cdot \bm{v} = 0$ for $\bm{v}$ tangent to $\mathcal{S}$, one finds
\begin{equation}
\bm{D}_{\bm{u}} \bm{v} = \bm{\mathscr{D}}_{\!\bm{u}} \bm{v} 
	-  \bm{L}\left( \bm{u}, \bm{v} \right) \, \bm{\mathscr{n}}  -  \bm{K}\left( \bm{u}, \bm{v} \right) \, \bm{\xi}
	\quad \quad (\text{on } B\mathcal{E} \text{ or } B\mathcal{G}).
\label{eq:CompareDerivativesB}
\end{equation}
There is an additional term relative to Eq.~(\ref{eq:CompareDerivativesE}).
Here
\[
\begin{aligned}
\bm{L}\left( \bm{u}, \bm{v} \right) &= - \bm{G}\left( \bm{u}, \bm{\mathscr{D}}_{\!\bm{v}} \bm{\chi} \right) 
	=  - \bm{u} \cdot  \bm{\mathscr{D}}_{\!\bm{v}} \underline{\bm{\chi}}, \\
\bm{K}\left( \bm{u}, \bm{v} \right) &= - \bm{G}\left( \bm{u}, \bm{\mathscr{D}}_{\!\bm{v}} \bm{\mathscr{n}} \right) 
	=  - \bm{u} \cdot  \bm{\mathscr{D}}_{\!\bm{v}} \underline{\bm{\mathscr{n}}}
\end{aligned}
	\quad \quad (\text{on } B\mathcal{E} \text{ or } B\mathcal{G})
\]
are extrinsic curvature tensors associated with $\underline{\bm{\chi}}$ and $\underline{\bm{\mathscr{n}}}$, describing their variation on $\mathcal{S}$. 
As on $\mathcal{E}$, they are symmetric with respect to arguments tangent to $\mathcal{S}$:
\[
\bm{L}\left( \bm{v}, \bm{u} \right) = \bm{L}\left( \bm{u}, \bm{v} \right), \quad \quad
\bm{K}\left( \bm{v}, \bm{u} \right) = \bm{K}\left( \bm{u}, \bm{v} \right).
\]
To extend the definition from $\mathcal{S}$ to $B\mathcal{E}$ or $B\mathcal{G}$, evaluate
\[
\begin{aligned}
\bm{L}\left( \bm{u}, \bm{v} \right) 
	&= \bm{L} \left( \overleftarrow{\bm{\gamma}}(\bm{u}), \overleftarrow{\bm{\gamma}}(\bm{v}) \right) 
	= - \overleftarrow{\bm{\gamma}}(\bm{u}) \cdot  \bm{\mathscr{D}}_{\!\overleftarrow{\bm{\gamma}}(\bm{v})} \underline{\bm{\chi}}, \\
\bm{K}\left( \bm{u}, \bm{v} \right) 
	&= \bm{K} \left( \overleftarrow{\bm{\gamma}}(\bm{u}), \overleftarrow{\bm{\gamma}}(\bm{v}) \right) 
	= - \overleftarrow{\bm{\gamma}}(\bm{u}) \cdot  \bm{\mathscr{D}}_{\!\overleftarrow{\bm{\gamma}}(\bm{v})} \underline{\bm{\mathscr{n}}}.
\end{aligned}
\]
The structure of the projection tensor $\overleftarrow{\bm{\gamma}}$ in Eq.~(\ref{eq:ProjectionFiducialB}) makes generalized accelerations from Eqs.~(\ref{eq:FiducialAccelerationB}), (\ref{eq:AdditionalAccelerationB}), and (\ref{eq:AlternativeAccelerationB}) appear, with the result
\begin{equation}
\begin{aligned}
\bm{L} &= 
	- \bm{\mathscr{D}}\underline{\bm{\chi}} - \underline{\bm{a}} \otimes \underline{\bm{\chi}}
	+ \frac{1}{2} \, \underline{\bm{a}} \otimes \underline{\bm{\xi}}
	+ \frac{1}{2} \, \underline{\bm{\xi}} \otimes \underline{\bm{a}}, \\
\bm{K} &= 
	- \bm{\mathscr{D}}\underline{\bm{\mathscr{n}}}
	- \frac{1}{2} \, \underline{\bm{a}} \otimes \underline{\bm{\mathscr{n}}}
	- \frac{1}{2} \, \underline{\bm{\mathscr{n}}} \otimes \underline{\bm{a}}
\end{aligned}
	\quad \quad (\text{on } B\mathcal{E} \text{ or } B\mathcal{G}),
\label{eq:ExtrinsicCurvaturesB}
\end{equation}
or in components
\begin{equation}
\begin{aligned}
L_{I J} &= 
	- \mathscr{D}_{\! J} \chi_I - a_I \chi_J
	+ \frac{1}{2} \, a_I \xi_J
	+ \frac{1}{2} \, \xi_I a_J, \\
K_{I J} &= 
	- \mathscr{D}_{\! J} n_I
	- \frac{1}{2} \, a_I n_J
	- \frac{1}{2} \, n_I a_J
\end{aligned}
	\quad \quad (\text{on } B\mathcal{E} \text{ or } B\mathcal{G}).
\label{eq:ExtrinsicCurvatureComponentsB}
\end{equation}
The identities 
\[
\bm{\mathscr{n}} \cdot \bm{\mathscr{D}}_{\!\bm{v}} \underline{\bm{\chi}}
	= - \bm{v} \cdot \bm{\mathscr{D}}_{\!\bm{\chi}} \underline{\bm{\mathscr{n}}},
\quad \quad
\bm{\xi} \cdot \bm{\mathscr{D}}_{\!\bm{v}} \underline{\bm{\chi}}
	= - \frac{1}{c^2} \, \bm{v} \cdot \bm{\mathscr{D}}_{\!\bm{\xi}} \underline{\bm{\mathscr{n}}},
\quad \quad
\bm{\xi} \cdot \bm{\mathscr{D}}_{\!\bm{v}} \underline{\bm{\mathscr{n}}}
	= - \frac{1}{c^2} \, \bm{v} \cdot \bm{\mathscr{D}}_{\!\bm{\xi}} \underline{\bm{\mathscr{n}}}
\]
have also been employed; these can be proved with the definition of $\underline{\bm{\mathscr{n}}}$ in terms of the gradient of coordinate $\mathscr{X}^4 = \eta$ in Eq.~(\ref{eq:FiducialObserverCovectorB}), the definition of the action vector in Eq.~(\ref{eq:ActionVector}), the norms in Eq.~(\ref{eq:NormsB}), and the fact that the connection $\bm{\mathscr{D}}$ is torsion-free. 
The traces are 
\[
\begin{aligned}
L &= - \bm{\mathscr{D}} \cdot \bm{\chi} = - \mathscr{D}_{\! A}  \chi^A, \\
K &= - \bm{\mathscr{D}} \cdot \bm{\mathscr{n}} = - \mathscr{D}_{\! A}  n^A
\end{aligned}
	\quad \quad (\text{on } B\mathcal{E} \text{ or } B\mathcal{G}).
\]
Again, these expressions for $\bm{K}$ and $\bm{L}$ extended from $\mathcal{S}$ to $B\mathcal{E}$ or $B\mathcal{G}$ are valid for evaluation on any vector fields $\bm{u}$ and $\bm{v}$ on $B\mathcal{E}$ or $B\mathcal{G}$.

While there is no a priori justification to assume that the extrinsic curvatures on $B\mathcal{E}$ or $B\mathcal{G}$ are the same as the $\bm{K}$ and $\bm{L}$ defined on $\mathcal{E}$, this does in fact turn out to be case---for both $\bm{K}$ and $\bm{L}$ in the case of $B\mathcal{E}$, and for $\bm{K}$ in the case of $B\mathcal{G}$.
First consider $\bm{K}$.
From Eq.~(\ref{eq:ExtrinsicCurvatureComponentsB}),
\begin{equation}
K_{I J} = {^{\bm{G}} \! {\Gamma^A}_{I J}} \, \mathscr{n}_A 
	- \frac{1}{2} \, a_I \mathscr{n}_J
	- \frac{1}{2} \, \mathscr{n}_I a_J
	\quad \quad (\text{on } B\mathcal{E} \text{ or } B\mathcal{G}).
\label{eq:ExtrinsicCurvature_K_ComponentsB}
\end{equation}
Because $\mathscr{n}_4$ is the only non-vanishing component of $\underline{\mathscr{n}}$, the result for the traditional spacetime indices $\mu, \nu \in \left\{ 0, 1, 2, 3 \right\}$ is
\begin{equation}
K_{\mu\nu} =  \mathscr{n}^A \, {^{\bm{G}} \! {\Gamma}_{A \mu \nu}}
	= n^\alpha \, {^{\bm{G}} \! {\Gamma}_{\alpha \mu \nu}}
	\quad \quad (\text{on } B\mathcal{E} \text{ or } B\mathcal{G}),
\label{eq:ExtrinsicCurvatureComponentsB}
\end{equation}
precisely the same as Eq.~(\ref{eq:ExtrinsicCurvatureComponentsE}) on $\mathcal{E}$.
As to the action components,
\[
K_{4 J} = K_{J 4} = - {^{\bm{G}} \! {\Gamma^4}_{J 4}} + \frac{1}{2} \, a_J = 0
	\quad \quad (\text{on } B\mathcal{E} \text{ or } B\mathcal{G}),
\]
using a connection coefficient already encountered in Sec.~\ref{sec:LeviCivitaB}.
Displayed together,
\begin{equation}
\left[ K_{I J} \right] = \begin{bmatrix} K_{\mu\nu} & 0_\mu \\ 0_\nu & 0 \end{bmatrix}
	\quad \quad (\text{on } B\mathcal{E} \text{ or } B\mathcal{G}),
\label{eq:ExtrinsicCurvatureMatrixB}
\end{equation}
and in this sense $\bm{K}$ is the same on $B\mathcal{E}$ or $B\mathcal{G}$ as on $\mathcal{E}$.
Moreover, at first glance it appears that, since the leaves of the foliation are of codimension 2, there are two independent extrinsic curvatures $\bm{L}$ and $\bm{K}$.
However, it turns out that the nature of the Bargmann spacetimes is such that there are not two independent extrinsic curvatures; instead,
\begin{equation}
\bm{L} = 
\left\{ \begin{aligned}
- \frac{1}{c^2} \, \bm{K} & & (\mathrm{on \ } B\mathcal{E}) \\[5pt]
0  & & (\mathrm{on \ } B\mathcal{G}).
\end{aligned} \right.
\label{eq:ExtrinsicCurvatureRelationB}
\end{equation}
Remarkably, the straightforward definition of Eq.~(\ref{eq:ExtrinsicCurvatureRelationE}) motivated by the collinearity of $\underline{\bm{\chi}}$ and $\underline{\bm{n}}$ on $\mathcal{E}$ is preserved on $B\mathcal{E}$ as a nontrivial relation.
On $B\mathcal{G}$, the extrinsic curvature $\bm{K}$ is the same as on $B\mathcal{E}$ and on $\mathcal{E}$, but $\bm{L}$ vanishes.
To prove Eq.~(\ref{eq:ExtrinsicCurvatureRelationB}), note that
\[
L_{IJ} = -\chi_I \, \partial_J \ln \alpha 
	+ \alpha \,  {^{\bm{G}} \! {\Gamma^0}_{I J}}
	- a_I \chi_J + \frac{1}{2} \, a_I \xi_J + \frac{1}{2} \, \xi_I a_J \\
	\quad \quad (\text{on } B\mathcal{E} \text{ or } B\mathcal{G})
 \]
follows from Eq.~(\ref{eq:ExtrinsicCurvatureComponentsB}) and the definition of $\underline{\sfchi}$ in Eq.~(\ref{eq:DualFiducialObserverCovectorB}).
Then
\[
\begin{aligned}
{^{\bm{G}} \! {\Gamma^0}_{I J}} &= G^{0A} \, {^{\bm{G}} \! {\Gamma}_{A I J}},
& & 
G^{0A} = \left\{ \begin{aligned}
		& \frac{1}{\alpha} \left( \xi^A - \frac{1}{c^2} \, \mathscr{n}^A \right) 
			& & (\mathrm{on \ } B\mathcal{E}) \\[5pt]
		& \frac{1}{\alpha} \, \xi^A  & & (\mathrm{on \ } B\mathcal{G}),
	\end{aligned} \right. \\[5pt]
G_{0A} &= - \chi_A, 
& &
 {^{\bm{G}} \! {\Gamma}_{4 I J}} = -\frac{1}{2} \left( \chi_J \, \partial_I \ln \alpha 
							+ \chi_I \, \partial_J \ln \alpha \right),
\end{aligned}
\]
together with Eq.~(\ref{eq:ActionVector}) for $\sfxi$ and Eq.~(\ref{eq:AccelerationPartialDerivativesE}) for $\underline{\mathsf{a}}$ and Eq.~(\ref{eq:ExtrinsicCurvature_K_ComponentsB}) for $\mathsf{K}$, produce the desired result. 

Using Eq.~(\ref{eq:ActionVector}), one consequence of Eq.~(\ref{eq:ExtrinsicCurvatureRelationB}) is that Eq.~(\ref{eq:CompareDerivativesB}), valid for vector fields $\bm{u}, \bm{v}$ tangent to $\mathcal{S}$, simplifies to 
\begin{equation}
\bm{D}_{\bm{u}} \bm{v} = \bm{\mathscr{D}}_{\!\bm{u}} \bm{v} 
	 -  \bm{K}\left( \bm{u}, \bm{v} \right) \, \bm{\chi}
\quad \quad (\text{on } B\mathcal{E} \text{ or } B\mathcal{G}),
\label{eq:CompareDerivativesB_2}
\end{equation}
of the same form as Eq.~(\ref{eq:CompareDerivativesE_2}) on $\mathcal{E}$ (though the vector version of $\bm{\chi}$ on $B\mathcal{E}$ or $B\mathcal{G}$ differs from that on $\mathcal{E}$, even though the covector versions $\underline{\bm{\chi}}$ are the same).
Referring back to Eq.~(\ref{eq:ExtrinsicCurvaturesB}), another consequence of Eq.~(\ref{eq:ExtrinsicCurvatureRelationB}) is that the equations for $\bm{\mathscr{D}}\underline{\bm{\chi}}$ and $\bm{\mathscr{D}}\underline{\bm{\mathscr{n}}}$ required later can be expressed
\begin{equation}
\begin{aligned}
\bm{\mathscr{D}}\underline{\bm{\chi}} &=
\left\{ \begin{aligned} 
	\frac{1}{c^2} \, \bm{K}  - \underline{\bm{a}} \otimes \underline{\bm{\chi}}
	+ \frac{1}{2} \, \underline{\bm{a}} \otimes \underline{\bm{\xi}}
	+ \frac{1}{2} \, \underline{\bm{\xi}} \otimes \underline{\bm{a}} 
		& & (\mathrm{on \ } B\mathcal{E}) \\[5pt]
	\phantom{\frac{1}{c^2} \, \bm{K} } - \underline{\bm{a}} \otimes \underline{\bm{\chi}}
	+  \frac{1}{2} \, \underline{\bm{a}} \otimes \underline{\bm{\xi}}
	+ \frac{1}{2} \, \underline{\bm{\xi}} \otimes \underline{\bm{a}} 
		& & (\mathrm{on \ } B\mathcal{G}), \\[5pt]
\end{aligned} \right. \\
\bm{\mathscr{D}}\underline{\bm{\mathscr{n}}}&= 
	- \bm{K} 
	- \frac{1}{2} \, \underline{\bm{a}} \otimes \underline{\bm{\mathscr{n}}}
	- \frac{1}{2} \, \underline{\bm{\mathscr{n}}} \otimes \underline{\bm{a}}
		\quad \quad (\text{on } B\mathcal{E} \text{ or } B\mathcal{G}).
\label{eq:GradientsFiducialCovectorsB}
\end{aligned}
\end{equation}
Moreover
\begin{equation}
\bm{\mathscr{D}}\underline{\bm{\xi}} =
	- \frac{1}{2} \, \underline{\bm{a}} \otimes \underline{\bm{\chi}}
	+ \frac{1}{2} \, \underline{\bm{\chi}} \otimes \underline{\bm{a}}
	\quad \quad (\text{on } B\mathcal{E} \text{ or } B\mathcal{G}),\label{eq:GradientActionCovectorB}
\end{equation}
thanks to Eq.~(\ref{eq:ActionVector}),
with trace
\begin{equation}
\bm{\mathscr{D}} \cdot \bm{\xi} = \mathscr{D}_A \, \xi^A = 0
	\quad \quad (\text{on } B\mathcal{E} \text{ or } B\mathcal{G}).
\label{eq:ActionTraceB}
\end{equation} 
And indeed $\bm{\mathscr{D}}\underline{\bm{\xi}} = \bm{\mathscr{D}}\underline{\bm{\chi}}$ in the case of $B\mathcal{G}$, as expected from Eq.~(\ref{eq:ActionVector}). 

As on $\mathcal{E}$, the time evolution of the 3-metric $\bm{\gamma}$ is once again given in terms of $\bm{K}$.
Using a calculation similar to that on $\mathcal{E}$ but using quantities appropriate to $B\mathcal{E}$ or $B\mathcal{G}$, the Lie derivative of $\bm{\gamma}$ along $\bm{\mathscr{m}}$,
\[
\mathscr{L}_{\!\bm{\mathscr{m}}}   \gamma_{I J} = 
	\mathscr{m}^A \, \mathscr{D}_A \gamma_{I J} +
	\gamma_{A J} \, \mathscr{D}_I \mathscr{m}^A +
	\gamma_{I A} \, \mathscr{D}_J \mathscr{m}^A
	\quad \quad (\text{on } B\mathcal{E} \text{ or } B\mathcal{G}),
\]
results in
\begin{equation}
\bm{\mathscr{L}}_{\!\bm{\mathscr{m}}}  \bm{\gamma} = -2 \, \alpha \, \bm{K}
	\quad \quad (\text{on } B\mathcal{E} \text{ or } B\mathcal{G})
\label{eq:MetricEvolutionB}
\end{equation}
as also found previously on $\mathcal{E}$.
(The `quantities appropriate to $B\mathcal{E}$ or $B\mathcal{G}$' are $\bm{\mathscr{D}}$ and $\bm{\mathscr{m}}$, instead of $\bm{\nabla}$ and $\bm{m}$ on $\mathcal{E}$.)
The similarly calculated result
\begin{equation}
\bm{\mathscr{L}}_{\!\bm{\mathscr{m}}}  \overleftarrow{\bm{\gamma}} = 0
	\quad \quad (\text{on } B\mathcal{E} \text{ or } B\mathcal{G})
\label{eq:ProjectionEvolutionB}
\end{equation}
found previously on $\mathcal{E}$ holds also on $B\mathcal{E}$ or $B\mathcal{G}$, yielding the same important consequence that for any tensor field $\bm{T}$ tangent to $\mathcal{S}$, the evolved tensor field $\bm{\mathscr{L}}_{\!\bm{\mathscr{m}}} \bm{T}$ is also tangent to $\mathcal{S}$.

It is worth emphasizing that, even though the leaves $\mathcal{S}$ are of codimension 2, such that in a generic case two independent extrinsic curvature tensors might have been expected, the Bargmann extensions of Poincar\'e and Galilei physics are such that the two extrinsic curvature tensors are related by a constant factor; there are no new degrees of freedom introduced, a result not unexpected in hindsight.
In the case of $B\mathcal{G}$ where $c \rightarrow \infty$, this constant factor is $0$, that is, $\bm{L}$ vanishes, even while $\bm{K}$ does not.
Therefore equations henceforth will be expressed in terms of $\bm{K}$ in this paper, in order to make factors of $c$ explicit so as to manifestly exhibit the relation between Poincar\'e and Galilei physics, as for example in Eq.~(\ref{eq:GradientsFiducialCovectorsB}).

\section{Obtaining Physical Laws on Bargmann Spacetimes}
\label{sec:PhysicalLawsBargmann}

The presentation thus far has considered the 4-metric $\bm{g}$ on an Einstein spacetime $\mathcal{E}$ and the 5-metric $\bm{G}$ on a Bargmann spacetime $B\mathcal{E}$ or $B\mathcal{G}$; the decompositions of these and other tensors according to a foliation of spacetime into spacelike position space leaves with 3-metric $\bm{\gamma}$; and 
multiple interpretations of the extrinsic curvature tensor $\bm{K}$, which has been shown here to be the same on $B\mathcal{E}$ or $B\mathcal{G}$ as on $\mathcal{E}$.
Apart from a few allusions to the motion of material particles, for the most part it is geometry that has been discussed to this point, in particular aspects related to spacetime foliations. 
Physics does not really enter the picture until physical laws are given in the form of tensor equations on spacetime.

How are physical laws embodying a unified perspective on Poincar\'e and Galilei physics on curved Bargmann spacetimes $B\mathcal{E}$ and $B\mathcal{G}$ to be obtained? 
Recall the general strategy mentioned in Sec.~\ref{sec:Introduction}: first, re-express a known physical law on the usual Einstein spacetime $\mathcal{E}$ as a law with the same physical content but in a form appropriate to Bargmann-Einstein spacetime $B\mathcal{E}$; and second, consider a $c \rightarrow \infty$ limit of that law appropriate to the closely related Bargmann-Galilei spacetime $B\mathcal{G}$.
The first step of this procedure---the \textit{encoding} on $B\mathcal{E}$ of known Poincar\'e physics on $\mathcal{E}$---might be accomplished by reverse engineering equations that express the \textit{decoding} of physics on $B\mathcal{E}$ back to physics on $\mathcal{E}$.
Such decoding expressions involve an operator $\overleftarrow{\bm{g}}^*$ that relates tensors on $B\mathcal{E}$ to tensors on $\mathcal{E}$, in a manner characterized by important comparisons and contrasts with the operators $\overleftarrow{\bm{\gamma}}^*$ in Eqs.~(\ref{eq:GeneralProjectionOperatorE}) and (\ref{eq:GeneralProjectionOperatorB}) which respectively project tensors from spacetime $\mathcal{E}$ or $B\mathcal{E}$ to a position space slice $\mathcal{S}$.

In order to understand these comparisons and contrasts, it is helpful first to review the several relationships between tensors on spacetimes $\mathcal{E}$ or $B\mathcal{E}$ and tensors on a leaf $\mathcal{S}$ of the spacetime foliation; this review then fosters a working understanding of the relationships between tensors on spacetime $\mathcal{E}$ and tensors on spacetime $B\mathcal{E}$.
The entire matter is unlocked by the following key insight: the structure of a Bargmann spacetime is such that it furnishes relationships between tensors on $\mathcal{E}$ and $B\mathcal{E}$ that are partly analogous to those associated with the embedding of the leaves $\mathcal{S}$ in $\mathcal{E}$ or $B\mathcal{E}$, but with the roles of $(p,0)$ tensors (including vectors) and $(0,q)$ tensors (including 1-forms) reversed.

Following the elucidation of these relationships and some properties of the decoding operator $\overleftarrow{\bm{g}}$, a proposed procedure for reverse engineering physical laws on Einstein spacetime $\mathcal{E}$ to yield tensor relations on Bargmann-Einstein spacetime $B\mathcal{E}$ is described and illustrated with an elementary example.

\subsection{Relationships between Tensors on Spacetime $\mathcal{E}$ or $B\mathcal{E}$ and Tensors on Position Space Leaves $\mathcal{S}$}
\label{sec:TensorRelationshipsEmbedding}

Here the several relationships between tensors on spacetimes $\mathcal{E}$ or $B\mathcal{E}$ and tensors on a leaf $\mathcal{S}$ of the spacetime foliation will be recalled. 
Begin with relationships that exist naturally by virtue of the embedding of $\mathcal{S}$ in $\mathcal{E}$ or $B\mathcal{E}$: the \textit{push-forward} of a vector field (or more generally any $(p,0)$ tensor field) from $\mathcal{S}$ to $\mathcal{E}$ or $B\mathcal{E}$, and the \textit{pull-back} of a 1-form (or more generally any $(0,p)$ tensor field) from $\mathcal{E}$ or $B\mathcal{E}$ to $\mathcal{S}$. 
Then consider the relationships in the opposite directions, which differ in that they make explicit reference to 1-forms and vector fields normal to $\mathcal{S}$.
These relations are the \textit{projection} of a vector field (or more generally any $(p,0)$ tensor field) from $\mathcal{E}$ or $B\mathcal{E}$ to $\mathcal{S}$, and the \textit{extension} of a 1-form (or more generally any $(0,q)$ tensor field) from $\mathcal{S}$ to $\mathcal{E}$ or $B\mathcal{E}$.
They exist thanks to the operator $\overleftarrow{\bm{\gamma}}$ in Eqs.~(\ref{eq:ProjectionFiducialE}) and (\ref{eq:ProjectionFiducialB}).
It will prove consistent to adopt a naming convention in which tensors that originate on $\mathcal{S}$ are denoted by the \textit{same} symbol when they are regarded as or extended to tensors on $\mathcal{E}$ or $B\mathcal{E}$, while tensors induced on or projected to $\mathcal{S}$ that originated on $\mathcal{E}$ or $B\mathcal{E}$ are given a \textit{different} symbol.

Consider first the push-forward of a vector field $\bm{v}$ on $\mathcal{S}$ to the spacetime $\mathcal{E}$ or $B\mathcal{E}$. 
Note that $\mathcal{S}$ may be regarded either in isolation as a manifold in its own right or as a submanifold embedded in $\mathcal{E}$ or $B\mathcal{E}$.
As a vector field on $\mathcal{S}$ regarded in isolation as a manifold in its own right, at any point $\bm{v}$ gives the tangent vector to some curve in $\mathcal{S}$, and is represented by a 3-column  $\left[ v^i \right]$.
When $\mathcal{S}$ is regarded as a submanifold embedded in $\mathcal{E}$ or $B\mathcal{E}$, that curve in $\mathcal{S}$ and its tangent vector are now a curve that still lies in the submanifold $\mathcal{S}$ and a vector in $\mathcal{E}$ or $B\mathcal{E}$ that now happens to be tangent to the submanifold $\mathcal{S}$. In coordinates adapted to the foliation, $\bm{v}$ regarded as a vector in spacetime is now represented by a 4-column (for $\mathcal{E}$) or 5-column (for $B\mathcal{E}$) with additional vanishing components, since the vector does not `point away' from $\mathcal{S}$:
\[
\left[ v^\mu \right] = \begin{bmatrix}  0 \\[5pt] v^i \end{bmatrix} 
	\quad\quad (\text{on } \mathcal{E}), \quad\quad
\left[ v^I \right] = \begin{bmatrix}  0 \\[5pt] v^i \\[5pt] 0 \end{bmatrix} 
	\quad\quad (\text{on } B\mathcal{E}).
\]
The convention is that the same symbol (in this case, $\bm{v}$) is used to denote the vector field on $\mathcal{S}$ in isolation and its push-forward to a vector field on $\mathcal{E}$ or $B\mathcal{E}$ that happens to be tangent to $\mathcal{S}$ embedded in $\mathcal{E}$ or $B\mathcal{E}$.

Consider next the pull-back of a 1-form $\bm{\omega}$ from $\mathcal{E}$ or $B\mathcal{E}$ to a leaf $\mathcal{S}$ of the foliation.
At any point, $\bm{\omega}$ is a real-valued function of vectors tangent to $\mathcal{E}$ or $B\mathcal{E}$, and is represented by a 4-row $\left[ \omega_\nu \right]$ (on $\mathcal{E}$) or 5-row $\left[ \omega_J \right]$ (on $B\mathcal{E}$).
The pull-back of $\bm{\omega}$ to $\mathcal{S}$ amounts to the restriction of its domain to vectors tangent to $\mathcal{S}$.  
As $\mathcal{S}$ may be considered in isolation as a manifold in its own right, it is conventional to denote this restricted function as a 1-form on $\mathcal{S}$ by a different symbol---say, for instance, $\bm{\lambda}$ in the present example---even though its components are precisely the same as the position space components of $\bm{\omega}$ with respect to coordinates adapted to the foliation:
\[
\left[ \lambda_j \right] = \left[ \omega_j \right] \quad\quad (\text{on } \mathcal{S}).
\]
That is, in coordinates adapted to the foliation, while the push-forward of a vector involves simply padding the time and action components with zeros, the pull-back of a form involves simply deleting the time and action components.
The obvious example of a pull-back of a form in the present work is the 3-metric $\bm{\gamma}$ on $\mathcal{S}$, whose components with respect to coordinates adapted to the foliation are simply $\left[ \gamma_{i j} \right] = \left[ g_{i j} \right]$ (for $\mathcal{E}$) or $\left[ \gamma_{i j} \right] = \left[ G_{i j} \right]$ (for $B\mathcal{E}$).
This is what has been meant by prior statements that the Einstein metric $\bm{g}$ or the Bargmann metric $\bm{G}$ on spacetime \textit{induces} a 3-metric $\bm{\gamma}$ on the leaves of the foliation.

Turn now to the first `opposite direction' operation, the projection of a vector field from $\mathcal{E}$ or $B\mathcal{E}$ to a leaf $\mathcal{S}$ of the foliation.
A vector field $\bm{A}$ on $\mathcal{E}$ or $B\mathcal{E}$ can be decomposed into pieces tangent to $\mathcal{S}$ and normal to $\mathcal{S}$:
\[
\bm{A} = a_{\bm{n}} \, \bm{n} + \bm{a} 
	\quad\quad (\text{on } \mathcal{E}), 
\quad\quad
\bm{A} = a_{\bm{\mathscr{n}}} \, \bm{\mathscr{n}} + \bm{a} + a_{\bm{\xi}} \, \bm{\xi} 
	 \quad\quad (\text{on } B\mathcal{E}).
\]
The vector field $\bm{A}$ is represented by the 4-column or 5-column
\[
\left[ A^\mu \right] 
	= \begin{bmatrix} 
		\frac{1}{\alpha} \, a_{\bm{n}} \\[5pt]
		a^i - \frac{1}{\alpha} \, a_{\bm{n}} \, \beta^i
	\end{bmatrix}	\quad\quad (\text{on } \mathcal{E}),
\quad\quad
\left[ A^I \right] 
	= \begin{bmatrix} 
		\frac{1}{\alpha} \, a_{\bm{\mathscr{n}}} \\[5pt]
		a^i - \frac{1}{\alpha} \, a_{\bm{\mathscr{n}}} \, \beta^i \\[5pt]
		- a_{\bm{\xi}}
	\end{bmatrix}	\quad\quad (\text{on } B\mathcal{E}).
\]
That is, use of the projection operator $\overleftarrow{\bm{\gamma}}$ given by Eqs.~(\ref{eq:ProjectionFiducialE}) or (\ref{eq:ProjectionFiducialB}) yields the vector field $\bm{a} = \overleftarrow{\bm{\gamma}} \cdot \bm{A}$ tangent to $\mathcal{S}$.
When $\mathcal{S}$ is regarded in isolation as a manifold in its own right, $\bm{a}$ is represented by the 3-column $\left[ a^i \right]$.
But when $\mathcal{S}$ is regarded as a submanifold of spacetime, $\bm{a}$ is represented by the 4-column or 5-column
\[
\left[ a^\mu \right] 
	= \begin{bmatrix} 
		0 \\[5pt]
		a^i
	\end{bmatrix}	\quad\quad (\text{on } \mathcal{E}),
\quad\quad
\left[ a^I \right] 
	= \begin{bmatrix} 
		0 \\[5pt]
		a^i \\[5pt]
		0
	\end{bmatrix}	\quad\quad (\text{on } B\mathcal{E}).
\]
In general the projection to $\mathcal{S}$ involves `information loss'; and as illustrated here, the convention is that different symbols are used to denote the original vector field in spacetime and its projection to a leaf of the foliation, unless the original vector field in spacetime already happens to be tangent to $\mathcal{S}$.

Turn finally to the second `opposite direction' operation, the extension of a 1-form on $\mathcal{S}$ to $\mathcal{E}$ or $B\mathcal{E}$.
A 1-form $\bm{\sigma}$ on $\mathcal{S}$ regarded in isolation as a manifold in its own right is represented by a 3-row $\left[ \sigma_j \right]$.
Regarded as a function on vectors, an extension of $\bm{\sigma}$ to an expanded domain beyond vectors on $\mathcal{S}$ requires additional information (in contrast to the case of projection, which generally deletes information). 
In the present case of extending $\bm{\sigma}$ to $\mathcal{E}$ or $B\mathcal{E}$ when $\mathcal{S}$ is regarded as a submanifold, the `extra information' is simply that the value of $\bm{\sigma}$ vanishes for vectors normal to $\mathcal{S}$.
This is accomplished by writing
\[
\bm{\sigma} = \bm{\sigma}\left( \overleftarrow{\bm{\gamma}} \left( . \right) \right)
	= \bm{\sigma} \cdot \overleftarrow{\bm{\gamma}},
\]
with $\bm{\sigma}$ on the left the extension to $\mathcal{E}$ or $B\mathcal{E}$ and the expression  on the right the original tensor on $\mathcal{S}$, with $\overleftarrow{\bm{\gamma}}$ guaranteeing only vector arguments tangent to $\mathcal{S}$.
Another way to think about this is to imagine $\bm{\sigma}$ on $\mathcal{S}$ being raised to the 3-vector field $\overleftarrow{\bm{\sigma}} = \overleftrightarrow{\bm{\gamma}} \cdot \bm{\sigma}$, pushed forward to $\mathcal{E}$ or $B\mathcal{E}$ (acquiring vanishing time and action components), and then lowered back to a 1-form with the spacetime metric, either $\bm{\sigma} = \bm{g} \cdot \overleftarrow{\bm{\sigma}}$ or $\bm{\sigma} = \bm{G} \cdot \overleftarrow{\bm{\sigma}}$.
On $\mathcal{E}$ for example, in components this sequence of operations corresponds to $\sigma_\nu = g_{\nu \alpha} \, \gamma^{\alpha b} \, \sigma_b = \sigma_b \, {\gamma^b}_\nu$.
Thus (see the components of $\overleftarrow{\bm{\gamma}}$ in Eqs.~(\ref{eq:ProjectionFiducialE}) and (\ref{eq:ProjectionFiducialB})) the extension is represented by the 4-row or 5-row
\begin{equation}
\left[ \sigma_\nu \right] = \begin{bmatrix} \sigma_a \, \beta^a & \sigma_j \end{bmatrix}
	\quad\quad (\text{on } \mathcal{E}),
\quad\quad
\left[ \sigma_J \right] = \begin{bmatrix} \sigma_a \, \beta^a & \sigma_j & 0 \end{bmatrix}
	\quad\quad (\text{on } B\mathcal{E}).
\label{eq:ExtensionS}
\end{equation}
The acquired non-vanishing time component ensures that $\bm{\sigma} \cdot \bm{n} = 0$, consistent with the extended $\bm{\sigma}$ remaining tangent to $\mathcal{S}$.
Note that the convention is that the same symbol is used for both the original 1-form on $\mathcal{S}$ and its extension to $\mathcal{E}$ or $B\mathcal{E}$.

\subsection{Relationships between Tensors on Spacetime $\mathcal{E}$ and Tensors on Spacetime $B\mathcal{E}$}
\label{sec:TensorRelationshipsBargmann}

With this review of the relationships between tensors on spacetimes $\mathcal{E}$ or $B\mathcal{E}$ and tensors on a leaf $\mathcal{S}$ of the spacetime foliation in mind, comparison and contrast with the relationships between tensors on spacetime $\mathcal{E}$ and tensors on spacetime $B\mathcal{E}$ is in order.
As mentioned above, the structure of a Bargmann spacetime is such that it furnishes relationships between tensors on $\mathcal{E}$ and $B\mathcal{E}$ that are partly analogous to those described above associated with the embedding of the leaves $\mathcal{S}$ in $\mathcal{E}$ or $B\mathcal{E}$, but with the roles of $(p,0)$ tensors (including vectors) and $(0,q)$ tensors (including 1-forms) reversed.

Consider first a push-forward type of relationship.
As described above, vector fields (and $(p,0)$ tensor fields generally) on $\mathcal{S}$ can also be regarded as tensor fields on $\mathcal{E}$ or $B\mathcal{E}$, and in practical terms are placed into this new setting by adding vanishing time components, and also action components in the case of $B\mathcal{E}$.
But as emphasized in Paper~I, the nature of the Bargmann groups is such that 1-forms (and $(0,q)$ tensor fields generally, that is, multilinear forms) on 4-dimensional spacetime are treated in precisely this way, that is, set directly in 5-dimensional Bargmann spacetime with vanishing action components, such that they retain their identity as the `same' tensors in the new setting.
Important examples from Paper~I include the Einstein metric $\bm{g}$ on $\mathbb{M}$ (and here, $\mathcal{E}$) and the time form $\bm{\tau}$ on $\mathbb{G}$, which retain their causality-governing function in the Bargmann spacetimes $B\mathbb{M}$ and $B\mathbb{G}$ (and here, the curved Bargmann spacetimes $B\mathcal{E}$ and $B\mathcal{G}$). 
The 2-forms of electrodynamics are additional examples from Paper~I.
Notice retention of the same symbol to denote the `same tensor'.

Consider next a pull-back type of relationship.
As described above, 1-forms (and $(0,q)$ tensor fields generally) on $\mathcal{E}$ or $B\mathcal{E}$ induce corresponding covariant tensor fields on $\mathcal{S}$ by restricting their domain to vectors tangent to $\mathcal{S}$, and in practical terms the components of these induced tensors are simply the position space components of the originating tensors on $\mathcal{E}$ or $B\mathcal{E}$.
But the nature of the Bargmann groups is such that vector fields (and $(p,0)$ tensor fields generally) on 5-dimensional Bargmann spacetime can be treated in precisely this way, that is, as inducing corresponding contravariant tensor fields on 4-dimensional spacetime whose components are simply the traditional spacetime components of the originating tensors on 5-dimensional spacetime.
Notable examples include the 5-velocity $\bm{\mathscr{U}}$ inducing the 4-velocity $\bm{U}$, and the inverse 5-metric $\overleftrightarrow{\bm{G}}$ inducing the inverse 4-metric $\overleftrightarrow{\bm{g}}$ (Poincar\'e physics) or degenerate inverse `4-metric' $\overleftrightarrow{\bm{\gamma}}$ (Galilei physics).
Notice the use of a different symbol to denote the induced tensor.

Turn now to a projective type of relationship, which in the partial analogue relating tensor fields on $B\mathcal{E}$ to tensor fields on $\mathcal{E}$ will be called `decoding' rather than `projection'.
As described above, in the reverse direction of the push-forward operation, vector fields (and $(p,0)$ tensor fields generally) on $\mathcal{E}$ or $B\mathcal{E}$ can be projected to $\mathcal{S}$ by contraction with the projection tensor $\overleftarrow{\bm{\gamma}}$, which is the mixed-index version of the extension of the induced 3-metric $\bm{\gamma}$ on $\mathcal{S}$ back to $\mathcal{E}$ or $B\mathcal{E}$.
It turns out that there exists a `decoding operator' $\overleftarrow{\bm{g}}$ partly analogous to the projection operator $\overleftarrow{\bm{\gamma}}$.
To begin to see this, notice first that the 5-metric $\bm{G}$ on $B\mathcal{E}$ `encodes' the push-forward of the 4-metric $\bm{g}$ on $\mathcal{E}$ as follows:
\[
\bm{G} = \bm{g} + c^2 \, \underline{\bm{\xi}} \otimes \underline{\bm{\xi}},
\quad\quad
G_{IJ} = g_{IJ} + c^2 \, \xi_I \, \xi_J,
\]
or
\begin{equation}
\bm{g} = \bm{G} - c^2 \, \underline{\bm{\xi}} \otimes \underline{\bm{\xi}},
\quad\quad
g_{IJ} = G_{IJ} - c^2 \, \xi_I \, \xi_J,
\label{eq:EinsteinMetricBE}
\end{equation}
a relation confirmed directly from the component expressions in Eq.~(\ref{eq:Metric4}), padded with vanishing action components, and Eqs.~(\ref{eq:Metric5}) and (\ref{eq:ActionCovector}). 
Raising the index yields
\begin{equation}
\overleftarrow{\bm{g}} = \bm{\delta} - c^2 \, \bm{\xi} \otimes \underline{\bm{\xi}},
\quad\quad
{g^I}_J = {\delta^I}_J - c^2 \, \xi^I \, \xi_J,
\label{eq:DecodingOperator}
\end{equation}
where in this context $\bm{\delta}$ is the identity tensor on $B\mathcal{E}$.
If $\bm{G}$ is said to be the `Bargmann encoding' of $\bm{g}$, then $\overleftarrow{\bm{g}}$ provides the decoding:
\begin{equation}
\bm{g} = \overrightarrow{\bm{g}} \cdot \bm{G} \cdot \overleftarrow{\bm{g}},
\quad\quad
g_{IJ} = {g_I}^A \, G_{AB} \, {g^B}_J = G_{AB} \, {g^A}_I \, {g^B}_J,
\label{eq:MetricDecoding}
\end{equation}
as confirmed by direct computation.
This is projection in part, in the sense that the action components vanish on the left-hand side ($\overleftarrow{\bm{g}}$ is a non-invertible operator); but instead of being simply deleted, the action components are recombined with the time components in a particular way.
Hence the designation `decoding operator' rather than `projection operator'.
The other fundamental example is the spacetime momentum.
Recall that the kinetic-energy--momentum--mass covector $\bm{\Pi}$ on $B\mathcal{E}$ is the Bargmann encoding of the push-forward of the total-energy--momentum $\bm{P}$ on $\mathcal{E}$, separating mass from kinetic energy:
\[
\bm{\Pi} = \bm{P} + m c^2 \, \underline{\bm{\xi}},
\quad\quad
\Pi_J = P_J + m c^2 \, \xi_J,
\]
or
\begin{equation}
\bm{P} = \bm{\Pi} - m c^2 \, \underline{\bm{\xi}},
\quad\quad
P_J = \Pi_J - m c^2 \, \xi_J,
\label{eq:MomentumDecoding}
\end{equation}
as confirmed from Eq.~(\ref{eq:EnergyMomentumE}), padded with vanishing action component, and Eqs.~(\ref{eq:EnergyMomentumB}) and (\ref{eq:ActionCovector}).
Once again $\overleftarrow{\bm{g}}$ provides the decoding:
\[
\bm{P} = \bm{\Pi} \cdot \overleftarrow{\bm{g}},
\quad\quad
P_J = \Pi_A \, {g^A}_J,
\] 
as confirmed by direct computation.
Note again the use of a different symbol to denote the projected or decoded tensor.

Turn finally to an extension type of relationship.
As described above, in the reverse direction of the pull-back operation, 1-forms (and $(0,q)$ tensor fields generally) on $\mathcal{S}$ can be extended to $\mathcal{E}$ or $B\mathcal{E}$ by applying the projection operator $\overleftarrow{\bm{\gamma}}$ to vector arguments before evaluation of the form, and defining the extension of the 1-form by absorbing the projection operator into the form itself. 
But the nature of Bargmann groups is such that it is vector fields (and $(p,0)$ tensor fields generally) that can be extended from $\mathcal{E}$ to $B\mathcal{E}$ in a similar way using the decoding operator $\overleftarrow{\bm{g}}$.
This can be conceptualized in a manner analogous to one of the ways extensions are discussed above.
Begin with a vector field $\bm{w}$ on $\mathcal{E}$ represented by a 4-column $\left[ w^\mu \right]$.
Its extension to $B\mathcal{E}$ is obtained by lowering $\bm{w}$ on $\mathcal{E}$ to the 1-form $\underline{\bm{w}} = \bm{g} \cdot \bm{w}$, pushing it forward to $B\mathcal{E}$ (acquiring vanishing action components), and then raising it back to the vector field $w = \overleftrightarrow{\bm{G}} \cdot \underline{\bm{w}}$ on $B\mathcal{E}$. 
In components this sequence of operations corresponds to $w^I = G^{IA} \, g_{A\beta} \, w^\beta = {g^I}_\beta \, w^\beta$.
Using the components of $\overleftarrow{\bm{g}}$ in Eq.~(\ref{eq:DecodingOperator}) along with Eq.~(\ref{eq:ActionCovector}) for $\underline{\bm{\xi}}$, the result is that the extension is represented by the 5-column
\[
\left[ w^I \right] = \begin{bmatrix} w^\mu \\[5pt] c^2 \, \chi_\alpha \, w^\alpha \end{bmatrix}.
\]
This is reminiscent of the extension of a 1-form from $\mathcal{S}$ to $\mathcal{E}$ or $B\mathcal{E}$ in Eq.~(\ref{eq:ExtensionS}), in that, unlike the push forward, the extension involves the acquisition of an additional nonzero component.
Note again that the same symbol is used to denote the original tensor on $\mathcal{E}$ and its extension to $B\mathcal{E}$.
As an example, consider the extensions of the fiducial observer 4-velocity $\bm{n}$ and general 4-velocity $\bm{U}$ from $\mathcal{E}$ to $B\mathcal{E}$, represented by the 5-columns
\begin{equation}
\left[ n^I \right] = \begin{bmatrix} n^\mu \\[5pt] c^2 \end{bmatrix},
\quad\quad
\left[ U^I \right] = \begin{bmatrix} U^\mu \\[5pt] c^2 \, \Lambda_{\bm{V}} \end{bmatrix}.
\label{eq:VelocityExtension}
\end{equation}
These differ in the last component from the 5-columns
\[
\left[ \mathscr{n}^I \right] = \begin{bmatrix} n^\mu \\[5pt] 0 \end{bmatrix},
\quad\quad
\left[ \mathscr{U}^I \right] = \begin{bmatrix} U^\mu \\[5pt] c^2 \, \left( \Lambda_{\bm{V}} - 1 \right)\end{bmatrix}
\]
representing the fiducial observer 5-velocity $\bm{\mathscr{n}}$ and general 5-velocity $\bm{\mathscr{U}}$, which is why it was necessary to use different symbols as insisted in Sec.~\ref{sec:MetricB}: the extension of a vector on $\mathcal{E}$ to $B\mathcal{E}$ is not the same as its Bargmann encoding on $B\mathcal{E}$. 
In fact the proper relationship between 4-velocities on $\mathcal{E}$ and 5-velocities on $B\mathcal{E}$ is one of decoding rather than extension:
\begin{equation}
\bm{n} = \overleftarrow{\bm{g}} \cdot \bm{\mathscr{n}},
\quad\quad
\bm{U} = \overleftarrow{\bm{g}} \cdot \bm{\mathscr{U}},
\label{eq:VelocityDecoding}
\end{equation}
as can be shown by direct calculation with the above component expressions.
(Of course, fiducial observer 4- or 5-velocities correspond to general 4- or 5-velocities for 3-velocity $\bm{V} = 0$.)
Using Eq.~(\ref{eq:DecodingOperator}) in Eq.~(\ref{eq:VelocityDecoding}) yields
\[
\bm{n} = \bm{\mathscr{n}} - c^2 \, \bm{\xi},
\]
and then Eq.~(\ref{eq:ActionVector}) gives
\[
\bm{n} = - c^2 \, \bm{\chi},
\]
thus confirming that a vector version of Eq.~(\ref{eq:DualCollinearE}) on $\mathcal{E}$ holds for the extension of the fiducial observer 4-velocity $\bm{n}$ on $\mathcal{E}$ to $B\mathcal{E}$.

\subsection{Properties of the Decoding Operator}
\label{sec:PropertiesDecodingOperator}

Some relations involving the decoding operator $\overleftarrow{\bm{g}}$ are worth noting.
It is idempotent:
\begin{equation}
\overleftarrow{\bm{g}} \cdot \overleftarrow{\bm{g}} = \overleftarrow{\bm{g}},
\quad\quad
{g^I}_A \, {g^A}_J = {g^I}_J.
\label{eq:DecodingIdempotence}
\end{equation}
It nullfies $\bm{\xi}$, making $\overleftarrow{\bm{g}}$ non-invertible:
\begin{equation}
\begin{aligned}
\underline{\bm{\xi}} \cdot \overleftarrow{\bm{g}} &= 0, \\[5pt]
\overleftarrow{\bm{g}} \cdot \bm{\xi} &=0,
\end{aligned}
\quad\quad
\begin{aligned}
\xi_A \, {g^A}_J &= 0, \\[5pt]
{g^I}_A \, \xi^A &= 0.
\end{aligned}
\label{eq:DecodingAction}
\end{equation}
It preserves $\bm{\chi}$:
\begin{equation}
\begin{aligned}
\underline{\bm{\chi}} \cdot \overleftarrow{\bm{g}} &= \underline{\bm{\chi}}, \\[5pt]
\overleftarrow{\bm{g}} \cdot \bm{\chi} &= \bm{\chi},
\end{aligned}
\quad\quad
\begin{aligned}
\chi_A \, {g^A}_J &= \chi_J, \\[5pt]
{g^I}_A \, \chi^A &= \chi^I.
\end{aligned}
\label{eq:DecodingDualFiducial}
\end{equation}
It transforms $\bm{\mathscr{n}}$ in a manner reminiscent of Eq.~(\ref{eq:DualCollinearE}):
\[
\begin{aligned}
\underline{\bm{\mathscr{n}}} \cdot \overleftarrow{\bm{g}} &= -c^2 \, \underline{\bm{\chi}}, \\[5pt]
\overleftarrow{\bm{g}} \cdot \bm{\mathscr{n}} &= -c^2 \, \bm{\chi},
\end{aligned}
\quad\quad
\begin{aligned}
\mathscr{n}_A \, {g^A}_J &= -c^2 \, \chi_J, \\[5pt]
{g^I}_A \, \mathscr{n}^A &= -c^2 \, \chi^I.
\end{aligned}
\]
These relations follow from Eq.~(\ref{eq:DecodingOperator}) for $\overleftarrow{\bm{g}}$ and the mutual contractions in Eqs.~(\ref{eq:NormsB}) and (\ref{eq:MutualContractionsB}).
Moreover its divergence vanishes,
\begin{equation}
\bm{\mathscr{D}} \cdot \overleftarrow{\bm{g}} = 0,
\quad\quad
\mathscr{D}_A \,{g^A}_J = 0,
\label{eq:DecodingDivergence}
\end{equation}
thanks to the vanishing divergence of $\bm{\xi}$ in Eq.~(\ref{eq:ActionTraceB}) and the fact that $\bm{\mathscr{D}} \underline{\bm{\xi}}$ in Eq.~(\ref{eq:GradientActionCovectorB}) has no action component. 

Similar to Eqs.~(\ref{eq:GeneralProjectionOperatorE}) and (\ref{eq:GeneralProjectionOperatorB}) for the generalized projection operator $\overleftarrow{\bm{\gamma}}^*$ for arbitrary tensors, these observations about decoding and extension using $\overleftarrow{\bm{g}}$ give rise to the generalized decoding operator
\begin{equation}
{\left( \overleftarrow{\bm{g}}^* \bm{\mathscr{T}} \right)^{I_1 \dots I_p}}_{J_1 \dots J_q}
	= {g^{I_1}}_{A_1} \dots {g^{I_p}}_{A_p} \; {\mathscr{T}^{A_1 \dots A_p}}_{B_1 \dots B_q}
		\; {g^{B_1}}_{J_1} \dots {g^{B_q}}_{J_q}
	\quad \quad (\text{on } B\mathcal{E})
\label{eq:GeneralProjectionOperatorBE}
\end{equation}
that gives the decoding to $\mathcal{E}$ of a $(p,q)$ tensor $\bm{\mathscr{T}}$ on $B\mathcal{E}$.

This generalized decoding operator appears in an important relation between tensor gradients on $B\mathcal{E}$ and $\mathcal{E}$, similar in spirit to Eq.~(\ref{eq:DerivativeProjection}) relating tensor gradients on spacetime to tensor gradients on position space leaves.
Consider a tensor $\bm{T}$ on $B\mathcal{E}$ that is decoded to $\mathcal{E}$, that is, such that $\overleftarrow{\bm{g}}^* \bm{T} = \bm{T}$, so that all possible contractions with $\bm{\xi}$ and $\underline{\bm{\xi}}$ vanish. 
Then the relation 
\begin{equation}
\bm{\nabla}\bm{T} = \overleftarrow{\bm{g}}^* \bm{\mathscr{D}}\bm{T}
\quad \quad (\text{on } B\mathcal{E})
\label{eq:DerivativeProjectionE}
\end{equation}
holds, where $\overleftarrow{\bm{g}}^*$ is given by Eq.~(\ref{eq:GeneralProjectionOperatorBE}). 
The reasoning is similar to that leading to Eq.~(\ref{eq:DerivativeProjection}): it must be shown that $\overleftarrow{\bm{g}}^* \bm{\mathscr{D}}\bm{g} = 0$, and then the uniqueness of the Levi-Civita connection can be invoked to deduce that $\overleftarrow{\bm{g}}^* \bm{\mathscr{D}} = \bm{\nabla}$.
Indeed it is the case that
\[
\left( \overleftarrow{\bm{g}}^* \bm{\mathscr{D}}\bm{g} \right)_{K I J} 
	= {g^C}_K \, {g^A}_I \, {g^B}_J \,
		\mathscr{D}_C \left( G_{A B} - c^2 \, \xi_A \xi_B  \right)
	= 0 
\quad \quad (\text{on } B\mathcal{E})
\]
thanks to $\bm{\mathscr{D}} \bm{G} = 0$ and Eq.~(\ref{eq:DecodingAction}).

\subsection{Reverse Engineering Poincar\'e Physics on $B\mathcal{E}$ from Poincar\'e Physics on $\mathcal{E}$}
\label{sec:ReverseEngineering}

Given physical laws on Einstein spacetime $\mathcal{E}$, the resources and a proposed procedure to be used for encoding them on Bargmann-Einstein spacetime $B\mathcal{E}$ can now be summarized.
First, there are three key encodings that arise in connection with the construction of Bargmann spacetime itself presented in Sec.~\ref{sec:MetricB}: the 5-velocity $\bm{\mathscr{U}}$ as the Bargmann encoding of the 4-velocity $\bm{U}$; the Bargmann metric $\bm{G}$ as the Bargmann encoding of the Einstein metric $\bm{g}$; and the kinetic-energy--momentum--mass $\bm{\Pi}$ as the Bargmann encoding of the total-energy--momentum $\bm{P}$.
(It is worth emphasizing again that the Bargmann encoding $\bm{\mathscr{U}}$ of $\bm{U}$ is not the same as the extension of $\bm{U}$, and that the push-forwards of $\bm{g}$ and $\bm{P}$ are not the same as their Bargmann encodings $\bm{G}$ and $\bm{\Pi}$.)
And second, there is the clear understanding developed above of the relationships between tensors on $\mathcal{E}$ and on $B\mathcal{E}$: the push-forward of $(0,q)$ tensor fields and the extension of $(p,0)$ tensors from $\mathcal{E}$ to $B\mathcal{E}$, and the pull-back of $(p,0)$ tensors and decoding of $(0,q)$ tensors from $B\mathcal{E}$ to $\mathcal{E}$.
The generalized decoding operator $\overleftarrow{\bm{g}}^*$ is key to these relationships.
The proposed procedure is to take a physical law on $\mathcal{E}$; use the known Bargmann-encoded entities to write it as the decoding of an expression on $B\mathcal{E}$ using the decoding operator $\overleftarrow{\bm{g}}^*$; and, with luck or a perhaps a bit of additional information, reverse engineer the resulting decoding to work out the encoding of the physical law on $B\mathcal{E}$.
What makes the reverse engineering possible is that the first term of $\overleftarrow{\bm{g}}$ in Eq.~(\ref{eq:DecodingOperator}) is the identity operator, which allows unprojected tensor expressions on $B\mathcal{E}$ to appear; the second term of $\overleftarrow{\bm{g}}$ gives terms that one might hope to explicitly compute.

Consider an elementary example as a warm-up exercise and by way of illustration: What is the Bargmann-encoded version on $B\mathcal{E}$ of the equation $\bm{g}\left(\bm{U},\bm{U}\right) = -c^2$ expressing the normalization of a 4-velocity $\bm{U}$ on $\mathcal{E}$?
It was already noted in Sec.~\ref{sec:MetricB} that the answer is $\bm{G}\left( \bm{\mathscr{U}}, \bm{\mathscr{U}} \right) = 0$, but this result is obtained here by using the decoding operator $\overleftarrow{\bm{g}}$.
The easily-confirmed relation
\begin{equation}
\underline{\bm{\xi}} \cdot \bm{\mathscr{U}} = 1, 
\quad\quad
\xi_A \, \mathscr{U}^A = 1
\label{eq:VelocityActionLemma}
\end{equation}
will be a useful lemma.
First, take the equation on $\mathcal{E}$ and set it in $B\mathcal{E}$ by pushing forward $\bm{g}$ and extending $\bm{U}$.
In so doing, $\bm{U}$ acquires an action component according to Eq.~(\ref{eq:VelocityExtension}), but it contributes nothing because the action components of the push-forward of $\bm{g}$ vanish:
\[
\begin{aligned}
-c^2 &= g_{\alpha\beta} \, U^\alpha \, U^\beta \\
	&= g_{A B} \, U^A \, U^B.
\end{aligned}
\]  
Next, introduce quantities appropriate to the Bargmann perspective by using Eqs.~(\ref{eq:MetricDecoding}) and (\ref{eq:VelocityDecoding}) to express $\bm{g}$ and $\bm{U}$ as the decodings of $\bm{G}$ and $\bm{\mathscr{U}}$:
\[
\begin{aligned}
-c^2 &= G_{C D} \, {g^C}_A \, {g^D}_B \, {g^A}_E \, {g^B}_F \, \mathscr{U}^E \, \mathscr{U}^F \\
	&= G_{C D} \, {g^C}_E \, {g^D}_F \, \mathscr{U}^E \, \mathscr{U}^F
\end{aligned}
\]
in which the idempotence property of Eq.~(\ref{eq:DecodingIdempotence}) has been used.
Now comes the reverse engineering step, enabled by Eq.~(\ref{eq:DecodingOperator}) for the decoding operator, along with the normalization of $\bm{\xi}$ from Eq.~(\ref{eq:NormsB}) and repeated application of the lemma of Eq.~(\ref{eq:VelocityActionLemma}):
\[
\begin{aligned}
-c^2 &= G_{C D} \left( {\delta^C}_E - c^2 \, \xi^C \, \xi_E \right) 
	\left( {\delta^C}_E - c^2 \, \xi^C \, \xi_E \right) \mathscr{U}^E \, \mathscr{U}^F \\
	&= G_{C D} \, \mathscr{U}^C \, \mathscr{U}^D - c^2,
\end{aligned}
\]
thus arriving at the expected result $\bm{G}\left( \bm{\mathscr{U}}, \bm{\mathscr{U}} \right) = 0$.
Again, notice that it is the identity tensor $\bm{\delta}$ in the first term of Eq.~(\ref{eq:DecodingOperator}) for the decoding operator $\overleftarrow{\bm{g}}$ that enables the desired reverse engineering of physical laws on $B\mathcal{E}$ from physical laws on $\mathcal{E}$, in this example by allowing the term $G_{C D} \, \mathscr{U}^C \, \mathscr{U}^D$ to appear.

\section{Dynamics of an Elementary Particle}
\label{sec:DynamicsParticles}

The dynamics of a free classical particle---that is, a non-quantum particle with a definite trajectory subject to no force other than gravitation implied by spacetime curvature---provides a first opportunity to explore the possibility of strong-field Galilei physics on $B\mathcal{G}$, deduced from Poincar\'e physics on $B\mathcal{E}$, derived in turn from Poincar\'e physics on $\mathcal{E}$.
Here the word `elementary' means that the particle has no internal degrees of freedom, and therefore in particular constant mass.
Spacetime equations and their 1+3 (on $\mathcal{E}$) and 1+3+1 (on $B\mathcal{E}$ and $B\mathcal{G}$) decompositions are derived.
(It is worth noting that the evolution equations for particle 3-momentum covector $\bm{p}$ presented here are considerably simpler than the equations for physical 3-velocity vector $\bm{V}$ presented for instance in \cite{Vincent201231-geodesic-equ}.)
The lapse function $\alpha$, shift vector $\bm{\beta}$, 3-metric $\bm{\gamma}$, and extrinsic curvature $\bm{K}$ are taken as given.
These may embody gravitational fields of arbitrary strength, appearing not only in the equations for a particle on $\mathcal{E}$ and on $B\mathcal{E}$, but in the very similar equations for a particle on $B\mathcal{G}$ as well.

\subsection{Dynamics of an Elementary Particle on Einstein Spacetime $\mathcal{E}$}
\label{sec:DynamicsParticlesE}

On Einstein spacetime $\mathcal{E}$ the worldline of a free particle satisfies the geodesic equation
\begin{equation}
\bm{\nabla}_{\!\overleftarrow{\bm{P}}} \bm{P} = 0
\quad \quad (\text{on } \mathcal{E}).
\label{eq:GeodesicE}
\end{equation}
Here $\bm{P}$ is the total-energy--momentum 1-form and $\overleftarrow{\bm{P}}$ is the inertia-momentum vector field along the particle worldline, whose 1+3 decompositions are given in Sec.~\ref{sec:MetricE}.
Specialize to an elementary material particle of constant mass $m$.
Because $\overleftarrow{\bm{P}} = m \, \bm{U}$, where $\bm{U}$ is the particle 4-velocity, the equation of motion can also be expressed
\begin{equation}
\bm{\nabla}_{\!\bm{U}} \bm{P} = 0
\quad \quad (\text{on } \mathcal{E}).
\label{eq:ParticleMotionE}
\end{equation}
Contraction with $-\bm{n}$ yields an equation of motion for the particle total energy $\mathscr{E}_{\bm{p}}$.
This will depend on the particle 3-momentum $\bm{p}$, whose equation of motion is given by projecting Eq.~(\ref{eq:ParticleMotionE}) to position space slices $\mathcal{S}$ by contraction with $\overleftarrow{\bm{\gamma}}$.
With $\bm{p}$ in hand, the particle trajectory $\left( x^i (t) \right)$ is determined by the kinematical relation
\begin{equation}
\frac{1}{\alpha}  \frac{\mathrm{d} x^i}{\mathrm{d}t}
	= \frac{c^2}{\mathscr{E}_{\bm{p}}} \, p^i - \frac{1}{\alpha} \, \beta^i
\quad \quad (\text{on } \mathcal{E}),
\label{eq:ParticleCoordinatesE}
\end{equation}
which follows from Eq.~(\ref{eq:ThreeVelocity}) relating the physical 3-velocity $\bm{V}$ to the coordinate 3-velocity $\bm{v}$, and the relation between $\bm{V}$ and $\overleftarrow{\bm{p}}$ in Eq.~(\ref{eq:FourVelocityThreeMomentum}).
Referring to Appendix~\ref{sec:DecompositionParticleE} for derivations, the evolution of the particle total energy and 3-momentum as measured by the fiducial observers turn out to be
\begin{align}
\frac{1}{\alpha} \frac{\mathrm{d} \mathscr{E}_{\bm{p}} }{\mathrm{d}t}
	&= - c^2 \, \overleftarrow{\bm{p}} \cdot \bm{D} \ln \alpha
		+ \frac{c^2}{\mathscr{E}_{\bm{p}}} 
			\, \bm{K}\left( \overleftarrow{\bm{p}}, \overleftarrow{\bm{p}} \right)
&(\text{on } \mathcal{E}) 
\label{eq:ParticleEnergyE}, \\[5pt]
\frac{1}{\alpha} \frac{\mathrm{d} p_j }{\mathrm{d}t}
	&= - \mathscr{E}_{\bm{p}} \, \frac{\partial \ln\alpha}{\partial x^j}  
		+ \frac{p_a}{\alpha} \, \frac{\partial \beta^a}{\partial x^j} 
		- \frac{c^2 \,  p_a \, p_b }{2 \, \mathscr{E}_{\bm{p}} } 
			\frac{\partial \gamma^{a b}}{\partial x^j}  
& (\text{on } \mathcal{E}).
\label{eq:ParticleMomentumE}
\end{align}
Thus, given the metric fields $\alpha$, $\beta^i$, $\gamma^{ij}$ as functions of $t$ along with initial conditions for $x^i$ and $p_j$, Eqs.~(\ref{eq:ParticleCoordinatesE}) and (\ref{eq:ParticleMomentumE}) are sufficient to determine the trajectory of a particle on $\mathcal{E}$; and with the addition of the extrinsic curvature components $K_{ij}$ implied by the metric fields, Eq.~(\ref{eq:ParticleEnergyE}) provides a convenient equation for the evolution of the particle energy.

\subsection{Dynamics of an Elementary Particle on Bargmann Spacetimes $B\mathcal{E}$ and $B\mathcal{G}$}
\label{sec:DynamicsParticlesB}

The equation determining the worldline of a particle in spacetime provides a first opportunity to demonstrate the strategy or procedure described in Sec.~\ref{sec:PhysicalLawsBargmann} for deducing physical laws in a theory of strong-field gravity consistent with Galilei relativity: first, translate a physical law on the usual Einstein spacetime $\mathcal{E}$ into a law with the same physical content but in a form appropriate to Bargmann-Einstein spacetime $B\mathcal{E}$; and second, consider the $c \rightarrow \infty$ limit of that law appropriate to the closely related Bagmann-Galilei spacetime $B\mathcal{G}$.

The first step in this example is to find the analogue of Eq.~(\ref{eq:GeodesicE}) for particles on $B\mathcal{E}$---an equation to be expressed in terms of the relative-energy--momentum--mass 1-form $\bm{\Pi}$ on $B\mathcal{E}$ from Sec.~\ref{sec:MetricB} instead of the total-energy--momentum 1-form $\bm{P}$ on $\mathcal{E}$, and the Levi-Civita connection $\bm{\mathscr{D}}$ associated with the Bargmann metric $\bm{G}$ instead of $\bm{\nabla}$ associated with the Einstein metric $\bm{g}$.
Noting the relations between $\bm{P}$ and $\bm{\Pi}$ in Eq.~(\ref{eq:MomentumDecoding}) and  
between $\bm{\nabla}$ and $\bm{\mathscr{D}}$ in Eq.~(\ref{eq:DerivativeProjectionE}), 
the desired translation will be given by 
\begin{equation}
0 = \bm{\nabla}_{\!\overleftarrow{\bm{P}}} \bm{P}
	= \overleftarrow{\bm{g}}^* \bm{\mathscr{D}}_{\overleftarrow{\bm{\Pi}}} 
		\left(  \bm{\Pi} \cdot \overleftarrow{\bm{g}} \right).
\label{eq:GeodesicDecoding}
\end{equation}
Use Eqs.~(\ref{eq:DecodingOperator}) and (\ref{eq:GeneralProjectionOperatorBE}) for the decoding operators and the idempotence of $\overleftarrow{\bm{g}}$ in Eq.~(\ref{eq:DecodingIdempotence}) to compute
\[
\begin{aligned}
0 &= {g^A}_B \, {g^C}_A \, \Pi^B \,
	\mathscr{D}_C \left( \Pi_D \, {g^D}_J \right) \\[5pt]
&=  {g^C}_B \, \Pi^B \,
	\mathscr{D}_C \left( \Pi_D \, {g^D}_J \right) \\[5pt]
&= \left( \Pi^C - m c^2 \, \xi^C \right) \mathscr{D}_C \left( \Pi_J - m c^2 \, \xi_J \right).
\end{aligned}
\] 
Two of the four terms here involve $\bm{\mathscr{D}} \underline{\bm{\xi}}$ given by Eq.~(\ref{eq:GradientActionCovectorB}). 
One of these, $m^2 c^4 \, \xi^C \, \mathscr{D}_C \xi_J$, vanishes because the $C = 4$ component of $\mathscr{D}_C \xi_J$ vanishes.
Noting also that
\[
- \xi^C \, \mathscr{D}_C \Pi_J 
	= \mathscr{D}_4 \Pi_J
	= - {^{\bm{G}} \! {\Gamma^A}_{J4}} \, \Pi_A
	= \Pi_A \, {^{\bm{G}} \! {\Gamma^A}_{J4}} \, \xi^A
	= \Pi_A \, \mathscr{D}_J \xi^A
	= \Pi^A \, \mathscr{D}_J \xi_A
\]
and that $\bm{\mathscr{D}} \underline{\bm{\xi}}$ is antisymmetric, the result is
\[
0 = \Pi^A \, \mathscr{D}_A \, \Pi_J - 2 \, m c^2 \, \Pi^A \, \mathscr{D}_A \, \xi_J
\]
or
\begin{equation}
\bm{\mathscr{D}}_{\overleftarrow{\bm{\Pi}}} \, \bm{\Pi}
	= m c^2 \, \left( \overleftarrow{\bm{\Pi}} \cdot \underline{\bm{a}} \right) \, \underline{\bm{\chi}} 
	- m c^2 \, \left( \overleftarrow{\bm{\Pi}} \cdot \underline{\bm{\chi}} \right) \underline{\bm{a}}
\quad \quad (\text{on } B\mathcal{E}).
\label{eq:NotGeodesicB}
\end{equation}
This is the equation for the wordline of an elementary particle on $B\mathcal{E}$ analogous to Eq.~(\ref{eq:GeodesicE}) for particles on $\mathcal{E}$.

The first thing to notice about Eq.~(\ref{eq:NotGeodesicB}) is that, unlike the situation on $\mathcal{E}$ in which the worldline is geodesic with respect to the Einstein spacetime metric $\bm{g}$, the particle worldline on $B\mathcal{E}$ is not geodesic with respect to the Bargmann spacetime metric $\bm{G}$---that is, the right-hand side does not vanish.
(That worldlines of \textit{massless} particles apparently \textit{are} geodesic with respect to $\bm{G}$ is a notable apparent exception, but not one of immediate present focus.)
In fact it seems that the translation to Bargmann spacetime involves a remarkable reversion of Einstein's perspective on gravitation to one more like Newton's, in this respect: whereas Einstein turned things upside down by saying that it is the fiducial observers who are gravitationally accelerated (see Eq.~(\ref{eq:FiducialAccelerationE})) while the freely falling particles they observe experience no gravitation (see Eq.~(\ref{eq:GeodesicE})), in Bargmann spacetime the fiducial observers are geodesic (see Eq.~(\ref{eq:FiducialAccelerationB})) and therefore analogous to Newton's inertial observers, while the particles they observe are subject to an explicit gravitational force (the right-hand side of Eq.~(\ref{eq:NotGeodesicB})).

This may seem like good news, a development in sympathy with the second stage of the procedure proposed here, which is to move from Poincar\'e physics on $B\mathcal{E}$ to Galilei physics on $B\mathcal{G}$.
But despite efforts to tame the presence of $c$ in Poincar\'e physics by encoding $\bm{g}$ as $\bm{G}$ and $\bm{P}$ as $\bm{\Pi}$ via the artifice of Bargmann spacetime, Eq.~(\ref{eq:NotGeodesicB}) presents an immediate apparent obstacle to a $c \rightarrow \infty$ limit, namely the factor of $c^2$.
This comes from the decoding operator $\overleftarrow{\bm{g}}$ given by Eq.~(\ref{eq:DecodingOperator}), and is ultimately traceable to the $- c^2 \, \alpha^2$ term in $\bm{g}$ reasserting itself (see also Eq.~(\ref{eq:EinsteinMetricBE})).
Regard this as a signal that the nature of the $c \rightarrow \infty$ but strong-field limit pursued here will need to be further clarified, but postpone this reckoning by drawing inspiration from the weak-field limit.
It is well known that the correspondence with Newton gravity is obtained with $\alpha = 1+ \phi / c^2$ with $\left| \phi \right| / c^2 \ll 1$, where $\phi$ is the Newton gravitational potential.
Allow for a strong-field generalization of this by writing
\begin{equation}
\alpha = \exp\left( \frac{\phi}{c^2} \right), 
\quad \quad \underline{\bm{a}} = \bm{D} \ln \alpha = \frac{1}{c^2} \, \bm{D}\phi,
\label{eq:StrongPotential}
\end{equation}
which corresponds to the weak-field limit when $\phi / c^2 \ll 1$ but is not limited to that.
In terms of this strong-field version of a gravitational potential $\phi$, and assuming $m \ne 0$ so as to write $\overleftarrow{\bm{\Pi}} = m \, \bm{\mathscr{U}}$ in the directional derivative, the particle worldline equation for material particles of constant mass becomes
\begin{equation}
\bm{\mathscr{D}}_{\bm{\mathscr{U}}} \, \bm{\Pi}
	= \left( \overleftarrow{\bm{\Pi}} \cdot \bm{D}\phi \right) \, \underline{\bm{\chi}} 
	-  \left( \overleftarrow{\bm{\Pi}} \cdot \underline{\bm{\chi}} \right) \bm{D}\phi
\quad \quad (\text{on } B\mathcal{E} \text{ or } B\mathcal{G}),
\label{eq:ParticleMotionB}
\end{equation}
analogous to Eq.~(\ref{eq:ParticleMotionE}) on $\mathcal{E}$. 
This now appears suitable for adoption on $B\mathcal{G}$ as well as on $B\mathcal{E}$.
It is the firstfruits of the present quest for physical laws featuring potentially strong gravity consistent with Galilei relativity. 

Turn now to the 1+3+1 decomposition of Eq.~(\ref{eq:ParticleMotionB}).
Contraction with $-\bm{\mathscr{n}}$ once again yields an energy equation, but one for the particle kinetic energy $\epsilon_{\bm{p}}$ which is well-defined on $B\mathcal{G}$ as well as on $B\mathcal{E}$ (and $\mathcal{E}$), rather than the particle total energy $\mathscr{E}_{\bm{p}}$ as was the case on $\mathcal{E}$.
Contraction with $\overleftarrow{\bm{\gamma}}$ again yields the evolution of 3-momentum $\bm{p}$, and the particle trajectory is then given by the kinematical relation
\begin{equation}
\frac{1}{\alpha}  \frac{\mathrm{d} x^i}{\mathrm{d}t}
	= \left\{ \begin{aligned}
		\frac{c^2}{\mathscr{E}_{\bm{p}}} \, p^i - \frac{1}{\alpha} \, \beta^i
				& & (\text{on } B\mathcal{E}), \\[5pt]
		\frac{1}{m} \, p^i - \frac{1}{\alpha} \, \beta^i
				& & (\text{on } B\mathcal{G}).
	 \end{aligned} \right.
\label{eq:ParticleCoordinatesB}
\end{equation}
For $B\mathcal{E}$ this is the same as on $\mathcal{E}$, and for $B\mathcal{G}$ it is the obvious $c^2 \rightarrow \infty$ limit.
Referring to Appendix~\ref{sec:DecompositionParticleB} for derivations, the evolution of the particle kinetic energy and 3-momentum as measured by the fiducial observers turn out to be
\begin{align}
\frac{1}{\alpha} \frac{\mathrm{d} \epsilon_{\bm{p}} }{\mathrm{d}t}
	&= \left\{ \begin{aligned}
		- \overleftarrow{\bm{p}} \cdot \bm{D} \phi
		+ \frac{c^2}{\mathscr{E}_{\bm{p}}} 
			\, \bm{K}\left( \overleftarrow{\bm{p}}, \overleftarrow{\bm{p}} \right)
		& & (\text{on } B\mathcal{E}), \\[5pt]
		- \overleftarrow{\bm{p}} \cdot \bm{D} \phi
		+ \frac{1}{m} 
			\, \bm{K}\left( \overleftarrow{\bm{p}}, \overleftarrow{\bm{p}} \right)
		& & (\text{on } B\mathcal{G}) 
	\end{aligned} \right.
\label{eq:ParticleEnergyB} \\[5pt]
\frac{1}{\alpha} \frac{ \mathrm{d} p_j }{\mathrm{d}t}
	&= \left\{ \begin{aligned}
		-  \frac{\mathscr{E}_{\bm{p}}}{c^2} \, \frac{\partial \phi}{\partial x^j}
		+ \frac{p_a}{\alpha} \, \frac{\partial \beta^a}{\partial x^j} 
		- \frac{c^2 \,  p_a \, p_b }{2 \, \mathscr{E}_{\bm{p}} } 
			\frac{\partial \gamma^{a b}}{\partial x^j}  
		& & (\text{on } B\mathcal{E}), \\[5pt]		
		 - m \, \frac{\partial \phi}{\partial x^j}
		+ \frac{p_a}{\alpha} \, \frac{\partial \beta^a}{\partial x^j} 
		- \quad \frac{ p_a \, p_b }{2 \, m } 
			 \frac{\partial \gamma^{a b}}{\partial x^j} 
		& & (\text{on } B\mathcal{G})	
	\end{aligned} \right.
\label{eq:ParticleMomentumB}
\end{align}
Given Eq.~(\ref{eq:StrongPotential}) and the fact that $\mathscr{E}_{\bm{p}} = \epsilon_{\bm{p}} + m c^2$ where $m c^2$ is constant, the equations on $B\mathcal{E}$ are the same as Eqs.~(\ref{eq:ParticleEnergyE}) and (\ref{eq:ParticleMomentumE}) on $\mathcal{E}$, and the equations on $B\mathcal{G}$ are the obvious $c \rightarrow \infty$ limits. 

There are a couple of loose ends to address.
First, while explicit factors of $c$ have been eliminated from the equations presented here for particle worldlines on $B\mathcal{G}$, from Eq.~(\ref{eq:StrongPotential}) it is clear that instances of $c^2$ persist through its presence in $\alpha$.
Noting that $\phi / c^2 \sim G M / c^2 R$, where $G$ is the gravitational constant and $M$ and $R$ are characteristic mass and length scales, characterize a strong-field $c \rightarrow \infty$ limit by the additional demand that $G \rightarrow \infty$ as well, in such a way that $G / c^2$ remains constant.
Finally, note that Eq.~(\ref{eq:ParticleMotionB}) has an additional component, the action component projected out by contraction with $\bm{\xi}$.
In the present case of constant particle mass $m$ this simply gives $0 = 0$.

\section{Dynamics of a Simple Fluid}
\label{sec:DynamicsFluids}

The dynamics of a simple fluid provides a second opportunity to explore strong-field Galilei physics on $B\mathcal{G}$, deduced from Poincar\'e physics on $B\mathcal{E}$, derived in turn from Poincar\'e physics on $\mathcal{E}$.
Here the word `simple' means that microscopically the fluid consists of classical material particles of a single type, and of constant mass $m$.
Spacetime equations are given, along with 1+3 (on $\mathcal{E}$) and 1+3+1 (on $B\mathcal{E}$ and $B\mathcal{G}$) decompositions in the form of balance equations for particle number, energy, and momentum.
The lapse function $\alpha$, shift vector $\bm{\beta}$, 3-metric $\bm{\gamma}$, and extrinsic curvature $\bm{K}$ are taken as given.
These embody gravitational fields of arbitrary strength, appearing not only in the equations for a fluid on $\mathcal{E}$ and on $B\mathcal{E}$, but in the very similar equations for a fluid on $B\mathcal{G}$ as well.

\subsection{Dynamics of a Simple Fluid on Einstein Spacetime $\mathcal{E}$}
\label{sec:DynamicsFluidsE}

Consider a simple fluid on $\mathcal{E}$, regarded macroscopically as a continuous medium, but which microscopically consists of classical particles of constant mass $m$.
Describe the fluid in terms of kinetic theory (see e.g. \cite{Cardall2019Minkowski-and-G}).

The fluid kinematics---the \textit{description} of where fluid elements are, and how fast they are moving---is given in terms of the particle number flux $\bm{N}$.
This is the 4-vector field
\begin{equation}
\begin{aligned}
\bm{N} &= \int f \, \overleftarrow{\bm{P}} \, \mathrm{d}P_m \\[5pt]
	&= \int f \, \left( \frac{\mathcal{E}_{\bm{p}}}{c^2} \, \bm{n} + \overleftarrow{\bm{p}} \right) 
		\, \frac{c^2 \, \mathrm{d}\bm{p} }{ (2\pi\hbar)^3 \, \mathcal{E}_{\bm{p}} } \\[5pt]
	&= \int f \, \left( \bm{n} + \bm{V}_{\bm{p}} \right) 
		\, \frac{ \mathrm{d}\bm{p} }{ (2\pi\hbar)^3 } 
\end{aligned}
\label{eq:ParticleFluxE}
\end{equation}
where $f ( X, p ) = f \left( X^\mu, p_j \right)$ is the Lorentz-invariant particle distribution function (particle density in phase space restricted to the mass shell), $\mathrm{d}P_m$ is the Lorentz-invariant 3-volume element in momentum space, and $\overleftarrow{\bm{P}}$ is the particle inertia-momentum given in Eq.~(\ref{eq:InertiaMomentumE}).
The 1+3 decomposition of $\bm{N}$ is immediate:
\begin{equation}
\bm{N} = N \, \bm{n} + N \, \bm{V},
\label{eq:ParticleFluxDecompositionE}
\end{equation}
where
\begin{equation}
N = \int f \, \frac{ \mathrm{d}\bm{p} }{ (2\pi\hbar)^3 },
\quad \quad
\bm{V} = \frac{1}{N} \, \int f \, \bm{V}_{\bm{p}} \, \frac{ \mathrm{d}\bm{p} }{ (2\pi\hbar)^3 }
\label{eq:DensityVelocityFiducial}
\end{equation}
are the particle number density and average 3-velocity of the fluid measured by fiducial observers: 
\[
N = \underline{\bm{\chi}} \cdot \bm{N},
\quad \quad
N \, \bm{V} = \overleftarrow{\bm{\gamma}} \cdot \bm{N}.
\]
Comoving observers---observers `riding along' with the fluid---have 4-velocity $\bm{U}$ proportional to $\bm{N}$ itself and given in terms of the average 3-velocity:
\begin{equation}
\bm{N} = n \, \bm{U},
\quad \quad
\bm{U} = \Lambda_{\bm{V}} \left( \bm{n} + \bm{V} \right).
\label{eq:NumberFluxComovingE}
\end{equation}
Here the scalar field $n$ is the particle number density measured by comoving observers:
noting the quantities
\[
\underline{\bm{\chi}}_{\bm{U}} = - \frac{1}{c^2} \underline{\bm{U}}, 
\quad \quad
\overleftarrow{\bm{\gamma}_{\bm{U}}} = \bm{\delta} - \bm{U} \otimes \underline{\bm{\chi}}_{\bm{U}}
\quad \quad (\text{on } \mathcal{E})
\]
analogous to their fiducial observer counterparts and useful for 1+3 decompositions according to comoving observers, it is evident that
\[
n = \underline{\bm{\chi}}_{\bm{U}} \cdot \bm{N}.
\]
Using Eq.~(\ref{eq:NumberFluxComovingE}) it is clear that
\begin{equation}
N = \underline{\bm{\chi}} \cdot \bm{N} = \Lambda_{\bm{V}} \, n
\quad \quad (\text{on } \mathcal{E})
\label{eq:FiducialDensityComovingE}
\end{equation}
relates the number densities $N$ and $n$ measured by fiducial and comoving observers.

Particle conservation is expressed by 
\begin{equation}
\bm{\nabla} \cdot \bm{N} = 0
\quad \quad (\text{on } \mathcal{E}).
\label{eq:ParticleConservationSpacetimeE}
\end{equation}
Thanks to the well-known identity ${^{\bm{g}} \! {\Gamma^\alpha}_{\nu\alpha}} = \partial_\nu \ln \sqrt{-g}$ and referring to Eq.~(\ref{eq:MetricDeterminant4}), this can be expressed in terms of components as 
\[
\begin{aligned}
0 = \bm{\nabla} \cdot \bm{N} 
	&= \frac{1}{\sqrt{-g}} \, \frac{\partial}{\partial X^\alpha} 
		\left( \sqrt{-g} \, N \left( n^\alpha + V^\alpha \right) \right) \\[5pt]
	&= \frac{1}{\alpha \sqrt{\gamma}} \, \frac{\partial}{\partial t} \left( \sqrt{\gamma} \, N \right)
		+ \frac{1}{\alpha \sqrt{\gamma}} \, \frac{\partial}{\partial x^a} 
			\left( \sqrt{\gamma} \, N \left( \alpha \, V^a - \beta^a \right) \right).
\end{aligned}
\]
More geometrically,
\begin{equation}
 \frac{1}{\alpha \sqrt{\gamma}} \, \frac{\partial}{\partial t} \left( \sqrt{\gamma} \, N \right)
 	+ \frac{1}{\alpha} \, \bm{D} \cdot \left( N \left( \alpha \, \bm{V} - \bm{\beta} \right) \right)
	= 0
\quad \quad (\text{on } \mathcal{E})
\label{eq:ParticleConservationE}
\end{equation}
expresses particle conservation on $\mathcal{E}$ in terms of the 3-divergence of the particle 3-flux tangent to the leaves $\mathcal{S}$ of the foliation.

The fluid dynamics---the \textit{prescription} of what governs the fluid element motion---is given in terms of the total-energy--momentum flux $\bm{T}$, the $(1,1)$ tensor field 
\begin{equation}
\bm{T} = \int f \, \left( \overleftarrow{\bm{P}} \otimes \bm{P} \right) \, \mathrm{d}P_m.
\label{eq:MomentumFluxE}
\end{equation}
This is manifestly self-adjoint; its raised-index and and lowered-index versions are symmetric.
Using Eq.~(\ref{eq:EnergyMomentumE}) for the energy-momentum $\bm{P}$ and Eq.~(\ref{eq:InertiaMomentumE}) for the inertia-momentum $\overleftarrow{\bm{P}}$, its 1+3 decomposition is
\begin{equation}
\bm{T} = - \mbox{$\mathfrak{E}$} \, \bm{n} \otimes \underline{\bm{\chi}}
		+ \bm{n} \otimes \bm{S}
		- \bm{\mathfrak{Q}} \otimes \underline{\bm{\chi}}
		- \bm{\Sigma},
\label{eq:EnergyMomentumFluxFiducialE}
\end{equation}
in which the total energy density scalar field $\mathfrak{E}$, the 3-momentum density 1-form $\bm{S}$, the total energy flux 3-vector field $\bm{\mathfrak{Q}}$, and the 3-stress $(1,1)$ tensor field $\bm{\Sigma}$ measured by fiducial observers are 
\begin{equation}
\begin{aligned}
\mathfrak{E} &= \int \, f \, \mathcal{E}_{\bm{p}} \, \frac{ \mathrm{d}\bm{p} }{ (2\pi\hbar)^3 },
\quad \quad 
& \bm{S} = \int \, f \, \bm{p} \, \frac{ \mathrm{d}\bm{p} }{ (2\pi\hbar)^3 }, 
\quad \quad \quad \quad \\[5pt]
\bm{\mathfrak{Q}} &= \int \, f \, \mathcal{E}_{\bm{p}} \, \bm{V}_{\bm{p}} \, \frac{ \mathrm{d}\bm{p} }{ (2\pi\hbar)^3 }, 
\quad \quad
&\bm{\Sigma} = - \int \, f \, \left( \bm{V}_{\bm{p}} \otimes \bm{p} \right) \, \frac{ \mathrm{d}\bm{p} }{ (2\pi\hbar)^3 }.
\end{aligned}
\label{eq:MomentumFluxDecompositionE}
\end{equation}
(Note that the `stress tensor' $\bm{\Sigma}$ is the negative of a `momentum tensor'.)
A consequence of $\bm{T}$ being self-adjoint is that
\begin{equation}
\bm{\mathfrak{Q}} = c^2 \, \overleftarrow{\bm{S}}, \quad \quad \bm{S} = \frac{1}{c^2} \, \underline{\bm{\mathfrak{Q}}}.
\label{eq:EnergyFluxMomentum}
\end{equation}
This relation between total energy flux and momentum density is a manifestation of the equivalence in Poincar\'e physics between energy and mass (up to a factor of $c^2$).

Because constitutive relations (e.g. an equation of state) relate quantities measured by comoving observers, it is important to consider the alternative decomposition
\begin{equation}
\bm{T} = - \mbox{$\mathfrak{e}$} \, \bm{U} \otimes \underline{\bm{\chi}}_{\bm{U}}
		+ \frac{1}{c^2} \, \bm{U} \otimes \underline{\bm{q}}
		- \bm{q} \otimes \underline{\bm{\chi}}_{\bm{U}}
		- \bm{\sigma},
\label{eq:EnergyMomentumFluxComovingE}
\end{equation}
in which the total energy density $\mathfrak{e}$, heat 3-flux $\bm{q}$, and 3-stress $\bm{\sigma}$ are measured by comoving observers.
In general a fluid is dissipative even without shocks; a perfect fluid is one for which, in the absence of shocks, there are no heat exchange and no shear stresses between neighboring comoving fluid elements (no thermal conductivity or shear viscosity), and no inelastic compression of comoving fluid elements (no bulk viscosity).
Mathematically this translates into a vanishing heat 3-flux, and a 3-stress that is isotropic according to comoving observers:
\[
\bm{q} = 0, \quad \quad \bm{\sigma} = - p \, \overleftarrow{\bm{\gamma}_{\bm{U}}}
\quad \quad (\text{perfect fluid}),
\]
where $p$ is the pressure.

Quantities measured by fiducial observers are projected out of Eq.~(\ref{eq:EnergyMomentumFluxFiducialE}) by appropriate contractions with $\underline{\bm{\chi}}$ and $\overleftarrow{\bm{\gamma}}$.
Substitution of Eq.~(\ref{eq:EnergyMomentumFluxComovingE}) into these contractions yields
\[
\begin{aligned}
\mathfrak{E} &= - \underline{\bm{\chi}} \cdot \bm{T} \cdot \bm{n} \\[5pt]
	&= {\Lambda_{\bm{V}}}^2 \, \mathfrak{e} 
		+ \frac{2}{c^2} \, \Lambda_{\bm{V}} \, \underline{\bm{V}} \cdot \bm{q}
		- \frac{1}{c^2} \, \underline{\bm{V}} \cdot \bm{\sigma} \cdot \bm{V}
\end{aligned}
\]
for the total energy,
\[
\begin{aligned}
\bm{S} &= \underline{\bm{\chi}} \cdot \bm{T} \cdot \overleftarrow{\bm{\gamma}} \\[5pt]
	&= \frac{1}{c^2} \left( {\Lambda_{\bm{V}}}^2 \, \mathfrak{e} \, \underline{\bm{V}}
		+ \Lambda_{\bm{V}} 
			\left( \underline{\bm{q}} \cdot \overleftarrow{\bm{\gamma}} \right)
		+ \frac{1}{c^2} \, \Lambda_{\bm{V}} \left( \underline{\bm{V}} \cdot \bm{q} \right)
			\underline{\bm{V}}
		- \underline{\bm{V}} \cdot \bm{\sigma} \cdot \overleftarrow{\bm{\gamma}} \right)
\end{aligned}
\]
for the 3-momentum density,
\[
\begin{aligned}
\bm{\mathfrak{Q}} &= -  \overleftarrow{\bm{\gamma}} \cdot \bm{T} \cdot \underline{\bm{\chi}} \\[5pt]
	&= {\Lambda_{\bm{V}}}^2 \, \mathfrak{e} \, \bm{V}
		+ \frac{1}{c^2} \, \Lambda_{\bm{V}} 
			\left( \underline{\bm{q}} \cdot \bm{V} \right) \bm{V}
		+ \Lambda_{\bm{V}} \left( \overleftarrow{\bm{\gamma}}\cdot \bm{q} \right)
		- \overleftarrow{\bm{\gamma}} \cdot \bm{\sigma} \cdot \bm{V}
\end{aligned}
\]
for the total energy 3-flux, and
\[
\begin{aligned}
\bm{\Sigma} &= - \overleftarrow{\bm{\gamma}} \cdot \bm{T} \cdot \overleftarrow{\bm{\gamma}} \\[5pt]
	&= - \frac{1}{c^2} \left( {\Lambda_{\bm{V}}}^2 \, \mathfrak{e} \, 
					\bm{V} \otimes \underline{\bm{V}} 	
		+ \Lambda_{\bm{V}} \, \bm{V} \otimes 
			\left(  \underline{\bm{q}} \cdot \overleftarrow{\bm{\gamma}} \right)		
		+ \Lambda_{\bm{V}} \left( \overleftarrow{\bm{\gamma}}\cdot \bm{q} \right)
			\otimes \underline{\bm{V}}
					\right)
		+ \overleftarrow{\bm{\gamma}} \cdot \bm{\sigma} \cdot \overleftarrow{\bm{\gamma}}
\end{aligned}
\]
for the 3-stress, or
\begin{equation}
\begin{aligned}
\mathfrak{E} &= {\Lambda_{\bm{V}}}^2 \, \left( \mathfrak{e} + p \right) - p, \\[5pt]
\bm{S} &= \frac{1}{c^2} \, {\Lambda_{\bm{V}}}^2 \, 
		\left( \mathfrak{e} + p \right) \, \underline{\bm{V}}, \\[5pt]
\bm{\mathfrak{Q}} &= {\Lambda_{\bm{V}}}^2 \,
		\left( \mathfrak{e} + p \right) \, \bm{V}, \\[5pt]
\bm{\Sigma} &= - \frac{1}{c^2} \, {\Lambda_{\bm{V}}}^2 \,
		\left( \mathfrak{e} + p \right) \, \bm{V} \otimes \underline{\bm{V}}
		- p \, \overleftarrow{\bm{\gamma}}
\end{aligned}
\quad \quad (\text{perfect fluid on } \mathcal{E})
\label{eq:FiducialEnergyMomentumB}
\end{equation}
in the case of a perfect fluid, relating quantities measured by fiducial observers to those measured by comoving observers.
From this point forward a perfect fluid will be assumed for simplicity.

In the absence of an external force (apart from gravitation implied by the spacetime curvature of $\mathcal{E}$), total-energy--momentum balance is expressed by
\begin{equation}
\bm{\nabla} \cdot \bm{T} = 0
\quad \quad (\text{on } \mathcal{E}).
\label{eq:FluidMotionE}
\end{equation}
Total energy balance is obtained by contraction with $-\bm{n}$, and balance of 3-momentum is obtained by contraction with $\overleftarrow{\bm{\gamma}}$, resulting in
\begin{align}
 \frac{1}{\alpha \sqrt{\gamma}} \, \frac{\partial}{\partial t} \left( \sqrt{\gamma} \, \mathfrak{E} \right)
 	+ \frac{1}{\alpha} \, \bm{D} \cdot \left( \alpha \, \bm{\mathfrak{Q}} - 
									\mbox{$\mathfrak{E}$} \, \bm{\beta} \right)
	&= - \overleftarrow{\bm{S}} \cdot \bm{D} \phi
		- \overrightarrow{\bm{\Sigma}} : \bm{K} 
& (\text{on }\mathcal{E})
\label{eq:TotalEnergyBalanceE} \\[5pt]
\frac{1}{\alpha \sqrt{\gamma}} \, 
		\frac{\partial}{\partial t} \left( \sqrt{\gamma} \, S_j \right)
		+ \frac{1}{\alpha} \, D_a \left( - \alpha \, {\Sigma^a}_j - \beta^a \, S_j \right)
		&= - \frac{\mathfrak{E}}{c^2} \, D_j \phi + \frac{1}{\alpha} \, S_a \, D_j \beta^a
& (\text{on }\mathcal{E})
\label{eq:MomentumBalanceE}
\end{align}
See Appendix~\ref{sec:DecompositionFluidE} for details.

To make contact with physics on Bargmann spacetimes $B\mathcal{E}$ and $B\mathcal{G}$ it will prove convenient to derive a balance equation for kinetic energy rather than total energy.
This is done by separating that portion of of energy associated only with the microscopic particles' very existence, that is, the energy due to the particle mass $m$.
The total energy density and 3-flux measured by fiducial observers separate into 
\begin{equation}
\begin{aligned}
\mathfrak{E} &= E + m c^2 \, N, \\[5pt]
\bm{\mathfrak{Q}} &= \bm{Q} + m c^2 \, N \,\bm{V}
\end{aligned}
\label{eq:TotalEnergySeparation}
\end{equation}
where
\begin{equation}
\begin{aligned}
E &= \int \, f \, \epsilon_{\bm{p}} \, \frac{ \mathrm{d}\bm{p} }{ (2\pi\hbar)^3 } \\[5pt]
\bm{Q} &= \int \, f \, \epsilon_{\bm{p}} \, \bm{V}_{\bm{p}} \, \frac{ \mathrm{d}\bm{p} }{ (2\pi\hbar)^3 } \\[5pt]
\end{aligned}
\label{eq:KineticEnergyDensity}
\end{equation}
are the kinetic energy density and 3-flux measured by fiducial observers.
Meanwhile the total energy density measured by comoving observers separates into
\[
\mathfrak{e} = e + m c^2 \, n,
\]
where $e$ is the kinetic energy measured by comoving observers, here called the internal energy density.
Thus in terms of the internal energy density $e$, pressure $p$, and particle number density $n$ measured by comoving observers, 
\begin{equation}
\begin{aligned}
E &= \mathfrak{E} - m c^2 \, N \\[5pt]
	&= {\Lambda_{\bm{V}}}^2 \, \left( e + p \right) - p
		+ m c^2 \, n  \, \Lambda_{\bm{V}} \left( \Lambda_{\bm{V}} - 1 \right)		
\end{aligned}
\quad \quad (\text{perfect fluid on } \mathcal{E})
\label{eq:FiducialKineticEnergyB}
\end{equation}
is the kinetic energy density and
\begin{equation}
\begin{aligned}
\bm{Q} &= \bm{\mathfrak{Q}} - m c^2 \, N \,\bm{V} \\[5pt]
	&= \left( {\Lambda_{\bm{V}}}^2 \, \left( e + p \right) 
		+ m c^2  \, n \, \Lambda_{\bm{V}} \left( \Lambda_{\bm{V}} - 1 \right) \right) \, \bm{V}
\end{aligned}
\quad \quad (\text{perfect fluid on } \mathcal{E})
\label{eq:FiducialKineticEnergyFluxB}
\end{equation}
is the kinetic energy 3-flux, both specialized here to a perfect fluid.
The terms in which $c^2$ appear in these expressions for $E$ and $\bm{Q}$ cause no trouble as $c \rightarrow \infty$ because they limit nicely to the Galilei bulk or macroscopic kinetic energy density $m \, n \, \bm{\gamma} \left( \bm{V}, \bm{V} \right) / 2$.
With constant particle mass $m$, subtraction of $m c^2$ times Eq.~(\ref{eq:ParticleConservationE}) for particle conservation from Eq.~(\ref{eq:TotalEnergyBalanceE}) for total energy balance 
yields the balance equation
\begin{equation}
 \frac{1}{\alpha \sqrt{\gamma}} \, \frac{\partial}{\partial t} \left( \sqrt{\gamma} \, E \right)
 	+ \frac{1}{\alpha} \, \bm{D} \cdot \left( \alpha \, \bm{Q} - E \, \bm{\beta} \right)
	= - \overleftarrow{\bm{S}} \cdot \bm{D} \phi
		- \overrightarrow{\bm{\Sigma}} : \bm{K}
\quad \quad (\text{on }\mathcal{E})
\label{eq:KineticEnergyBalanceE}
\end{equation}
for the kinetic energy density.
This is the same as Eq.~(\ref{eq:TotalEnergyBalanceE}) for total energy balance but with total energy density $\mathfrak{E}$ replaced by the kinetic energy density $E$ and the total energy 3-flux $\bm{\mathfrak{Q}}$ replaced by the kinetic energy 3-flux $\bm{Q}$; the geometric source terms are unchanged.

\subsection{Dynamics of a Simple Fluid on Bargmann Spacetimes $B\mathcal{E}$ and $B\mathcal{G}$}
\label{sec:DynamicsFluidsB}

The kinematics and dynamics of a simple fluid on $\mathcal{E}$ were given in the previous subsection by the divergences of a particle number 4-vector field $\bm{N}$ and total-energy--momentum flux $(1,1)$ tensor field $\bm{T}$.
Analogous objects and their divergences will similarly give the kinematics and dynamics of a simple fluid on $B\mathcal{E}$ or $B\mathcal{G}$.

The fluid kinematics on $B\mathcal{E}$ or $B\mathcal{G}$ is given in terms of a particle number flux 5-vector field
\begin{equation}
\begin{aligned}
\bm{\mathscr{N}} &= \int f \, \overleftarrow{\bm{\Pi}} \, \mathrm{d}P_m \\[5pt]
	&= N \, \bm{\mathscr{n}} + N \, \bm{V} - Z \, \bm{\xi}
\end{aligned}
\quad \quad (\text{on } B\mathcal{E} \text{ or } B\mathcal{G}).
\label{eq:ParticleFluxB}
\end{equation}
This is similar to Eqs.~(\ref{eq:ParticleFluxE}) and (\ref{eq:ParticleFluxDecompositionE}) for $\bm{N}$ on $\mathcal{E}$, but with the extended inertia--momentum--kinetic-energy $\overleftarrow{\bm{\Pi}}$ given by Eq.~(\ref{eq:InertiaMomentumB}) replacing the inertia-momentum $\overleftarrow{\bm{P}}$ given by Eq.~(\ref{eq:InertiaMomentumE}).
Because $\overleftarrow{\bm{P}} = m \, \bm{U}$ and $\overleftarrow{\bm{\Pi}} = m \, \bm{\mathscr{U}}$, it is clear from Eq.~(\ref{eq:VelocityDecoding}) that $\bm{\mathscr{N}}$ is the Bargmann encoding of $\bm{N}$ and that the decoding relation
\begin{equation}
\bm{N} = \overleftarrow{\bm{g}} \cdot \bm{\mathscr{N}},
\quad\quad
N^I = {g^I}_A \, \mathscr{N}^A
\label{eq:NumberFluxDecoding}
\end{equation}
holds.
The particle number density $N = \underline{\bm{\chi}} \cdot \bm{\mathscr{N}}$ and physical 3-velocity $\bm{V} = N^{-1} \left( \overleftarrow{\bm{\gamma}} \cdot \bm{\mathscr{N}} \right)$ measured by fiducial observers have the same meaning here as in Eq.~(\ref{eq:DensityVelocityFiducial}) on $\mathcal{E}$.
The particle number density measured by fiducial observers is related to that measured by comoving observers by
\[
N = \left\{ \begin{aligned}
		\Lambda_{\bm{V}} \, n & & (\text{on } B\mathcal{E}) \\[5pt]
		n & & (\text{on } B\mathcal{G}),
	\end{aligned} \right.
\]
the relation on $B\mathcal{E}$ agreeing with Eq.~(\ref{eq:FiducialDensityComovingE}) on $\mathcal{E}$.
There is also an additional action component
\[
Z = \left\{ \begin{aligned} 
		\int \, f \, \frac{c^2 \, \epsilon_{\bm{p}}}{\mathscr{E}_{\bm{p}}} \,
			\frac{\mathrm{d}\bm{p}}{(2\pi\hbar)^3}	
			\quad\quad (\text{on } B\mathcal{E}), \\[5pt]
		\int \, f \, \frac{\epsilon_{\bm{p}}}{m} \,
			\frac{\mathrm{d}\bm{p}}{(2\pi\hbar)^3}	
			\quad\quad (\text{on } B\mathcal{G}), 
	\end{aligned} \right.
\quad \quad \quad
Z = - \underline{\bm{\mathscr{n}}} \cdot \bm{\mathscr{N}}
\quad \quad (\text{on } B\mathcal{E} \text{ or } B\mathcal{G}).
\]
The scalar field $Z$ is not a quantity encountered in fluid dynamics as traditionally formulated.
It will play an intermediate role in the 3-momentum equation below, but here it disappears from the particle conservation law on $B\mathcal{E}$ or $B\mathcal{G}$.
This is most easily seen in the direct computation
\[
\begin{aligned}
\bm{\mathscr{D}} \cdot \bm{\mathscr{N}} 
	&= \frac{1}{\sqrt{-G}} \, \frac{\partial}{\partial \mathscr{X}^A} 
		\left( \sqrt{-G} \, \mathscr{N}^A \right) \\[5pt]
	&= \frac{1}{\sqrt{-g}} \, \frac{\partial}{\partial X^\alpha} 
		\left( \sqrt{-g} \, N^\alpha \right) \\[5pt]
	&= \bm{\nabla} \cdot \bm{N} = 0.
\end{aligned}
\]
Note that Eqs.~(\ref{eq:MetricDeterminant4}) and (\ref{eq:MetricDeterminant5}) for the metric determinants $g$ and $G$, and the fact that partial derivatives with respect to action coordinate $\mathscr{X}^4$ vanish, have been employed.
Therefore 
\begin{equation}
\bm{\mathscr{D}} \cdot \bm{\mathscr{N}} = 0
\quad \quad (\text{on } B\mathcal{E} \text{ or } B\mathcal{G})
\label{eq:ParticleConservationSpacetimeB}
\end{equation}
is the spacetime particle conservation law on $B\mathcal{E}$ or $B\mathcal{G}$.
It is clear from the above derivation that its 1+3+1 decomposition is
\begin{equation}
 \frac{1}{\alpha \sqrt{\gamma}} \, \frac{\partial}{\partial t} \left( \sqrt{\gamma} \, N \right)
 	+ \frac{1}{\alpha} \, \bm{D} \cdot \left( N \left( \alpha \, \bm{V} - \bm{\beta} \right) \right)
	= 0
\quad \quad (\text{on } B\mathcal{E} \text{ or } B\mathcal{G}),
\label{eq:ParticleConservationB}
\end{equation}
precisely the same as Eq.~(\ref{eq:ParticleConservationE}) on $\mathcal{E}$.
As a warm-up for energy-momentum conservation, and by way of illustrating some examples of what happens in these sorts of calculations, consider an alternative derivation using the more formal procedure deduced in this paper for translating physics on $\mathcal{E}$ to physics on $B\mathcal{E}$, articulated in Sec.~\ref{sec:PhysicalLawsBargmann} and demonstrated in Sec.~\ref{sec:DynamicsParticles} in connection with particle dynamics.
Use the relation between $\bm{\mathscr{D}}$ and $\bm{\nabla}$ in Eq.~(\ref{eq:DerivativeProjectionE}), and recall from Eq.~(\ref{eq:NumberFluxDecoding}) that the vector field $\bm{N}$ is the decoding of $\bm{\mathscr{N}}$:
\[
0 = \bm{\nabla} \cdot \bm{N}
	= \overleftarrow{\bm{g}}^* \bm{\mathscr{D}} \cdot 
		\left( \overleftarrow{\bm{g}} \cdot \bm{\mathscr{N}} \right),
\] 
or
\[
\begin{aligned}
0 &= {g^B}_A \, 
		\mathscr{D}_B \left( {g^A}_C \, \mathscr{N}^C \right) \\[5pt]
	&= \left( {\delta^B}_A - c^2 \, \xi^B \, \xi_A \right) 
		\mathscr{D}_B \left( \left( {\delta^A}_C - c^2 \, \xi^A \, \xi_C \right)  \mathscr{N}^C \right).
\end{aligned}
\]
This gives four terms, only one of which survives.
The first is
\[
{\delta^B}_A \, \mathscr{D}_B \left( {\delta^A}_C \,  \mathscr{N}^C \right)
	= \mathscr{D}_A \,  \mathscr{N}^A.
\]
The second term is
\[
\begin{aligned}
- {\delta^B}_A \, \mathscr{D}_B \left( \xi^A \, \xi_C \, \mathscr{N}^C\right)
	&= - \mathscr{D}_A \left( \xi^A \, \xi_C \, \mathscr{N}^C \right) \\[5pt]
	&= - \, \xi_C \, \mathscr{N}^C \, \mathscr{D}_A \, \xi^A 
		- \xi^A \, \mathscr{D}_A \left( \, \xi_C \, \mathscr{N}^C \right) \\[5pt]
	&= 0,
\end{aligned}
\]
because $\mathscr{D}_A \, \xi^A = 0$ according to Eq.~(\ref{eq:ActionTraceB}) and $\xi^A \, \mathscr{D}_A \left( \, \xi_C \, \mathscr{N}^C \right) = -\partial_4 \left( \, \xi_C \, \mathscr{N}^C \right) = 0$.
The third term is
\[
\begin{aligned}
-c^2 \, \xi^B \, \xi_A \, \mathscr{D}_B \left( {\delta^A}_C \, \mathscr{N}^C \right)
	&=  - c^2 \, \xi^B \, \mathscr{D}_B \left( \xi_A \, \mathscr{N}^A \right)
		+ c^2 \, \mathscr{N}^A \, \xi^B \, \mathscr{D}_B \, \xi_A \\[5pt]
	&= 0
\end{aligned}
\]
because the first term is just a vanishing partial derivative with respect to $\mathscr{X}^4$, and $\mathscr{D}_B \, \xi_A$ in the second term has no $B = 4$ component according to Eq.~(\ref{eq:GradientActionCovectorB}).
The final term is
\[
\begin{aligned}
c^4 \, \xi^B \, \xi_A \, \mathscr{D}_B \left( \xi^A \, \xi_C \,  \mathscr{N}^C \right) 
	&= c^4 \, \xi^B \, \mathscr{D}_B \left( \xi_A \, \xi^A \, \xi_C \,  \mathscr{N}^C \right) 
		- c^2 \, \xi_C \,  \mathscr{N}^C \, \xi^B \, \xi^A \, \mathscr{D}_B \, \xi_A \\[5pt]
	&= 0
\end{aligned}
\]
because the first term is yet another vanishing partial derivative with respect to $\mathscr{X}^4$, and in the second term again $\mathscr{D}_B \, \xi_A$ has no $B = 4$ component (and is also antisymmetric, not to mention that also $\xi^A \, \mathscr{D}_B \, \xi_A = \mathscr{D}_B \left( \xi_A \, \xi^A \right) / 2 = 0$ since $\xi_A \, \xi^A = 1/c^2$, a constant).
The first term $\mathscr{D}_A \,  \mathscr{N}^A$ above being the only one that survives, Eq.~(\ref{eq:ParticleConservationSpacetimeB}) as the Bargmann spacetime encoding of particle number conservation is confirmed by this alternative calculation.

The fluid dynamics on $B\mathcal{E}$ or $B\mathcal{G}$ is given in terms of a kinetic-energy--momentum--mass-density $(1,1)$ tensor field
\begin{equation}
\begin{aligned}
\bm{\mathscr{T}} &= \int f \, \left( \overleftarrow{\bm{\Pi}} \otimes \bm{\Pi} \right) 
		\, \mathrm{d}P_m \\[5pt]
	&= - \ E \, \bm{\mathscr{n}} \otimes \underline{\bm{\chi}}
		\  +  \ \bm{\mathscr{n}} \otimes \bm{S}
		\  +  \ m \, N \, \bm{\mathscr{n}} \otimes \underline{\bm{\mathscr{n}}} \\[5pt]
	& \quad \ \ - \bm{Q} \otimes \underline{\bm{\chi}}
		\ \ - \ \quad \bm{\Sigma} \ \ \
		\ + \ m \, N \, \bm{V} \otimes \underline{\bm{\mathscr{n}}} \\[5pt]	
	& \quad \  + Y \, \bm{\xi} \otimes \underline{\bm{\chi}}	
		\ - \ \bm{\xi} \otimes \underline{\bm{Q}}
		\  -  \ m \, Z \, \bm{\xi} \otimes \underline{\bm{\mathscr{n}}}.
\end{aligned}
\label{eq:MomentumFluxB}
\end{equation}
This is similar to Eqs.~(\ref{eq:MomentumFluxE}) and (\ref{eq:EnergyMomentumFluxFiducialE}) for $\bm{T}$ on $\mathcal{E}$, but with the extended inertia--momentum--kinetic-energy $\overleftarrow{\bm{\Pi}}$ given by Eq.~(\ref{eq:InertiaMomentumB}) replacing the inertia-momentum $\overleftarrow{\bm{P}}$ given by Eq.~(\ref{eq:InertiaMomentumE}), and the kinetic-energy--momentum--mass $\bm{\Pi}$ of Eq.~(\ref{eq:EnergyMomentumB}) replacing the total-energy--momentum $\bm{P}$ of Eq.~(\ref{eq:EnergyMomentumE}).
Similar to $\bm{\mathscr{N}}$ being the Bargmann encoding of $\bm{N}$, so also here $\bm{\mathscr{T}}$ is the Bargmann encoding of $\bm{T}$, such that the decoding relation
\begin{equation}
\bm{T} = \overleftarrow{\bm{g}} \cdot \bm{\mathscr{T}} \cdot \overleftarrow{\bm{g}},
\quad\quad
{T^I}_J = {g^I}_A \, {\mathscr{T}^A}_B \, {g^B}_J
\label{eq:MomentumFluxDecoding}
\end{equation}
holds.
In addition to $N$ and $\bm{V}$ and $Z$ which have already appeared in the particle density 5-flux $\bm{\mathscr{N}}$, the 3-momentum density $\bm{S}$ and 3-stress $\bm{\Sigma}$ are precisely the same as in Eq.~(\ref{eq:MomentumFluxDecompositionE}) on $\mathcal{E}$; the latter are related to quantities measured by comoving observers by
\[
\begin{aligned}
\bm{S} &= \left\{ \begin{aligned}
			{\Lambda_{\bm{V}}}^2 \, 
			\left( m \, n \, + \frac{1}{c^2} \left( e + p \right) \right) \, \underline{\bm{V}}
			& & (\text{perfect fluid on } B\mathcal{E}), \\[5pt]
	m \, n \,  \underline{\bm{V}}
			& & (\text{perfect fluid on } B\mathcal{G}),
		\end{aligned} \right. 
\\[10pt]
\bm{\Sigma} &= \left\{ \begin{aligned}
			- {\Lambda_{\bm{V}}}^2 \,
				\left( m \, n \, + \frac{1}{c^2} \left( e + p \right) \right) \, \bm{V} \otimes \underline{\bm{V}}
				- p \, \overleftarrow{\bm{\gamma}}
			& & (\text{perfect fluid on } B\mathcal{E}), \\[5pt]
			- m \, n  \, \bm{V} \otimes \underline{\bm{V}}
				- p \, \overleftarrow{\bm{\gamma}}
			& & (\text{perfect fluid on } B\mathcal{G}), \\[5pt]
		\end{aligned} \right.
\end{aligned}
\]
agreeing with Eq.~(\ref{eq:FiducialEnergyMomentumB}) on $\mathcal{E}$ in the case of $B\mathcal{E}$.
But instead of the total energy density $\mathfrak{E}$ and 3-flux $\bm{\mathfrak{Q}}$ from Eq.~(\ref{eq:MomentumFluxDecompositionE}), the kinetic energy density $E$ and 3-flux $\bm{Q}$ already introduced in Eq.~(\ref{eq:KineticEnergyDensity}) appear naturally, the latter once as a vector $\bm{Q}$ in the combination $\bm{Q} \otimes \underline{\bm{\chi}}$ and once as a 1-form $\underline{\bm{Q}}$ in the combination $\bm{\xi} \otimes \underline{\bm{Q}}$, both of which will prove important.
These are related to quantities measured by a comoving observer by
\[
\begin{aligned}
E &= \left\{ \begin{aligned}
		{\Lambda_{\bm{V}}}^2 \, \left( e + p \right) - p  
		+ m \, n \, c^2 \, \Lambda_{\bm{V}} \left( \Lambda_{\bm{V}} - 1 \right) 
			& & (\text{perfect fluid on } B\mathcal{E}), \\[5pt]
		e + \frac{1}{2} \, m \, n \, \bm{\gamma}\left( \bm{V}, \bm{V} \right)
			& & (\text{perfect fluid on } B\mathcal{G}),
	\end{aligned} \right.
\\[10pt]
\bm{Q} &= \left\{ \begin{aligned}
	\left( {\Lambda_{\bm{V}}}^2 \, \left( e + p \right) 
		+ m  \, n \, c^2 \, \Lambda_{\bm{V}} \left( \Lambda_{\bm{V}} - 1 \right) \right) \, \bm{V}
			& & (\text{perfect fluid on } B\mathcal{E}), \\[5pt]
	\left( e + p + \frac{1}{2} \, m \, n \, \bm{\gamma}\left( \bm{V}, \bm{V} \right) \right) \, \bm{V}
			& & (\text{perfect fluid on } B\mathcal{G}),
	\end{aligned} \right.
\end{aligned}
\]
agreeing with Eqs.~(\ref{eq:FiducialKineticEnergyB}) and (\ref{eq:FiducialKineticEnergyFluxB}) on $\mathcal{E}$ in the case of $B\mathcal{E}$.
And just as a new quantity $Z$ was introduced above in connection with $\bm{\mathscr{N}}$, the new quantity
\[
Y = \left\{ \begin{aligned} 
		\int \, f \, \frac{c^2 \, {\epsilon_{\bm{p}}}^2 }{\mathscr{E}_{\bm{p}}} \,
			\frac{\mathrm{d}\bm{p}}{(2\pi\hbar)^3}	
			\quad\quad (\text{on } B\mathcal{E}), \\[5pt]
		\int \, f \, \frac{{\epsilon_{\bm{p}}}^2}{m} \,
			\frac{\mathrm{d}\bm{p}}{(2\pi\hbar)^3}	
			\quad\quad (\text{on } B\mathcal{G}), 
	\end{aligned} \right.
\quad \quad \quad
Y = \underline{\bm{\mathscr{n}}} \cdot \bm{\mathscr{T}} \cdot \bm{\mathscr{n}}
\quad \quad (\text{on } B\mathcal{E} \text{ or } B\mathcal{G})
\]
enters here.
As with $Z$, the scalar field $Y$ is not a quantity encountered in fluid dynamics as traditionally formulated.
Together with $Z$ it will turn out to play an intermediate role in the 3-momentum equation through the combination
\begin{equation}
E = 
	\left\{ \begin{aligned}
		\frac{1}{c^2} \, Y + m \, Z & & (\text{on } B\mathcal{E}), \\[5pt]
		m \, Z  & & (\text{on } B\mathcal{G}),
	\end{aligned} \right.
\label{eq:Y_Z_Combination}
\end{equation}
which follows from their definitions in terms of integrals over momentum space.
Finally, from the display of the 1+3+1 decomposition of $\bm{\mathscr{T}}$ in Eq.~(\ref{eq:MomentumFluxB}), it is apparent that the `time column' $- \bm{\mathscr{T}} \cdot \bm{\mathscr{n}}$ is a kinetic energy vector, the `position space columns' $\bm{\mathscr{T}} \cdot \overleftarrow{\bm{\gamma}}$ are 3-momentum vectors, and the `action column' 
\begin{equation}
\bm{\mathscr{T}} \cdot \bm{\xi} = m \, \bm{\mathscr{N}},
\quad \quad
{\mathscr{T}^I}_{\!A} \, \xi^A = m \, \mathscr{N}^I
\label{eq:MassFluxB}
\end{equation}
is a mass flux vector given in terms of the particle number flux $\bm{\mathscr{N}}$ already considered.

Consider next the spacetime law on $B\mathcal{E}$ satisfied by the kinetic-energy--momentum--mass-density tensor field $\bm{\mathscr{T}}$ obtained by reverse engineering the law on $\mathcal{E}$. 
Use again the relation between $\bm{\mathscr{D}}$ and $\bm{\nabla}$ in Eq.~(\ref{eq:DerivativeProjectionE}), and recall from Eq.~(\ref{eq:MomentumFluxDecoding}) that the $(1,1)$ tensor field $\bm{T}$ is the decoding of $\bm{\mathscr{T}}$:
\begin{equation}
0 = \bm{\nabla} \cdot \bm{T}
	= \overleftarrow{\bm{g}}^* \bm{\mathscr{D}} \cdot 
		\left( \overleftarrow{\bm{g}} \cdot \bm{\mathscr{T}} 
			\cdot \overleftarrow{\bm{g}} \right),
\label{eq:EnergyMomentumConservationDecoding}
\end{equation}
or
\[
\begin{aligned}
0 &= {g^B}_A \, 
		\mathscr{D}_B \left( {g^A}_C \, {\mathscr{T}^C}_D
			\, {g^D}_J \right) \\[5pt]
	&= \left( {\delta^B}_A - c^2 \, \xi^B \, \xi_A \right) 
		\mathscr{D}_B \left( \left( {\delta^A}_C - c^2 \, \xi^A \, \xi_C \right)  {\mathscr{T}^C}_D 
			\left( {\delta^D}_J - c^2 \, \xi^D \, \xi_J \right) \right).
\end{aligned}
\]
For temporary convenience write
\[
{\tilde{\mathscr{T}}^I}_{\ \ \ J} 
	= {\mathscr{T}^I}_A \,  {g^A}_J
	=  {\mathscr{T}^I}_A \left( {\delta^A}_J - c^2 \, \xi^A \, \xi_J \right)
	= {\mathscr{T}^I}_J - m c^2 \, \mathscr{N}^I \, \xi_J,
\]
noting the use of Eq.~(\ref{eq:MassFluxB}).
Then the above equation reads more compactly as
\[
0 = \left( {\delta^B}_A - c^2 \, \xi^B \, \xi_A \right) 
		\mathscr{D}_B \left( {\tilde{\mathscr{T}}^A}_{\ \ \ J} 
			- c^2 \, \xi^A \, \xi_C \, {\tilde{\mathscr{T}}^C}_{\ \ \ J} \right).
\]
Consider the two terms arising from the two terms inside the covariant derivative.
Take the second term first, as it ends up vanishing.
It is
\[
- c^2 \, \left( {\delta^B}_A - c^2 \, \xi^B \, \xi_A \right) 
		\mathscr{D}_B \left( \xi^A \, \xi_C \, {\tilde{\mathscr{T}}^C}_{\ \ \ J} \right).
\]
The first part of this second term is
\[
\begin{aligned}
- c^2 \, {\delta^B}_A \, \mathscr{D}_B \left( \xi^A \, \xi_C \, {\tilde{\mathscr{T}}^C}_{\ \ \ J} \right)
	&= - c^2 \, \xi^A \, \mathscr{D}_A \left( \xi_C \, {\tilde{\mathscr{T}}^C}_{\ \ \ J} \right) \\[5pt]
	&= c^2 \, \mathscr{D}_4 \left( \xi_C \, {\tilde{\mathscr{T}}^C}_{\ \ \ J} \right)
\end{aligned}
\]
where $\mathscr{D}_A \, \xi^A = 0$ was used in the first line.
The second part of the second term is
\[
\begin{aligned}
c^4 \, \xi^B \, \xi_A \, \mathscr{D}_B \left( \xi^A \, \xi_C \, {\tilde{\mathscr{T}}^C}_{\ \ \ J} \right)
	&= - c^4 \, \mathscr{D}_4 \left( \xi_A \, \xi^A \, \xi_C \, {\tilde{\mathscr{T}}^C}_{\ \ \ J} \right)
		+ c^4 \, \xi^A \, \xi_C \, {\tilde{\mathscr{T}}^C}_{\ \ \ J} \, \mathscr{D}_4 \, \xi_A \\[5pt]
	&= - c^2 \, \mathscr{D}_4 \left( \xi_C \, {\tilde{\mathscr{T}}^C}_{\ \ \ J} \right)
\end{aligned}
\]
because $\xi_A \, \xi^A = 1/c^2$ and $\mathscr{D}_4 \, \xi_A = 0$.
Thus these two parts of the second term sum to zero.
Return then to the first term:
\[
\left( {\delta^B}_A - c^2 \, \xi^B \, \xi_A \right) \mathscr{D}_B \, {\tilde{\mathscr{T}}^A}_{\ \ \ J}
	= {\delta^B}_A \, \mathscr{D}_B \, {\tilde{\mathscr{T}}^A}_{\ \ \ J}
		- c^2 \, \xi^B \, \xi_A \, \mathscr{D}_B \, {\tilde{\mathscr{T}}^A}_{\ \ \ J}.
\]
The first part of this first term is
\[
\begin{aligned}
 {\delta^B}_A \, \mathscr{D}_B \, {\tilde{\mathscr{T}}^A}_{\ \ \ J}
 	&= \mathscr{D}_A \left( {\mathscr{T}^A}_J - m c^2 \, \mathscr{N}^A \, \xi_J \right) \\[5pt]
	&= \mathscr{D}_A \, {\mathscr{T}^A}_J -  m c^2 \, \mathscr{N}^A \, \mathscr{D}_A \, \xi_J
\end{aligned}
\]
because $m$ is constant and $\mathscr{D}_A \, \mathscr{N}^A = 0$ (particle conservation).
The second part of the first term is
\[
\begin{aligned}
- c^2 \, \xi^B \, \xi_A \, \mathscr{D}_B \, {\tilde{\mathscr{T}}^A}_{\ \ \ J}
	&= c^2 \, \xi_A \, \mathscr{D}_4 \, {\tilde{\mathscr{T}}^A}_{\ \ \ J} \\[5pt]
	&= - c^2 \, \xi_A \, {^{\bm{G}} \! \Gamma^B}_{J 4} \, {\tilde{\mathscr{T}}^A}_{\ \ \ B} \\[5pt]
	&= c^2 \, \xi_A \,  {\tilde{\mathscr{T}}^A}_{\ \ \ B} \, \mathscr{D}_J \, \xi^B \\[5pt]
	&= m c^2 \, \mathscr{N}^B \,  \mathscr{D}_J \, \xi_B
\end{aligned}
\]
thanks to substitution for ${\tilde{\mathscr{T}}^A}_{\ \ \ B}$, the fact that $\xi_B \, \mathscr{D}_J \, \xi^B = 0$, the symmetry of $\mathscr{T}^{AB}$, and Eq.~(\ref{eq:MassFluxB}).
From the antisymmetry of $\mathscr{D}_J \, \xi_B$ in Eq.~(\ref{eq:GradientActionCovectorB}), the first and second parts of the first term combine so that
\[
0 = \mathscr{D}_A \, {\mathscr{T}^A}_J - 2 \, m c^2 \, \mathscr{N}^A \, \mathscr{D}_A \, \xi_J,
\]
or
\begin{equation}
\bm{\mathscr{D}} \cdot \bm{\mathscr{T}}
	= m c^2 \, \left( \bm{\mathscr{N}} \cdot \underline{\bm{a}} \right) \, \underline{\bm{\chi}} 
		- m c^2 \, \left( \bm{\mathscr{N}} \cdot \underline{\bm{\chi}} \right) \, \underline{\bm{a}},
\label{eq:FluidMotionBAcceleration}
\end{equation}
or even
\begin{equation}
\bm{\mathscr{D}} \cdot \bm{\mathscr{T}}
	= m \, \left( \bm{\mathscr{N}} \cdot \bm{D}\phi \right) \, \underline{\bm{\chi}} 
		- m \, \left( \bm{\mathscr{N}} \cdot \underline{\bm{\chi}} \right) \, \bm{D}\phi 
\quad \quad (\text{on } B\mathcal{E} \text{ or } B\mathcal{G}),
\label{eq:FluidMotionB}
\end{equation}
in which Eq.~(\ref{eq:StrongPotential}) renders the result suitable for $B\mathcal{G}$ as well as $B\mathcal{E}$.
This equation for fluid dynamics is a second example of a spacetime law featuring potentially strong gravity consistent with Galilei relativity.
As seen previously in the case of particle motion, there is in the Bargmann perspective an explicit gravitational source term on the right-hand side, not coincidentally of the same structure as that appearing in Eq.~(\ref{eq:ParticleMotionB}).

Details of the 1+3+1 decomposition of Eq.~(\ref{eq:FluidMotionB}) are given in Appendix~\ref{sec:DecompositionFluidB}.
Kinetic energy balance is obtained by contraction with $-\bm{\mathscr{n}}$, yielding
\begin{equation}
 \frac{1}{\alpha \sqrt{\gamma}} \, \frac{\partial}{\partial t} \left( \sqrt{\gamma} \, E \right)
 	+ \frac{1}{\alpha} \, \bm{D} \cdot \left( \alpha \, \bm{Q} - E \, \bm{\beta} \right)
	= - \overleftarrow{\bm{S}} \cdot \bm{D} \phi
		- \overrightarrow{\bm{\Sigma}} : \bm{K}
\quad \quad (\text{on } B\mathcal{E} \text{ or } B\mathcal{G}),
\label{eq:KineticEnergyBalanceB}
\end{equation}
precisely the same as Eq.~(\ref{eq:KineticEnergyBalanceE}), obtained somewhat more artificially on $\mathcal{E}$ where the balance of total energy in Eq.~(\ref{eq:TotalEnergyBalanceE}) is more natural.
Balance of 3-momentum is obtained by contraction with $\overleftarrow{\bm{\gamma}}$, resulting in
\begin{equation}
\begin{aligned}
\frac{1}{\alpha \sqrt{\gamma}} \, 
		\frac{\partial}{\partial t} \left( \sqrt{\gamma} \, S_j \right)
		&+ \frac{1}{\alpha} \, D_a \left( - \alpha \, {\Sigma^a}_j - \beta^a \, S_j \right) \\[5pt]
		&= \left\{ \begin{aligned}	
		- \left( m \, N + \frac{E}{c^2} \right)\, D_j \phi
		+ \frac{1}{\alpha} \, S_a \, D_j \beta^a
		& & (\text{on } B\mathcal{E}), \\[5pt]
		\phantom{ - \left( \frac{E}{c^2} + \right) } - m \, N \, D_j \phi
		+ \frac{1}{\alpha} \, S_a \, D_j \beta^a
		& & (\text{on } B\mathcal{G}).
	\end{aligned} \right.
\label{eq:MomentumBalanceB}
\end{aligned}
\end{equation}
Taking into account the decomposition of the total energy density $\mathfrak{E}$ in Eq.~(\ref{eq:TotalEnergySeparation}), this is precisely the same as Eq.~(\ref{eq:MomentumBalanceE}) on $\mathcal{E}$.
In the present case of a simple fluid composed of classical particles of constant mass $m$, the action component of Eq.~(\ref{eq:FluidMotionB}) obtained by contraction with $\bm{\xi}$ turns out to be a mass conservation law redundant with particle conservation.

\section{Conclusion}
\label{sec:Conclusion}

If consideration of the motion of a material particle in 3-dimensional position space led to Galilei relativity, and investigation of the propagation of light led to Poincar\'e relativity understood in the context of 4-dimensional spacetime, a retrospective reconsideration of the motion of a material particle leads to a more unified perspective on Poincar\'e and Galilei relativity on a 5-dimensional extended spacetime---Bargmann-Einstein spacetime $B\mathcal{E}$ in the case of Poincar\'e physics, and Bargmann-Galilei spacetime $B\mathcal{G}$ in the case of Galilei physics.
The extra dimension plays no independent physical role implying new degrees of freedom, but, being associated with the kinetic energy per unit mass of a material particle (see Eq.~(\ref{eq:FifthVelocityComponent})), enables Galilei physics to be expressed in terms of a spacetime tensor formalism that respects the separation of mass and kinetic energy. 
This paper builds on Paper~I by working this out in curved spacetime, extending the usual $1+3$ formulation of Poincar\'e general relativity on Einstein spacetime $\mathcal{E}$ to a $1+3+1$ setting suitable for both $B\mathcal{E}$ and $B\mathcal{G}$.
Indeed the basic strategy throughout is to translate known Poincar\'e physics on a curved 4-dimensional spacetime into a 5-dimensional setting, where the corresponding Galilei physics can be deduced by a $c \rightarrow \infty$ limit.

A prime benefit of this `Bargmann' perspective is that the geometry (here, including curvature) of both $B\mathcal{E}$ and $B\mathcal{G}$ is governed by a 5-metric $\bm{G}$ (see Eq.~(\ref{eq:Metric5})), conferring the several blessings a spacetime metric affords: metric duality of tensors, a Levi-Civita connection, and a Levi-Civita volume form. 
It is worth reiterating, however, that the forms $\bm{g}$ (for Poincar\'e physics) and $\bm{\tau}$ (for Galilei physics) that govern causality and the measurement of proper time in a 4-dimensional setting retain this role in the 5-dimensional setting.
That the 4-metric $\bm{g}$ on $\mathcal{E}$ governs not only spacetime geometry but causality and time measurement as well, while on $B\mathcal{E}$ and $B\mathcal{G}$ these responsibilities are divided between the 5-metric $\bm{G}$ and either $\bm{g}$ or $\bm{\tau}$ respectively, is one way in which the Bargmann approach requires Poincar\'e physics to `give something up' for the sake of a more unified perspective yielding greater insight into Galilei physics.

A foundational example of the way the Bargmann perspective enables a tensor formalism for Galilei physics is the unification of energy and momentum (and mass) in a covector or 1-form.
As noted in Paper~I, a 4-velocity, or (multiplying by particle mass) a vector 4-momentum in the form of inertia-momentum, is not a problem for Galilei physics.
But taking the metric dual (by $\bm{g}$) to obtain the covector or 1-form 4-momentum $\bm{P}$---the total-energy--momentum of Eq.~(\ref{eq:EnergyMomentumE})---is a feat of Poincar\'e physics that Galilei physics cannot replicate on a 4-dimensional spacetime.
The magic of the Bargmann approach is that taking the metric dual (by $\bm{G}$) of the inertia--momentum--kinetic-energy 5-vector $m \, \bm{\mathscr{U}}$ of Eq.~(\ref{eq:InertiaMomentumB}) yields the kinetic-energy--momentum--mass covector or 1-form $\bm{\Pi}$ of Eq.~(\ref{eq:EnergyMomentumB}) in which mass is disentangled from kinetic energy by removing it from the first component and moving it to the fifth component without a factor of $c^2$.
Of course, that the equivalence up to a factor of $c^2$ of inertia and total energy is no longer  manifest is a second way in which the Bargmann approach requires Poincar\'e physics to `give something up' for the sake of a more unified perspective yielding greater insight into Galilei physics.

While physical laws expressed in terms of tensor fields on spacetime embody relativistic invariance, comparison with experiment requires tensor decompositions consistent with the way humans experience time evolution in position space.
This is of course at the heart of the $1+3$ formalism on $\mathcal{E}$ and the $1+3+1$ formalism on $B\mathcal{E}$ or $B\mathcal{G}$ featuring a foliation of spacetime into position space leaves.
Prominent tensor fields associated with the foliation and tensor decomposition include the 4-velocity field $\bm{n}$ (on $\mathcal{E}$, see Eq.~(\ref{eq:FiducialObserverVectorE})) or 5-velocity field $\bm{\mathscr{n}}$ (on $B\mathcal{E}$ or $B\mathcal{G}$, see Eq.~(\ref{eq:FiducialObserverVectorB})) of fiducial observers, orthogonal to position space leaves in a timelike direction; and the 1-form $\underline{\bm{\chi}}$ (see Eqs.~(\ref{eq:DualFiducialObserverCovectorE}) and (\ref{eq:DualFiducialObserverCovectorB})) dual to these, in the sense that $\underline{\bm{\chi}} \cdot \bm{n} = 1$ and $\underline{\bm{\chi}} \cdot \bm{\mathscr{n}} = 1$.
An important difference between $\bm{n}$ on $\mathcal{E}$ and $\bm{\mathscr{n}}$ on $B\mathcal{E}$ or $B\mathcal{G}$ is manifest in their directional derivatives along themselves (Eqs.~(\ref{eq:FiducialAccelerationE}) and (\ref{eq:FiducialAccelerationB}) respectively): fiducial observers are accelerated on $\mathcal{E}$, while fiducial observers are geodesic on $B\mathcal{E}$ or $B\mathcal{G}$.
On $B\mathcal{E}$ or $B\mathcal{G}$ the vector field $\bm{\xi}$ (anti-)parallel to the new action coordinate axis (see Eq.~(\ref{eq:ActionVector})) also points away from the position space leaves,  satisfying $\underline{\bm{\mathscr{n}}} \cdot \bm{\xi} = 1$ but also $\underline{\bm{\chi}} \cdot \bm{\xi} = 0$.
These vector fields and 1-forms appear in the operator $\overleftarrow{\bm{\gamma}}$ (see Eqs.~(\ref{eq:ProjectionFiducialE}) and (\ref{eq:ProjectionFiducialB})) that projects vector fields and 1-forms to the position space leaves.
This appears in a generalized projection operator $\overleftarrow{\bm{\gamma}}^*$ for all tensors on $\mathcal{E}$ (see Eq.~(\ref{eq:GeneralProjectionOperatorE})) or on $B\mathcal{E}$ or $B\mathcal{G}$ (see Eq.~(\ref{eq:GeneralProjectionOperatorB})), including by relating spacetime gradients of tensors tangent to position space leaves to gradients tangent to the leaves (see Eq.~(\ref{eq:DerivativeProjection})).

In this projective relationship between tensor gradients on spacetime and tensor gradients on the position space leaves, the piece---or pieces, in the case of Bargmann spacetimes---that are projected out serve to define extrinsic curvature (see Eqs.~(\ref{eq:CompareDerivativesE}) and (\ref{eq:CompareDerivativesB})), which is also related to the gravitational `kinematics' referenced in the title of this paper.
The label `kinematics' refers to the fact that the standard 1+3 formalism of Poincar\'e general relativity locates the gravitational degrees of freedom in the 3-metric $\bm{\gamma}$ on the position space leaves, with the extrinsic curvature tensor $\bm{K}$ serving as a kind of `velocity' of these gravitational degrees of freedom (see Eq.~(\ref{eq:MetricEvolutionE})), a relationship confirmed in the 1+3+1 formalism undertaken here (see Eq.~(\ref{eq:MetricEvolutionB})).  
Meanwhile the evolution of the projection operator normal to the leaves vanishes (see Eqs.~(\ref{eq:ProjectionEvolutionE}) and (\ref{eq:ProjectionEvolutionB})), with the important consequence that tensors tangent to the leaves remain tangent to the leaves.
Turning back to geometry, extrinsic curvature tensors carry information about the way the leaves of a foliation are embedded in the ambient manifold, in that they describe the variation along the leaves of vector fields or 1-forms normal to the leaves; the resulting relations in Eq.~(\ref{eq:GradientsFiducialCovectorsE}) and Eqs.~(\ref{eq:GradientsFiducialCovectorsB}) and (\ref{eq:GradientActionCovectorB}) are key to the 1+3 and 1+3+1 decomposition of physical laws expressed in terms of spacetime tensor fields.
In the general case of a foliation of a manifold into leaves of codimension 2 one might expect two independent extrinsic curvature tensors corresponding to the two directions normal to the leaves.
Remarkably---or perhaps inevitably, in hindsight---even though the position space leaves of $B\mathcal{E}$ and $B\mathcal{G}$ are of codimension 2, the geometry of Bargmann spacetimes is constrained in such a way that the two extrinsic curvature tensors are essentially the same (related by a constant factor, see Eq.~(\ref{eq:ExtrinsicCurvatureRelationB})); and moreover, with $\bm{K}$ on $B\mathcal{E}$ or $B\mathcal{G}$ precisely matching $\bm{K}$ on $\mathcal{E}$ (compare Eqs.~(\ref{eq:ExtrinsicCurvatureComponentsB}) and (\ref{eq:ExtrinsicCurvatureMatrixB}) with Eq.~(\ref{eq:ExtrinsicCurvatureComponentsE})). 
The similarity of Eq.~(\ref{eq:CompareDerivativesB_2}) on $B\mathcal{E}$ or $B\mathcal{G}$ to Eq.~(\ref{eq:CompareDerivativesE_2}) on $\mathcal{E}$, in contrast to the apparent difference between Eqs.~(\ref{eq:CompareDerivativesB}) and (\ref{eq:CompareDerivativesE}), is one manifestation of there effectively being only one extrinsic curvature tensor. 
This outcome, and the associated consistency of gravitational kinematics on Bargmann spacetimes with that seen in the standard 1+3 formulation of Poincar\'e relativity, is one of the signal results of this paper.

The working out of gravitational `dynamics' in the context of the Bargmann spacetimes $B\mathcal{E}$ and $B\mathcal{G}$---the constraint equations, and the evolution of the extrinsic curvature $\bm{K}$, as these result from the Einstein equations on $\mathcal{E}$ encoded on $B\mathcal{E}$---is left for the next installment in this series.
\ADD{Therefore, without equations relating the components of $\bm{G}$ on $B\mathcal{G}$ to the energy-momentum-mass content thereon, comparisons of the conjectured strong-field Galilei general relativity with particular applications of Poincar\'e general relativity are not yet possible; such considerations are outside the scope of this paper, and are left for future work.}
 
But this paper prepares for that next step by proposing a procedure, described in Sec.~\ref{sec:PhysicalLawsBargmann}, through which known physical laws on $\mathcal{E}$ can be translated to $B\mathcal{E}$, which might be hoped or expected to be amenable to a $c \rightarrow \infty$ limit yielding physical laws on $B\mathcal{G}$.
The procedure involves a `decoding' operator $\overleftarrow{\bm{g}}$ (see Eq.~(\ref{eq:DecodingOperator})) and its generalization $\overleftarrow{\bm{g}}^*$ (see Eq.~(\ref{eq:GeneralProjectionOperatorBE})) relating tensors on $B\mathcal{E}$ to tensors on $\mathcal{E}$.
They are partly analogous to the projection operators $\overleftarrow{\bm{\gamma}}$ and $\overleftarrow{\bm{\gamma}}^*$ discussed previously, which project tensor fields on spacetime to the position space leaves.
The basic idea is to express a known physical law on $\mathcal{E}$ as the `decoding' of an expression on $B\mathcal{E}$, and then `reverse engineer' this expression to obtain the `encoding' of this physical law on $B\mathcal{E}$.
The term `decoding' rather than `projection' is coined because action components are not simply deleted in going from $B\mathcal{E}$ back to $\mathcal{E}$, but recombined with the time component---a reversal of the kind of separation effected in the Bargmann approach of, for instance, mass from total energy, or the lapse function from the 4-metric.

By way of example, and as an application of the $1+3+1$ curved Bargmann spacetime geometry proposed here, assuming that the lapse function $\alpha$, shift vector $\beta$, 3-metric $\bm{\gamma}$, and extrinsic curvature $\bm{K}$ associated with the 4-metric $\bm{g}$ and 5-metric $\bm{G}$ are given, this reverse engineering procedure is applied to physical laws for two systems: the dynamics of an elementary particle, and the dynamics of a simple fluid.

For the dynamics of a material particle, the physical law in terms of spacetime tensors can be expressed as Eq.~(\ref{eq:ParticleMotionE}) on $\mathcal{E}$ and by Eq.~(\ref{eq:ParticleMotionB}) on $B\mathcal{E}$ or $B\mathcal{G}$. 
On the right-hand side in the latter case, Eq.~(\ref{eq:StrongPotential}), inspired by (but by no means limited to) the weak-field limit, has been employed.  
The $1+3$ decomposition on $\mathcal{E}$ and the $1+3+1$ decomposition on $B\mathcal{E}$ give the same results (in a simpler formulation than that presented for instance in  \cite{Vincent201231-geodesic-equ}), confirming the physical equivalence of the Bargmann encoding of this dynamical law.  
It also gives sensible results for $B\mathcal{G}$ without restriction on the strength of the gravitational fields, although it appears that retaining nonlinear terms would require a limit in which also $G \rightarrow \infty$ as $c \rightarrow \infty$ in such a way that $G / c^2$ remains constant.
But despite matching results for the decomposed equations, the spacetime tensor laws of Eqs.~(\ref{eq:ParticleMotionE}) and (\ref{eq:ParticleMotionB}) reflect very different perspectives: 
in referencing $\underline{\bm{a}} = \bm{D}\phi = \bm{D} \ln \alpha / c^2$ on the right-hand side, the Bargmann-encoded Eq.~(\ref{eq:ParticleMotionB}) reverts the Einstein perspective of accelerated fiducial observers (see Eq.~(\ref{eq:FiducialAccelerationE})) and geodesic material particles (see Eq.~(\ref{eq:GeodesicE})) to a Newton-like perspective of geodesic fiducial observers (see Eq.~(\ref{eq:FiducialAccelerationB}))---analogous to Newton's inertial observers---and accelerated material particles subject to a gravitational force.
In the case of Poincar\'e physics, strictly speaking Eq.~(\ref{eq:ParticleMotionB}) should not be regarded as an invariant spacetime tensor law, since $\underline{\bm{a}}$ arises from the foliation which is supposed to be freely chosen by virtue of coordinate freedom.  
This would be a third way in which the Bargmann approach requires Poincar\'e physics to `give something up' for the sake of a more unified perspective yielding greater insight into Galilei physics.
But in the case of Galilei physics with expectations of absolute time, it would not be surprising for the foliation to turn out to be fixed, such that Eq.~(\ref{eq:ParticleMotionB}) actually is an invariant spacetime tensor law.

The other example concerns the dynamics of a simple fluid composed of a single type of microscopic particle of constant mass, with the 4-fluxes (on $\mathcal{E}$) or 5-fluxes (on $B\mathcal{E}$ or $B\mathcal{G}$) characterizing the fluid given by momentum moments of a scalar distribution function.
The particle number flux $\bm{N}$ on $\mathcal{E}$ (see Eq.~(\ref{eq:ParticleFluxE})) and $\bm{\mathscr{N}}$ on $B\mathcal{E}$ or $B\mathcal{G}$ (see Eq.~(\ref{eq:ParticleFluxB})), and the $1+3$ and $1+3+1$ decompositions (see Eqs.~(\ref{eq:ParticleConservationE}) and (\ref{eq:ParticleConservationB})) of their vanishing divergences (see Eqs.~(\ref{eq:ParticleConservationSpacetimeE}) and (\ref{eq:ParticleConservationSpacetimeB})), are essentially kinematical in nature, in that they define and describe the fluid and its motion. 
They basically are to the fluid what the definition of 4-velocity $\bm{U}$ (on $\mathcal{E}$) or 5-velocity $\bm{\mathscr{U}}$, and the trajectory relation (Eqs.~(\ref{eq:ParticleCoordinatesE}) and (\ref{eq:ParticleCoordinatesB})), are to a material particle; indeed, the particle flux defines the fluid 4-velocity or 5-velocity, the proportionality factor being the particle density measured by a comoving observer.
The fluid dynamics is given in terms of the 4-momentum flux $\bm{T}$ on $\mathcal{E}$ (see Eq.~(\ref{eq:MomentumFluxE})) and 5-momentum flux $\bm{\mathscr{T}}$ on $B\mathcal{E}$ or $B\mathcal{G}$ (see Eq.~(\ref{eq:MomentumFluxB})); the latter might be expected to appear in the encoding to $B\mathcal{E}$ of the Einstein equations on $\mathcal{E}$, to be considered in the next installment in this series.
The spacetime dynamical law---the divergence of these $(1,1)$ tensor fields, given by Eq.~(\ref{eq:FluidMotionE}) on $\mathcal{E}$ and by Eq.~(\ref{eq:FluidMotionB}) on $B\mathcal{E}$ or $B\mathcal{G}$---is analogous to Eqs.~(\ref{eq:ParticleMotionE}) and (\ref{eq:ParticleMotionB}) for a material particle mentioned previously.
Indeed the gravitational force appearing on the right-hand side of Eq.~(\ref{eq:FluidMotionB}) on $B\mathcal{E}$ or $B\mathcal{G}$ is like that for a material particle in Eq.~(\ref{eq:ParticleMotionB}).
And as was the case for a material particle, the $1+3$ decomposition on $\mathcal{E}$ and the $1+3+1$ decomposition on $B\mathcal{E}$ give the same results, again confirming the physical equivalence of the Bargmann encoding of this dynamical law.  

A few words are in order on how the approach taken here differs from or relates to previous work.
Previous work in a 5-dimensional spacetime setting has allowed weak-field Newton gravity to be incorporated into the Levi-Civita connection associated with a spacetime 5-metric $\bm{G}$ \cite{Duval1985Bargmann-struct,de-Saxce2016Galilean-Mechan,de-Saxce20175-Dimensional-T}.
In the schematic diagram in Fig.~1 of Paper~I the spacetime of that theory is denoted $B\mathcal{N}$; and based on a comparison of their differing 5-metrics, the spacetime $B\mathcal{G}$ being explored in this series is indicated in that figure to be qualitatively different.

More can now be said about an important distinction between the weak-field, linear gravitation of $B\mathcal{N}$ and the potentially strong-field, nonlinear gravitation of $B\mathcal{G}$.
There is freedom in the choice of spacetime connection (covariant derivative) when one generalizes from flat spacetime to curved spacetime. In the mathematical language of the reduction of a frame bundle, a spacetime symmetry group---here, the Poincar\'e group or Galilei group---acts `vertically' within each fiber of the frame bundle (relating bases of the spacetime tangent space at a single point of spacetime), while the connection acts `horizontally' (relating bases at neighboring spacetime points).
A natural choice is to constrain the connection by requiring that it be `compatible' with tensors that are invariant under the action of the symmetry group, in the sense of requiring that their covariant derivatives vanish. 
In the present case, the invariant tensors under the Bargmann-Poincar\'e and Bargmann-Galilei groups are the metric $\bm{G}$ and the action vector $\bm{\xi}$, so that compatibility for both would require $\bm{\mathscr{D}}\bm{G} = 0$ (the condition defining a Levi-Civita connection) and also $\bm{\mathscr{D}}\bm{\xi} = 0$.
This is the choice that leads to $B\mathcal{N}$.
The spacetime $B\mathcal{G}$ explored here also features a Levi-Civita connection ($\bm{\mathscr{D}}\bm{G} = 0$), but the 5-metric derived from particle kinematics in the 1+3 formulation of Poincar\'e relativity yields $\bm{\mathscr{D}}\bm{\xi} \ne 0$; see Eq.~(\ref{eq:GradientActionCovectorB}).
The conjecture here is that this relaxation of a `compatibility' requirement on $\bm{\xi}$ may allow a Levi-Civita connection in the 5-dimensional setting to embody strong-field Galilei gravitation.

This may turn out to be related to work generalizing standard weak-field Newton-Cartan Galilei gravitation on 4-dimensional spacetime to strong-field Galilei gravitation by allowing the connection---not a Levi-Civita connection, since there is no spacetime metric---to include torsion (e.g. \cite{Geracie2015Curved-non-rela,Van-den-Bleeken2017Torsional-Newto,van-den-Bleeken2022Torsional-Newto,Hansen2019Action-Principl,Cariglia2018General-theory-,Hansen2020Non-relativisti}).  
The Frobenius condition for a 4-dimensional spacetime to be foliated into a family of spacelike hypersurfaces is $\bm{\tau} \wedge \bm{\mathrm{d}}\bm{\tau} = 0$, where $\bm{\tau}$ is the time form mentioned above.
Standard weak-field Newton-Cartan theory uses a torsion-free connection, which happens to be equivalent to $\bm{\mathrm{d}}\bm{\tau} = 0$ \cite{Kunzle1972Galilei-and-Lor}. 
This is of course stronger than necessary to satisfy the Frobenius condition.
In the works cited above, the requirement that the connection be torsion-free is relaxed to allow so-called `twistless torsion' compatible with the Frobenius condition. 
Whether this is directly related to the present work is not explored here, but it is intriguing to note that the antisymmetry of the right-hand side of Eq.~(\ref{eq:GradientActionCovectorB}), together with the fact that $\bm{\chi} \rightarrow \bm{\xi}$ on $B\mathcal{G}$, gives Eq.~(\ref{eq:GradientActionCovectorB}) a rotational (and in that sense `torsional') character.

\ADD{As noted briefly in Sec.~\ref{sec:Introduction}, a strong-field Galilei general relativity could serve as a useful approximation in astrophysical scenarios such as core-collapse supernovae.
In this system gravity associated with the nascent neutron star is enhanced at the 10-20\% level by energy density and pressure, along with nonlinearity; perhaps this could be accommodated while enjoying the simplifications of setting aside `Minkowski' bulk fluid flow and the back-reaction of gravitational radiation.
(This approach might conceptualized as `microscopically Poincar\'e' but `macroscopically Galilei'.)
And while it remains to be seen what the equations governing curvature (the `Einstein equations') on $B\mathcal{G}$ turn out to be in the sequel to this paper, strong-field Galilei gravitation encoded in twistless torsion on 4-dimensional spacetime gives an indication of what might be expected.
For example, the $c \rightarrow \infty$ but strong-field formalism of \cite{Van-den-Bleeken2017Torsional-Newto} is argued in \cite{van-den-Bleeken2022Torsional-Newto} to be a low-speed expansion about the `static sector' of Poincar\'e general relativity, in effect a resummation with respect to gravitational strength of the usual (weak-field) post-Newtonian series.
While not expressed exactly this way in those works, inspection of the strong-field but `Galilei' reinterpretation of the Schwarzschild geometry in \cite{Van-den-Bleeken2017Torsional-Newto}, and the strong-field but `Galilei' expansion of the Kerr geometry and Oppenheimer-Snyder collapse in \cite{van-den-Bleeken2022Torsional-Newto}, shows that indeed they involve limits in which both $G \rightarrow \infty$ and $c \rightarrow \infty$ such that $G / c^2$ remains constant, as deduced here.}

\vspace{6pt} 

\funding{This work was supported by the U.S. Department of Energy, Office of Science, Office of Nuclear Physics under contract number DE-AC05-00OR22725.}

\dataavailability{This article describes entirely theoretical research. No new data were created or analyzed in this study. Data sharing is not applicable to this article.}



\acknowledgments{This manuscript has been authored by UT-Battelle, LLC, under contract DE-AC05-00OR22725 with the US Department of Energy (DOE). The US government retains and the publisher, by accepting the article for publication, acknowledges that the US government retains a nonexclusive, paid-up, irrevocable, worldwide license to publish or reproduce the published form of this manuscript, or allow others to do so, for US government purposes. DOE will provide public access to these results of federally sponsored research in accordance with the DOE Public Access Plan (\texttt{http://energy.gov/downloads/doe-public-access-plan}).}

\conflictsofinterest{The author declares no conflict of interest. The funders had no role in the design of the study; in the collection, analyses, or interpretation of data; in the writing of the manuscript, or in the decision to publish the results.}



%

\appendixtitles{yes} 
\appendixstart
\appendix

\section[\appendixname~\thesection]{Decomposition of Elementary Particle Dynamics}
\label{sec:DecompositionParticle}

\subsection[\appendixname~\thesubsection]{Decomposition of Elementary Particle Dynamics on $\mathcal{E}$}
\label{sec:DecompositionParticleE}

Consider first the energy equation, given by
\[
\begin{aligned}
0 &= - \left( \bm{\nabla}_{\!\bm{U}} \bm{P} \right) \cdot \bm{n} \\[5pt]
	&= - \bm{\nabla}_{\!\bm{U}} \left( \bm{P} \cdot \bm{n} \right)
		+ \overleftarrow{\bm{P}} \cdot  \bm{\nabla}_{\!\bm{U}} \underline{\bm{n}}.
\end{aligned}
\]
Address the first and second terms in turn.
Using Eq.~(\ref{eq:dtdtau}) for $\mathrm{d}t / \mathrm{d}\tau$ along the worldline in order to obtain evolution with respect to the coordinate $t$ native to the 1+3 foliation, the first term is
\[
\bm{\nabla}_{\!\bm{U}} \mathscr{E}_{\bm{p}}
	= \frac{\mathrm{d}  \mathscr{E}_{\bm{p}} }{\mathrm{d}\tau}
	= \frac{\Lambda_{\bm{V}}}{\alpha} \frac{\mathrm{d}  \mathscr{E}_{\bm{p}} }{\mathrm{d}t}
	= \frac{\mathscr{E}_{\bm{p}}}{m c^2} \frac{1}{\alpha} 
		\frac{\mathrm{d} \mathscr{E}_{\bm{p}} }{\mathrm{d}t}.
\]
Thanks to Eq.~(\ref{eq:GradientsFiducialCovectorsE}) for $\bm{\nabla} \underline{\bm{n}}$ and Eq.~(\ref{eq:AccelerationComponentsE}) for $\underline{\bm{a}}$, the second term in the energy equation is
\[
 \frac{1}{m} \, \overleftarrow{\bm{P}} \cdot \left( - \bm{K} 
	- \underline{\bm a} \otimes \underline{\bm{n}} \right) \cdot  \overleftarrow{\bm{P}}
		=  - \frac{1}{m} \bm{K}\left( \overleftarrow{\bm{p}}, \overleftarrow{\bm{p}} \right)
			+ \frac{\mathscr{E}_{\bm{p}}}{m} 
				\, \overleftarrow{\bm{p}} \cdot \bm{D} \ln \alpha, 
\]
noting that $\bm{K}$ and $\underline{\bm{a}} = \bm{D} \ln \alpha$ are tangent to $\mathcal{S}$.
Putting both terms together yields
\[
\frac{1}{\alpha} \frac{\mathrm{d} \mathscr{E}_{\bm{p}} }{\mathrm{d}t}
	= - c^2 \, \overleftarrow{\bm{p}} \cdot \bm{D} \ln \alpha
		+ \frac{c^2}{\mathscr{E}_{\bm{p}}} 
			\, \bm{K}\left( \overleftarrow{\bm{p}}, \overleftarrow{\bm{p}} \right)
\quad \quad (\text{on } \mathcal{E})
\]
for the evolution of the particle total energy as measured by the fiducial observers.

Turning to the momentum equation,
\[
\begin{aligned}
0 &=  \left( \bm{\nabla}_{\!\bm{U}} \bm{P} \right) \cdot \overleftarrow{\bm{\gamma}} \\[5pt]
	&= \bm{\nabla}_{\!\bm{U}} \left( \bm{P} \cdot \overleftarrow{\bm{\gamma}} \right)
		- \overleftarrow{\bm{P}} \cdot  \bm{\nabla}_{\!\bm{U}} \bm{\gamma} \\[5pt]
	&=  \bm{\nabla}_{\!\bm{U}} \bm{p} 
		+ \left( \overleftarrow{\bm{P}} \cdot  \bm{\nabla}_{\!\bm{U}} \underline{\bm{n}} \right) 				\underline{\bm{\chi}} 
		+  \left( \overleftarrow{\bm{P}} \cdot  \underline{\bm{n}} \right) 
			\bm{\nabla}_{\!\bm{U}} \underline{\bm{\chi}},
\end{aligned}
\]
thanks to Eq.~(\ref{eq:ExtendedThreeMetricE}) for $\bm{\gamma}$ and the fact that $\bm{\nabla}\bm{g} = 0$ for the Levi-Civita connection $\bm{\nabla}$.
The three position space components of this equation provide the information of interest; noting that $\chi_j = 0$ in the middle term, there remain
\[
0 =  \left( \bm{\nabla}_{\!\bm{U}} \bm{p} \right)_j
	- \mathscr{E}_{\bm{p}} \left( \bm{\nabla}_{\!\bm{U}} \underline{\bm{\chi}} \right)_j.
\]
Address the first and second terms in turn.
The first term is
\[
\begin{aligned}
\left( \bm{\nabla}_{\!\bm{U}} \bm{p} \right)_j
	&= U^\alpha \, \nabla_{\!\alpha} p_j \\[5pt]
	&= U^\alpha \, \partial_\alpha p_j 
		- U^\alpha \, {^{\bm{g}} \! {\Gamma^\beta}_{j\alpha}} \, p_\beta.
\end{aligned}
\]
Similar to what was seen in the energy equation above, the first part of the first term is simply
\[
U^\alpha \, \partial_\alpha p_j = \frac{\mathscr{E}_{\bm{p}}}{m c^2} \frac{1}{\alpha} 
		\frac{\mathrm{d} p_j }{\mathrm{d}t}.
\]
The second part of the first term is
\[
\begin{aligned}
- U^\alpha \, {^{\bm{g}} \! {\Gamma^\beta}_{j\alpha}} \, p_\beta
	&= - \frac{1}{m} \, \left( \frac{\mathscr{E}_{\bm{p}}}{c^2} \, n^\alpha + p^\alpha \right)
		 {^{\bm{g}} \! {\Gamma^\beta}_{j\alpha}} \, p_\beta \\[5pt]
	&= - \frac{\mathscr{E}_{\bm{p}}}{m c^2} 
			\left( \nabla_{\! j} n^\beta -  \partial_j n^\beta\right) \, p_\beta
	 	- \frac{1}{m} \, {^{\bm{g}} \! {\Gamma}_{\beta j \alpha}} \, p^\beta \, p^\alpha \\[5pt]
	&= \frac{\mathscr{E}_{\bm{p}}}{m c^2} \left( p^a \, K_{a j} 
								- \frac{1}{\alpha} \, p_a \, \partial_j \beta^a \right)
		+ \frac{1}{2 m} \, p_a \, p_b \, \partial_j \gamma^{a b}.
\end{aligned}
\]
As to the second term in the momentum component equation,
\[
\begin{aligned}
- \mathscr{E}_{\bm{p}} \left( \bm{\nabla}_{\!\bm{U}} \underline{\bm{\chi}} \right)_j
	&= -  \mathscr{E}_{\bm{p}} \, U^\alpha \, \nabla_{\!\alpha} \chi_j \\[5pt]
	&= \frac{\mathscr{E}_{\bm{p}}}{m c^2} 
		\left(  \mathscr{E}_{\bm{p}} \, \partial_j \ln\alpha - p^a \, K_{a j} \right).
\end{aligned}
\]
Combining the pieces of the first and second terms yields
\[
\frac{1}{\alpha} \frac{\mathrm{d} p_j }{\mathrm{d}t}
	= - \mathscr{E}_{\bm{p}} \, \frac{\partial \ln\alpha}{\partial x^j}  
		+ \frac{p_a}{\alpha} \, \frac{\partial \beta^a}{\partial x^j} 
		- \frac{c^2 \,  p_a \, p_b }{2 \, \mathscr{E}_{\bm{p}} } 
			\frac{\partial \gamma^{a b}}{\partial x^j}  
\quad \quad (\text{on } \mathcal{E})
\]
for the evolution of the particle 3-momentum components as measured by the fiducial observers.

\subsection[\appendixname~\thesubsection]{Decomposition of Elementary Particle Dynamics on $B\mathcal{E}$ and $B\mathcal{G}$}
\label{sec:DecompositionParticleB}

Begin with the energy equation obtained by contraction of Eq.~(\ref{eq:ParticleMotionB}) with $-\bm{\mathscr{n}}$:
\[
- \left( \bm{\mathscr{D}}_{\bm{\mathscr{U}}} \, \bm{\Pi} \right) \cdot \bm{\mathscr{n}} 
	= - \overleftarrow{\bm{p}} \cdot \bm{D}\phi,
\]
where only the first term of the right-hand side of Eq.~(\ref{eq:ParticleMotionB}) contributes.
On the left-hand side,
\[
\begin{aligned}
- \left( \bm{\mathscr{D}}_{\bm{\mathscr{U}}} \, \bm{\Pi} \right) \cdot \bm{\mathscr{n}} 
	&= - \bm{\mathscr{D}}_{\bm{\mathscr{U}}} \left( \bm{\Pi} \cdot \bm{\mathscr{n}} \right)
		+ \overleftarrow{\bm{\Pi}} 
			\cdot \bm{\mathscr{D}}_{\bm{\mathscr{U}}} \underline{\bm{\mathscr{n}}} \\[5pt]
	&= \left\{ \begin{aligned}
		\frac{\mathscr{E}_{\bm{p}}}{m c^2} \frac{1}{\alpha} 
			\frac{\mathrm{d} \epsilon_{\bm{p}} }{\mathrm{d}t} 
				- \frac{1}{m} \bm{K}\left( \overleftarrow{\bm{p}}, \overleftarrow{\bm{p}} \right)
				+ \frac{\epsilon_{\bm{p}}}{m c^2} \, \overleftarrow{\bm{p}} \cdot \bm{D}\phi
			& & (\text{on } B\mathcal{E}), \\[5pt]
		\phantom{ \frac{\mathscr{E}_{\bm{p}}}{m c^2} } \frac{1}{\alpha} 
			\frac{\mathrm{d} \epsilon_{\bm{p}} }{\mathrm{d}t}  
				- \frac{1}{m} \bm{K}\left( \overleftarrow{\bm{p}}, \overleftarrow{\bm{p}} \right)
				\phantom{ + \frac{\epsilon_{\bm{p}}}{m c^2}\, \overleftarrow{\bm{p}} \cdot \bm{D}\phi}
			\ \ & & (\text{on } B\mathcal{G}),
		\end{aligned} \right.
\end{aligned}
\]
in which Eq.~(\ref{eq:GradientsFiducialCovectorsB}) for $\bm{\mathscr{D}} \underline{\bm{\mathscr{n}}}$ has been employed, and on $B\mathcal{G}$ the $\epsilon_{\bm{p}}/m c^2$ term has been dropped after substitution for $a_j$ from Eq.~(\ref{eq:StrongPotential}).
Put the left- and right-hand sides together to find
\[
\frac{1}{\alpha} \frac{\mathrm{d} \epsilon_{\bm{p}} }{\mathrm{d}t}
	= \left\{ \begin{aligned}
		- \overleftarrow{\bm{p}} \cdot \bm{D} \phi
		+ \frac{c^2}{\mathscr{E}_{\bm{p}}} 
			\, \bm{K}\left( \overleftarrow{\bm{p}}, \overleftarrow{\bm{p}} \right)
		& & (\text{on } B\mathcal{E}), \\[5pt]
		- \overleftarrow{\bm{p}} \cdot \bm{D} \phi
		+ \frac{1}{m} 
			\, \bm{K}\left( \overleftarrow{\bm{p}}, \overleftarrow{\bm{p}} \right)
		& & (\text{on } B\mathcal{G}) 
	\end{aligned} \right.
\]
for the evolution of the particle kinetic energy as measured by the fiducial observers.

The momentum equation is obtained from the contraction of Eq.~(\ref{eq:ParticleMotionB}) with $\overleftarrow{\bm{\gamma}}$:
\[
\left( \bm{\mathscr{D}}_{\bm{\mathscr{U}}} \, \bm{\Pi} \right) \cdot \overleftarrow{\bm{\gamma}}
	= \left\{ \begin{aligned} 
		-  \frac{\mathscr{E}_{\bm{p}}}{m c^2} \, m \, \bm{D}\phi
		& & (\text{on } B\mathcal{E}), \\[5pt]		
		\phantom{ \frac{\mathscr{E}_{\bm{p}}}{m c^2} } - m \, \bm{D}\phi
		& & (\text{on } B\mathcal{G}),	
	\end{aligned} \right.
\]
where only the second term of the right-hand side of Eq.~(\ref{eq:ParticleMotionB}) contributes.
On the left-hand side,
\[
\begin{aligned}
\left( \bm{\mathscr{D}}_{\bm{\mathscr{U}}} \, \bm{\Pi} \right) \cdot \overleftarrow{\bm{\gamma}}\	&= \bm{\mathscr{D}}_{\bm{\mathscr{U}}} \left( 
	\bm{\Pi} \cdot \overleftarrow{\bm{\gamma}} \right)
		- \overleftarrow{\bm{\Pi}} 
			\cdot \bm{\mathscr{D}}_{\bm{\mathscr{U}}} \bm{\gamma} \\[5pt]
	&= \bm{\mathscr{D}}_{\bm{\mathscr{U}}} \bm{p}
		 + \left( \overleftarrow{\bm{\Pi}} 
			\cdot \bm{\mathscr{D}}_{\bm{\mathscr{U}}} \bm{\underline{\bm{\mathscr{n}}}} \right) 
				\underline{\bm{\chi}}
		 + \left( \overleftarrow{\bm{\Pi}} 
			\cdot  \underline{\bm{\mathscr{n}}} \right) 
				 \bm{\mathscr{D}}_{\bm{\mathscr{U}}} \bm{\underline{\bm{\chi}}}
		 + \left( \overleftarrow{\bm{\Pi}} 
			\cdot \bm{\mathscr{D}}_{\bm{\mathscr{U}}} \bm{\underline{\bm{\xi}}} \right) 
				\underline{\bm{\mathscr{n}}}
		 + \left( \overleftarrow{\bm{\Pi}} 
			\cdot  \underline{\bm{\xi}} \right) 
				 \bm{\mathscr{D}}_{\bm{\mathscr{U}}} \bm{\underline{\bm{\mathscr{n}}}}
\end{aligned}
\]
thanks to Eq.~(\ref{eq:ExtendedThreeMetricB}) for $\bm{\gamma}$ and the fact that $\bm{\mathscr{D}}\bm{G} = 0$ for the Levi-Civita connection $\bm{\mathscr{D}}$.
Because $\chi_j = 0$ and $\mathscr{n}_j = 0$, the only terms that contribute to the position space components of interest are
\[
\left( \overleftarrow{\bm{\gamma}} \cdot \bm{\mathscr{D}}_{\bm{\mathscr{U}}} \, \bm{\Pi} \right)_j
	=  \left( \bm{\mathscr{D}}_{\bm{\mathscr{U}}} \bm{p} \right)_j
	- \epsilon_{\bm{p}} \left( \bm{\mathscr{D}}_{\bm{\mathscr{U}}} \bm{\underline{\bm{\chi}}} \right)_j
	+ m	\left( \bm{\mathscr{D}}_{\bm{\mathscr{U}}} \bm{\underline{\bm{\mathscr{n}}}} \right)_j.
\]
New additional terms vanishing, the first term of the left-hand side ends up giving the same result as on $\mathcal{E}$ (in the case of $B\mathcal{E}$):
\[
 \left( \bm{\mathscr{D}}_{\bm{\mathscr{U}}} \bm{p} \right)_j
 	= \left\{ \begin{aligned}
	\frac{\mathscr{E}_{\bm{p}}}{m c^2} \left( \frac{1}{\alpha} \frac{\mathrm{d} p_j }{\mathrm{d}t}
		 +  p^a \, K_{a j} - \frac{1}{\alpha} \, p_a \, \partial_j \beta^a\right) 
		+ \frac{1}{2 m} \, p_a \, p_b \, \partial_j \gamma^{a b}
		& & (\text{on } B\mathcal{E}), \\[5pt]
	\phantom{ \frac{\mathscr{E}_{\bm{p}}}{m c^2}  }
		  \frac{1}{\alpha} \frac{\mathrm{d} p_j }{\mathrm{d}t}
		 + p^a \, K_{a j} - \frac{1}{\alpha} \, p_a \, \partial_j \beta^a \ \ \ 
		+ \frac{1}{2 m} \, p_a \, p_b \, \partial_j \gamma^{a b}
		& & (\text{on } B\mathcal{G}).
	\end{aligned} \right.
\]
The second and third terms of the left-hand side combine to give
\[
- \epsilon_{\bm{p}} \left( \bm{\mathscr{D}}_{\bm{\mathscr{U}}} \bm{\underline{\bm{\chi}}} \right)_j
+ m	\left( \bm{\mathscr{D}}_{\bm{\mathscr{U}}} \bm{\underline{\bm{\mathscr{n}}}} \right)_j
	= \left\{ \begin{aligned}
		- \frac{\mathscr{E}_{\bm{p}}}{m c^2} 
			\left(  p^a \, K_{a j} + \frac{\epsilon_{\bm{p}}}{c^2}  \right)
		& & (\text{on } B\mathcal{E}), \\[5pt]
		\phantom{ \frac{\mathscr{E}_{\bm{p}}}{m c^2} } - p^a \, K_{a j} 
		\phantom{   \frac{\epsilon_{\bm{p}}}{c^2} \, \partial_j \phi } 
		& & (\text{on } B\mathcal{G}). \\[5pt]
	\end{aligned} \right.
\]
Put the left- and right-hand sides together to find
\[
\frac{1}{\alpha} \frac{ \mathrm{d} p_j }{\mathrm{d}t}
	= \left\{ \begin{aligned}
		-  \frac{\mathscr{E}_{\bm{p}}}{c^2} \, \frac{\partial \phi}{\partial x^j}
		+ \frac{p_a}{\alpha} \, \frac{\partial \beta^a}{\partial x^j} 
		- \frac{c^2 \,  p_a \, p_b }{2 \, \mathscr{E}_{\bm{p}} } 
			\frac{\partial \gamma^{a b}}{\partial x^j}  
		& & (\text{on } B\mathcal{E}), \\[5pt]		
		 - m \, \frac{\partial \phi}{\partial x^j}
		+ \frac{p_a}{\alpha} \, \frac{\partial \beta^a}{\partial x^j} 
		- \quad \frac{ p_a \, p_b }{2 \, m } 
			 \frac{\partial \gamma^{a b}}{\partial x^j} 
		& & (\text{on } B\mathcal{G})	
	\end{aligned} \right.
\]
for the evolution of the particle 3-momentum components as measured by the fiducial observers.

\section[\appendixname~\thesection]{Decomposition of Simple Fluid Dynamics}
\label{sec:DecompositionFluid}

\subsection[\appendixname~\thesubsection]{Decomposition of Simple Fluid Dynamics on $\mathcal{E}$}
\label{sec:DecompositionFluidE}

Total energy balance is obtained by contraction of Eq.~(\ref{eq:FluidMotionE}) with $-\bm{n}$:
\[
\begin{aligned}
0 &= - \left( \nabla_{\!\alpha} {T^\alpha}_\beta \right) \, n^\beta \\[5pt]
	&= - \nabla_{\!\alpha} \left( {T^\alpha}_\beta \, n^\beta \right) 
		+ {T^\alpha}_\beta \, \nabla_{\!\alpha} n^\beta.
\end{aligned}
\]
The first term is
\[
- \nabla_{\!\alpha} \left( {T^\alpha}_\beta \, n^\beta \right)
	= \frac{1}{\alpha \sqrt{\gamma}} \, 
		\frac{\partial}{\partial t} \left( \sqrt{\gamma} \, \mathfrak{E} \right)
		+ \frac{1}{\alpha \sqrt{\gamma}} \, \frac{\partial}{\partial x^a} 
			\left( \sqrt{\gamma} 
				\, \left( \alpha \, \mathfrak{Q}^a - \mathfrak{E} \, \beta^a \right) \right)
\]
and the second term is
\[
{T^{\alpha \beta}} \, \nabla_{\!\alpha} n_\beta
	= \Sigma^{a b} \, K_{b a} +  S^a \, \frac{\partial \phi}{\partial x^a}.
\]
More geometrically,
\[
 \frac{1}{\alpha \sqrt{\gamma}} \, \frac{\partial}{\partial t} \left( \sqrt{\gamma} \, \mathfrak{E} \right)
 	+ \frac{1}{\alpha} \, \bm{D} \cdot \left( \alpha \, \bm{\mathfrak{Q}} - 
									\mbox{$\mathfrak{E}$} \, \bm{\beta} \right)
	= - \overleftarrow{\bm{S}} \cdot \bm{D} \phi
		- \overrightarrow{\bm{\Sigma}} : \bm{K}
\quad \quad (\text{on }\mathcal{E})
\]
expresses total energy balance on $\mathcal{E}$ in terms of the 3-divergence of a 3-flux tangent to the leaves $\mathcal{S}$ of the foliation.

The balance of 3-momentum is obtained by contraction of Eq.~(\ref{eq:FluidMotionE}) with $\overleftarrow{\bm{\gamma}}$:
\[
\begin{aligned}
0 &= \left( \nabla_{\!\alpha} {T^\alpha}_\beta \right) \, {\gamma^\beta}_\nu \\[5pt]
	&= \nabla_{\!\alpha} \left( {T^\alpha}_\beta \, {\gamma^\beta}_\nu \right) 
		- {T^\alpha}_\beta \, \nabla_{\!\alpha} {\gamma^\beta}_\nu.
\end{aligned}
\]
The position space components $\nu = j$ contain the information of interest.
The first term is
\[
\begin{aligned}
\nabla_{\!\alpha} \left( {T^\alpha}_\beta \, {\gamma^\beta}_j \right)
	&= \frac{1}{\alpha \sqrt{\gamma}} \, 
		\frac{\partial}{\partial t} \left( \sqrt{\gamma} \, S_j \right)
		+ \frac{1}{\alpha \sqrt{\gamma}} \, \frac{\partial}{\partial x^a} 
			\left( \sqrt{\gamma} 
				\, \left( - \alpha \, {\Sigma^a}_j - \beta^a \, S_j \right) \right) \\[5pt]
	& \quad - \, {^{\bm{g}} \! {\Gamma^\beta}_{j \alpha} } \, n^\alpha \, S_\beta
		+ \, {^{\bm{g}} \! {\Gamma^\beta}_{j \alpha} } \, {\Sigma^\alpha}_\beta.
\end{aligned}
\]
The first connection coefficient term is
\[
\begin{aligned}
- \, {^{\bm{g}} \! {\Gamma^\beta}_{j \alpha} } \, n^\alpha \, S_\beta
	&= - \left( \nabla_{\!\nu} n^\beta - \partial_\nu n^\beta \right) S_\beta \\[5pt]
	&= S^a \, K_{a j} - \frac{1}{\alpha} \, S_a \, \partial_j \beta^a.
\end{aligned}
\]
The second connection coefficient term is
\[
\begin{aligned}
{^{\bm{g}} \! {\Gamma^\beta}_{j \alpha} } \, {\Sigma^\alpha}_\beta
	&= {\Gamma}_{b j a} \, \Sigma^{a b} \\[5pt]
	&= \frac{1}{2} \, {\Sigma^a}_b \, \gamma^{b c} \, \partial_j \gamma_{c a}.
\end{aligned}
\]
The second term of the above 3-momentum balance equation is
\[
- T^{ \alpha \beta} \, \nabla_{\!\alpha} \gamma_{\beta j}
 = - S^a \, K_{a j} + \frac{\mathfrak{E}}{c^2} \, \partial_j \phi.
\]
Putting the pieces of the first and second terms of the 3-momentum balance equation together yields
\[
\begin{aligned}
\frac{1}{\alpha \sqrt{\gamma}} \, 
		\frac{\partial}{\partial t} \left( \sqrt{\gamma} \, S_j \right)
		 + \frac{1}{\alpha \sqrt{\gamma}} \, \frac{\partial}{\partial x^a} 
			& \left( \sqrt{\gamma} 
				\, \left( - \alpha \, {\Sigma^a}_j - \beta^a \, S_j \right) \right) \\[5pt]
	&=  - \frac{\mathfrak{E}}{c^2} \, \frac{\partial \phi}{\partial x^j}
		+ \frac{1}{\alpha} \, S_a \, \frac{\partial \beta^a}{\partial x^j}
		- \frac{1}{2} \, {\Sigma^a}_b \, \gamma^{b c} \, \frac{\partial \gamma_{c a}}{\partial x^j},
\end{aligned}
\]
expressed completely in terms of partial derivatives.
More geometrically,
\[
\frac{1}{\alpha \sqrt{\gamma}} \, 
		\frac{\partial}{\partial t} \left( \sqrt{\gamma} \, S_j \right)
		+ \frac{1}{\alpha} \, D_a \left( - \alpha \, {\Sigma^a}_j - \beta^a \, S_j \right)
		= - \frac{\mathfrak{E}}{c^2} \, D_j \phi + \frac{1}{\alpha} \, S_a \, D_j \beta^a
\quad \quad (\text{on }\mathcal{E})
\]
expresses 3-momentum balance on $\mathcal{E}$ in terms of the 3-divergence of a 3-flux tangent to the leaves $\mathcal{S}$ of the foliation.

\subsection[\appendixname~\thesubsection]{Decomposition of Simple Fluid Dynamics on $B\mathcal{E}$}
\label{sec:DecompositionFluidB}

Kinetic energy balance is obtained by contraction of Eq.~(\ref{eq:FluidMotionB}) with $-\bm{\mathscr{n}}$:
\[
- \left( \bm{\mathscr{D}} \cdot \bm{\mathscr{T}} \right) \cdot \bm{\mathscr{n}}
	= - m \, N \, \bm{V} \cdot \bm{D} \phi,
\]
where only the first term of the right-hand side of Eq.~(\ref{eq:FluidMotionB}) contributes. 
As to the left-hand side,
\[
\begin{aligned}
- \left( \bm{\mathscr{D}} \cdot \bm{\mathscr{T}} \right) \cdot \bm{\mathscr{n}}
	 &= - \left( \mathscr{D}_{A} {\mathscr{T}^A}_B \right) \, \mathscr{n}^B \\[5pt]
	&= - \mathscr{D}_{A} \left( {\mathscr{T}^A}_B \, \mathscr{n}^B \right) 
		+ {\mathscr{T}^A}_B \, \mathscr{D}_{A} \mathscr{n}^B.
\end{aligned}
\]
The first term on the left-hand side is
\[
- \mathscr{D}_{A} \left( {\mathscr{T}^A}_B \, \mathscr{n}^B \right) 
	= \frac{1}{\alpha \sqrt{\gamma}} \, 
		\frac{\partial}{\partial t} \left( \sqrt{\gamma} \, E \right)
		+ \frac{1}{\alpha \sqrt{\gamma}} \, \frac{\partial}{\partial x^a} 
			\left( \sqrt{\gamma} 
				\, \left( \alpha \, Q^a - E \, \beta^a \right) \right).
\]
In the second term on the left-hand side, it is interesting to compare $\bm{\mathscr{D}}\underline{\bm{\mathscr{n}}}$ in Eq.~(\ref{eq:GradientsFiducialCovectorsB}) on $B\mathcal{E}$ or $B\mathcal{G}$ with $\bm{\nabla}\underline{\bm{n}}$ in Eq.~(\ref{eq:GradientsFiducialCovectorsE}) on $\mathcal{E}$.
The former has two terms involving the acceleration $\underline{\bm{a}}$ whereas the latter only has one.
On $B\mathcal{E}$ or $B\mathcal{G}$, one of those acceleration terms contracts with the $\bm{Q} \otimes \underline{\bm{\chi}}$ term in Eq.~(\ref{eq:MomentumFluxB}) and the other contracts with the $\bm{\xi} \otimes \underline{\bm{Q}}$ term, each contributing equally.
The result for the second term on the left-hand side is
\[
{\mathscr{T}^{A B}} \, \mathscr{D}_{A} \mathscr{n}_B
	= \Sigma^{a b} \, K_{b a} +  \frac{1}{c^2} \, Q^a \, \frac{\partial \phi}{\partial x^a}.
\]
Putting the first and second terms of the left-hand side together with the right-hand side yields
\[
 \frac{1}{\alpha \sqrt{\gamma}} \, \frac{\partial}{\partial t} \left( \sqrt{\gamma} \, E \right)
 	+ \frac{1}{\alpha} \, \bm{D} \cdot \left( \alpha \, \bm{Q} - E \, \bm{\beta} \right)
	= - \overleftarrow{\bm{S}} \cdot \bm{D} \phi
		- \overrightarrow{\bm{\Sigma}} : \bm{K}
\quad \quad (\text{on } B\mathcal{E} \text{ or } B\mathcal{G}), 
\]
precisely the same as Eq.~(\ref{eq:KineticEnergyBalanceE}), obtained somewhat more artificially on $\mathcal{E}$ where the balance of total energy in Eq.~(\ref{eq:TotalEnergyBalanceE}) is more natural.
The contravariant 3-momentum $\overleftarrow{\bm{S}}$ emerges on the right-hand side thanks to Eq.~(\ref{eq:TotalEnergySeparation}) relating $\bm{Q}$ and $N \, \bm{V}$ to $\bm{\mathfrak{Q}}$, and Eq.~(\ref{eq:EnergyFluxMomentum}) relating $\bm{\mathfrak{Q}}$ to $\overleftarrow{\bm{S}}$. 

The balance of 3-momentum is obtained by contraction of Eq.~(\ref{eq:FluidMotionB}) with $\overleftarrow{\bm{\gamma}}$:
\[
\left( \bm{\mathscr{D}} \cdot \bm{\mathscr{T}} \right) \cdot \overleftarrow{\bm{\gamma}}
	= - m \, N \,  \bm{D} \phi,
\]
where only the second term of the right-hand side of Eq.~(\ref{eq:FluidMotionB}) contributes. 
As to the left-hand side,
\[
\begin{aligned}
\left( \left( \bm{\mathscr{D}} \cdot \bm{\mathscr{T}} \right) 
	\cdot \overleftarrow{\bm{\gamma}} \right)_J
	 &= \left( \mathscr{D}_{A} {\mathscr{T}^A}_B \right) \, {\gamma^B}_J \\[5pt]
	&= \mathscr{D}_{A} \left( {\mathscr{T}^A}_B \, {\gamma^B}_J \right) 
		- {\mathscr{T}^A}_B \, \mathscr{D}_{A} {\gamma^B}_J.
\end{aligned}
\]
The position space components $J = j$ contain the information of interest.
The first term of this left-hand side is
\[
\begin{aligned}
\mathscr{D}_{A} \left( {\mathscr{T}^A}_B \, {\gamma^B}_j \right)
	&= \frac{1}{\alpha \sqrt{\gamma}} \, 
		\frac{\partial}{\partial t} \left( \sqrt{\gamma} \, S_j \right)
		+ \frac{1}{\alpha \sqrt{\gamma}} \, \frac{\partial}{\partial x^a} 
			\left( \sqrt{\gamma} 
				\, \left( - \alpha \, {\Sigma^a}_j - \beta^a \, S_j \right) \right) \\[5pt]
	& \quad - \, {^{\bm{G}} \! {\Gamma^B}_{j A} } \, \mathscr{n}^A \, S_B
		+ \, {^{\bm{G}} \! {\Gamma^B}_{j A} } \, {\Sigma^A}_B
		+ \, {^{\bm{G}} \! {\Gamma^B}_{j A} } \, \xi^A \, Q_B.
\end{aligned}
\]
The first and second connection coefficient terms give the same results as seen in the analogous terms on $\mathcal{E}$, and the third connection coefficient term ends up vanishing: 
\[
{^{\bm{G}} \! {\Gamma^B}_{j A} } \, \xi^A \, Q_B = Q^B \, \mathscr{D}_j \xi_B = 0.
\]
The second term on the left-hand side of the above 3-momentum balance equation turns out to be
\[
- \mathscr{T}^{A B} \, \mathscr{D}_{A} \gamma_{B j}
 = - S^a \, K_{a j} + E \, a_j 
 = \left\{ \begin{aligned}
 	- S^a \, K_{a j} + \frac{E}{c^2} \, \partial_j \phi
		& & (\text{on } B\mathcal{E}), \\[5pt]
 	- S^a \, K_{a j} \phantom{ \ \ + \frac{E}{c^2} \, \partial_j \phi }
		& & (\text{on } B\mathcal{G}),
	\end{aligned} \right.
\]
where on $B\mathcal{G}$ the $E/c^2$ term has been dropped after substitution for $a_j$ from Eq.~(\ref{eq:StrongPotential}).
On $B\mathcal{E}$ this is like the corresponding term on $\mathcal{E}$, but with total energy density $\mathfrak{E}$ replaced here by kinetic energy density $E$. 
Half of this term is due to the combination of the scalar fields $Y$ and $Z$ shown in Eq.~(\ref{eq:Y_Z_Combination}).
Putting the pieces of the first and second terms of the left-hand side of the 3-momentum balance equation together with the right-hand side yields
\[
\begin{aligned}
\frac{1}{\alpha \sqrt{\gamma}} \, 
		\frac{\partial}{\partial t} \left( \sqrt{\gamma} \, S_j \right)
		 & + \frac{1}{\alpha \sqrt{\gamma}} \, \frac{\partial}{\partial x^a} 
			 \left( \sqrt{\gamma} 
				\, \left( - \alpha \, {\Sigma^a}_j - \beta^a \, S_j \right) \right) \\[5pt]
	&=  \left\{ \begin{aligned}	
		- \left( m \, N + \frac{E}{c^2} \right)\, \frac{\partial \phi}{\partial x^j}
		+ \frac{1}{\alpha} \, S_a \, \frac{\partial \beta^a}{\partial x^j}
		- \frac{1}{2} \, {\Sigma^a}_b \, \gamma^{b c} \, \frac{\partial \gamma_{c a}}{\partial x^j}
		& & (\text{on } B\mathcal{E}), \\[5pt]
		\phantom{ - \left( \frac{E}{c^2} + \right) } - m \, N \, \frac{\partial \phi}{\partial x^j}
		+ \frac{1}{\alpha} \, S_a \, \frac{\partial \beta^a}{\partial x^j}
		- \frac{1}{2} \, {\Sigma^a}_b \, \gamma^{b c} \, \frac{\partial \gamma_{c a}}{\partial x^j}
		& & (\text{on } B\mathcal{G})
	\end{aligned} \right.
\end{aligned}
\]
expressed completely in terms of partial derivatives.
More geometrically,
\[
\begin{aligned}
\frac{1}{\alpha \sqrt{\gamma}} \, 
		\frac{\partial}{\partial t} \left( \sqrt{\gamma} \, S_j \right)
		&+ \frac{1}{\alpha} \, D_a \left( - \alpha \, {\Sigma^a}_j - \beta^a \, S_j \right) \\[5pt]
		&= \left\{ \begin{aligned}	
		- \left( m \, N + \frac{E}{c^2} \right)\, D_j \phi
		+ \frac{1}{\alpha} \, S_a \, D_j \beta^a
		& & (\text{on } B\mathcal{E}), \\[5pt]
		\phantom{ - \left( \frac{E}{c^2} + \right) } - m \, N \, D_j \phi
		+ \frac{1}{\alpha} \, S_a \, D_j \beta^a
		& & (\text{on } B\mathcal{G})
	\end{aligned} \right.
\end{aligned}
\]
expresses 3-momentum balance on $B\mathcal{E}$ or $B\mathcal{E}$ in terms of the 3-divergence of a 3-flux tangent to the leaves $\mathcal{S}$ of the foliation.
Taking into account the decomposition of the total energy density $\mathfrak{E}$ in Eq.~(\ref{eq:TotalEnergySeparation}), this is precisely the same as Eq.~(\ref{eq:MomentumBalanceE}) on $\mathcal{E}$.

Finally, Eq.~(\ref{eq:FluidMotionB}) contains an additional component, the action component obtained by contraction with $\bm{\xi}$.
As was the case with elementary particle motion, for constant particle mass $m$ the action component yields $0 = 0$.
The contraction of Eq.~(\ref{eq:FluidMotionB}) with $\bm{\xi}$ yields
\[
\left( \bm{\mathscr{D}} \cdot \bm{\mathscr{T}} \right) \cdot \bm{\xi}
	= 0,
\]
where the right-hand side of Eq.~(\ref{eq:FluidMotionB}) makes a vanishing contribution.
As to the left-hand side,
\[
\left( \bm{\mathscr{D}} \cdot \bm{\mathscr{T}} \right) \cdot \bm{\xi}
	= \bm{\mathscr{D}} \cdot \left( \bm{\mathscr{T}} \cdot \bm{\xi} \right)
		- \overrightarrow{\bm{\mathscr{T}}} : \bm{\mathscr{D}} \underline{\bm{\xi}}.
\]
The first term on the left-hand side is
\[
\bm{\mathscr{D}} \cdot \left( \bm{\mathscr{T}} \cdot \bm{\xi} \right)
	= \bm{\mathscr{D}} \cdot \left( m \, \bm{\mathscr{N}} \right)
	= m \, \bm{\mathscr{D}} \cdot \bm{\mathscr{N}}
	= 0
\]
thanks to constant $m$, and particle conservation.
Along with this, the second term on the left-hand side also vanishes:
\[
- \overrightarrow{\bm{\mathscr{T}}} : \bm{\mathscr{D}} \underline{\bm{\xi}}
	= - \mathscr{T}^{A B} \, \mathscr{D}_{A} \xi_B = 0
\]
because $\overrightarrow{\bm{\mathscr{T}}}$ is symmetric and $\bm{\mathscr{D}} \underline{\bm{\xi}}$ is antisymmetric.
Thus, in the present case of a simple fluid composed of classical particles of constant mass $m$, the action component of Eq.~(\ref{eq:FluidMotionB}) turns out to be a mass conservation law redundant with particle conservation.

%
%

\begin{adjustwidth}{-\extralength}{0cm}

\reftitle{References}


\def\prd{Phys. Rev. D}

\PublishersNote{}
\end{adjustwidth}
\end{document}